\documentclass[prd,preprintnumbers,aps]{revtex4-2}
\usepackage{graphicx}% Include figure files
\usepackage{dcolumn}% Align table columns on decimal point
\usepackage{bm}% bold math
\usepackage{epsfig,amsmath}
\usepackage{amssymb}
\usepackage{subfigure}
\usepackage{float}
\usepackage{url}
\usepackage{ulem}
\usepackage[usenames,dvipsnames]{color}
\usepackage[labelfont=bf]{caption}
\usepackage[pagebackref=false, colorlinks=true]{hyperref}

\definecolor{redish}{rgb}{0.7,0.2,0.0}  % color defined in (r=red,g=green,b=blue) model
\definecolor{bluish}{rgb}{0.2,0.5,0.8}

\hypersetup{linkcolor=redish,  % color of internal links
  citecolor=blue,% color of links to bibliography
  filecolor=magenta,  % color of file links
  urlcolor=bluish}  % color of the url

\DeclareFontFamily{U}{rsfs}{} % Formal Script%
\DeclareFontShape{U}{rsfs}{m}{n}{<5> rsfs5 <6><7> rsfs7  %
  <8><9><10><10.95><12><14.4><17.28><20.74><24.88> rsfs10}{} %
\DeclareMathAlphabet{\mathfs}{U}{rsfs}{m}{n}

\def \O{\Omega}

\def \f{\frac}

\def \O{\Omega}

\def \S{\Sigma}

\def \th{\theta}
\def \c{\chi}

\begin{document}

\title{Tilted thin accretion disks in the full Kerr spacetime and their implications}
\author{K. S. Sruthy}
\email{sruthy.mcnsmpl2023@learner.manipal.edu}
\affiliation{Manipal Centre for Natural Sciences, Manipal Academy of Higher Education, Manipal 576104, India}

\author{Chandrachur Chakraborty}
\email{chandrachur.c@manipal.edu}
\affiliation{Manipal Centre for Natural Sciences, Manipal Academy of Higher Education, Manipal 576104, India}

\author{Sudip Bhattacharyya}
\email{sudip@tifr.res.in}
\affiliation{Department of Astronomy and Astrophysics,
Tata Institute of Fundamental Research, Mumbai 400005, India}
\affiliation{MIT Kavli Institute for Astrophysics and Space Research,
Massachusetts Institute of Technology, Cambridge, Massachusetts, 02139, USA}

\begin{abstract}
We derive a steady-state warped-disk equation in the full Kerr spacetime to study the tilt dynamics of a thin, viscous accretion disk around a spinning collapsed object. The formulation, based on Pringle's framework, remains valid for all values of the Kerr parameter $a$, thereby encompassing both Kerr black holes (BHs; $0 < a \le 1$) and Kerr naked singularities ($a > 1$). By incorporating the exact Keplerian and Lense-Thirring precession frequencies, we analytically obtain the radial tilt equations of the disk without invoking slow-spin or weak-field approximations. Numerical solutions of the resulting equations, obtained under realistic boundary conditions, reveal significant deviations from slow-spin approximations, particularly in the inner disk where the relativistic effects dominate. In the diffusive regime, we find that for Kerr naked singularities the tilt profile exhibits distinct inner hump(s) near the radius where the specific angular momentum vanishes---a feature absent for Kerr BHs.
Consideration of a tilt in the inner disk could significantly influence the interpretations from observed X-ray spectral, timing, and polarization features, which are crucial to probe the strong gravity regime and to infer the spin of the central object. While such a distinct hump feature alone does not uniquely distinguish Kerr BHs from Kerr naked singularities, their interpretation in conjunction with constraints on the disk regime may provide a potential observational handle on the nature of the accreting collapsed object.
\end{abstract}

\maketitle
\section{Introduction}
Soon after Einstein proposed General Relativity, Lense and Thirring suggested that a rotating spacetime drags the local inertial frames associated with it \cite{Lense_1918}. As a result, the orbital plane of a test particle placed around a Kerr black hole (BH) precesses with a frequency known as the Lense-Thirring (LT) precession (or nodal plane precession) frequency.
Later, Bardeen and Petterson proposed that if the accretion disk around a spinning BH is tilted, the LT effect causes the inner part of the disk to align with the BH’s equatorial plane, while the outer disk remains misaligned. This phenomenon is known as the Bardeen–Petterson (BP) effect \cite{Bardeen_J_M_1975}. The effect arises from the interplay between LT torque and viscous torque within the disk. Together with radiative torques \cite{Pringle_1996} and tidal interaction effects due to a binary companion \cite{Larwood_1996}, the BP effect provides a theoretical framework for understanding warped accretion disks around astrophysical compact objects.

Numerous studies have been conducted to understand the physics governing warped accretion disks around spinning BHs. Bardeen and Petterson \cite{Bardeen_J_M_1975}, along with subsequent works \cite{Petterson_J_A_1977, Petterson_J_A_1977_1, Hatchett_S_P_1981}, modeled disk viscosity using the Shakura-Sunyaev prescription. However, these studies are limited by their incomplete treatment of angular momentum conservation. The first self-consistent analysis of a thin, warped accretion disk was carried out by Papaloizou and Pringle \cite{Papaloizou_1983}, who introduced a formalism incorporating two distinct viscosity parameters associated with azimuthal and vertical shear. Then Pringle obtained a general equation for a warped disk surrounding a compact object \cite{Pringle_1992}, which is valid for arbitrary values of warp and viscosity. He also considered the BP effect as a plausible mechanism of the disk's tilt, and investigated its profile resulting from the LT effect.

Building on the earlier work of Papaloizou \cite{Papaloizou_1983}, Ogilvie \cite{Ogilvie_1999} extended the analysis of warped accretion disks into the nonlinear regime. In a subsequent study, Scheuer and Feiler \cite{Scheuer_P_A_G_1996} derived the radial profile in the steady-state limit and used it to estimate the alignment timescale of the BH. They demonstrated that, during the alignment process driven by the BP effect, the BH precesses about the angular momentum direction of the outer, tilted accretion disk,  and that the precession timescale is equal to the alignment timescale. Notably, this study did not account for the contribution of the inner disk region. Following this, Natarajan and Pringle \cite{P_Natarajan_1999} estimated the alignment timescale using parameters appropriate for active galactic nuclei (AGNs), and introduced radial dependence in the viscosity coefficients. Subsequently, Martin et al. \cite{F_G_Martin_2007} derived the steady-state profile of a warped accretion disk,  incorporating power-law variations in both viscosity and surface density. Their results indicated that, under such conditions, the precession time scale is greater than the alignment time scale. Later, Chen et al. \cite{Chen_L_2009} obtained an analytical solution to the warped disk equation, treating the two viscosity parameters as independent power-law functions of radius. It is important to note that all of these studies neglected the contribution of the inner disk region, assuming it to be aligned due to the action of the BP effect.

There exist earlier works on misaligned inner accretion disks in the context of a thick disk or low viscous disk ($\alpha<H/R$), in which warping disturbances are transmitted through waves (not diffusively as in a thin viscous disk) \cite{Lubow_2002, Nealon_2015, Demianski_1997, Ivanov_1997, Drewes_2021}. It was demonstrated that an accretion disk in that regime need not align with the BH equatorial plane at small radii, and that discontinuities in the tilt profile can propagate both inward and outward in a wave-like manner \cite{Lubow_2002}. These analytical predictions were later confirmed by high-resolution three-dimensional smoothed particle hydrodynamics simulations \cite{Nealon_2015}, which reproduced the steady-tilt oscillations reported in earlier works \cite{Demianski_1997, Ivanov_1997}. More recently, Drewes and Nixon \cite{Drewes_2021} showed that such oscillatory tilt structures can persist over long timescales in global three-dimensional hydrodynamical simulations. In contrast, the present work is restricted entirely to the diffusive regime ($\alpha>H/R$), where warp evolution is governed by viscous diffusion and is generally expected to be monotonic \cite{Banerjee_2019, Banerjee_2019_1, Sen_2024}. The opposite regime $\alpha < H/R$, in which warping disturbances propagate as bending waves \cite{Lubow_2002, Ivanov_1997}, lies outside the scope of the present work \cite{Banerjee_2019}.

In between these, Lodato and Pringle \cite{Lodato_G_2006} adopted a numerical approach, based on the method proposed by Pringle \cite{Pringle_1992} to solve the time-dependent warped disk equation. Their results showed deviations from those obtained in \cite{Scheuer_P_A_G_1996} in certain parameter regimes, suggesting the possibility of misalignment in the inner accretion disk. Subsequently, Zhuravlev et al. \cite{Zhuravlev2014} investigated the dynamics of tilted accretion disks  using numerical methods and three-dimensional general relativistic magnetohydrodynamic (GRMHD) simulations. They reported a partial misalignment in disks undergoing retrograde motion. A global relativistic simulation incorporating magneto-rotational turbulence as the primary mechanism for angular momentum transport was carried out in \cite{Fragile_P_C_2007}. Their results indicated that the disk remains tilted throughout the simulation, suggesting the absence of the BP effect. The first study to include the self-gravity of the disk in a tilted BH–torus system was performed by \cite{Mewes_V_2016}. This study confirmed significant differential rotation induced by the LT effect, though no evidence of the BP effect was observed. The first GRMHD simulation of a thin accretion disk ($H/R \approx 0.015-0.05$) in the viscous regime ($\alpha > H/R $) around a rapidly spinning BH ($a \approx 0.9$) with a tilt angle of $45$ degrees to $65$ degrees was conducted in \cite{Liska2021}. Their results demonstrate that, under certain conditions, the BP effect can occur in thin disks. They also found that the alignment radius increases as the disk aspect ratio ($H/R$) decreases. Later, the work by \cite{Liska2019} showed that the inner part of the accretion disk $(r \lesssim 5r_g)$ aligns with the BH equator for $a = 0.9375$ and $H/R \lesssim 0.03$, exhibiting both BP-driven and global alignment.
 Furthermore, studies of quasiperiodic oscillations (QPOs) from the Galactic accreting BH H1743--322~\cite{Ingram_A_2016, Ingram_A_2016_1} suggest the presence of a tilted, precessing inner accretion flow within a truncated disk geometry. Such a configuration can remain consistent with the BP picture provided that the truncation radius exceeds the alignment radius and the inner flow satisfies \(\alpha < H/R \).

An analytical solution to the full steady-state warped disk equation—accounting for contributions from the inner accretion disk—was first obtained in \cite{Chakraborty_C_2017}, using the formalism developed by Scheuer and Feiler \cite{Scheuer_P_A_G_1996}. Subsequently, \cite{Banerjee_2019} applied Pringle’s formalism \cite{Pringle_1992} to solve the warped disk equation for a prograde accretion disk around a slowly spinning Kerr BH ($a \ll 1$), without making any a priori assumption regarding the alignment of the inner accretion disk.  In their analysis, the LT precession frequency ($\Omega_{\rm LT}$) was taken as $\O_{\rm LT} \approx {2acR_g^2}/{R^3}$, where $R$ is the radial distance from the central object of gravitational radius $R_g$. Their results indicate that the inner disk can maintain a nonzero tilt over a broad range of parameter values, including the spin parameter $a$, BH mass $M$, and vertical viscosity parameter $\nu_2$.
While the expression $\O_{\rm LT} \approx {2acR_g^2}/{R^3}$ is strictly valid for slowly spinning BHs, it has been widely used in the literature even for high-spin Kerr BHs \cite{Banerjee_2019, Caproni_2006, Fragile_P_C_2007, Fragile_2005, Martin_2008_1, P_Natarajan_1999, Banerjee_2019_1, Nelson_2000, Li_2015, Li_2013}. This is because the approximation ($\O_{\rm LT} \approx {2acR_g^2}/{R^3}$) can remain valid in regimes where $a \ll R/R_g$ holds true, as noted in \cite{Sen_2024}. Therefore, even for rapidly spinning BHs, $\O_{\rm LT} \approx {2acR_g^2}/{R^3}$ could still be applicable at large radial distances, where the ratio $a/(R/R_g)$ remains negligibly small.
However, our primary interest lies in the innermost regions of the accretion disk around a Kerr collapsed object of arbitrary spin parameter ($a$), where the gravitational field is significantly stronger and the condition $a \ll R/R_g$ is no longer valid. In such cases, the slow-spin approximation breaks down, and an exact treatment becomes essential to accurately capture the disk dynamics near the central collapsed object—be it a Kerr BH (\( 0 < a \leq 1 \)) or a Kerr naked singularity (NaS . \( a > 1 \)) \cite{CKJ, Chakraborty_C_2017_1}. The present work constitutes a natural extension of our earlier series of studies \cite{ Banerjee_2019, Banerjee_2019_1, Sen_2024}, in which we developed the theoretical framework for warped accretion disks--accounting for contributions from the inner accretion disk--in the diffusive regime and analyzed tilt profiles and associated timescales under the slow-rotation and weak-field approximations to the LT effect. The current work advances this program by employing the exact formulation in the full Kerr spacetime, thereby capturing strong-field effects that are not accessible within the approximate treatments. According to Penrose's cosmic censorship hypothesis, gravitational singularities are expected to remain concealed within event horizons. However, subsequent theoretical investigations have revealed the intriguing possibility that NaSs might exist in nature (see~\cite{Chakraborty_C_2017_1, GH, CKJ, JoshiPNaS, CCssm, CBJDarkcore, prl06, Joshi_2014} and references therein). The existence of such objects remains an open and fundamental question, recently explored in connection with the astrophysical sources Sgr~A*~\cite{EHT2022SgrA}, M87*~\cite{mgn, bambi}, and GRO~J1655--40~\cite{Chak_18, CC_19}. In this paper, we propose a new, physically motivated mechanism to distinguish a Kerr BH from a Kerr NaS by analyzing the tilt behavior of the inner region of the accretion disk surrounding the Kerr collapsed object, in the diffusive regime.

As noted in some previous papers \cite{Banerjee_2019, Banerjee_2019_1, Sen_2024, Chakraborty_C_2017, Chak_18}, a tilted disk study is very important from the observational perspective, to probe the fundamental aspects and to measure the spin of the collapsed object. The disk acts as a natural tool to probe the strong gravity region near the central object. Its tilt can significantly affect the observed spectral and timing behavior \cite{Motta, Ingram_A_2016}, such as a reflection spectrum including a broad relativistic spectral iron emission line \cite{Miller_2005}, and QPOs \cite{Stella, Schnittman_2006}. Further, the tilt could affect the X-ray polarization \cite{Cheng_2016}, which could be probed using the {\it IXPE} satellite data when modeled with an appropriate model. All these X-ray features are excellent tools to probe the strong gravity regime and to measure the spin of the central collapsed object. Thus, if tilt is not included in models, then that could introduce errors in the interpreted results and spin values. This paper provides, for the first time, to the best of our knowledge, a way to analytically and exactly include the effects of a tilted disk (in Kerr spacetime) in the theoretical calculations of observed features in the future, which will be very useful to correctly interpret observations.

In this work, we employ the formalism developed by Pringle \cite{Pringle_1992} to derive and analyze the equation governing a steady-state, tilted, thin accretion disk in the viscous regime around a spinning Kerr compact object with an arbitrary Kerr parameter. Unlike earlier studies, we do not impose any approximations corresponding to slow rotation or weak gravitational fields. The paper is organized as follows. Sec. \ref{section_Kerr} provides a brief overview of the Kerr metric and discusses the peculiar behavior of the exact LT precession in the strong-gravity regime. In Section \ref{section_formalism}, we derive the angular momentum density expression by incorporating the exact form of the LT precession. The governing equation for the disk tilt is derived in Sec. \ref{Section_derivation}. The results and their implications are discussed in Sec. \ref{Section_results}, and our conclusions are summarized in Sec. \ref{Section_conclusion}.

\section{Brief description of Kerr spacetime \label{section_Kerr}}
\subsection{Kerr metric : Black hole and naked singularity \label{subsection_Kerr}}
The Kerr metric \cite{Kerr_R_P_1963} in Boyer-Lindquist coordinates \cite{Boyer_R_H_1967} is expressed as
\begin{eqnarray} \nonumber
dS^2=-\left(1-\f{2R_g R}{\rho^2 }\right)c^2dt^2-\f{4bR_gR \sin^2\theta}{\rho^2 } c~d\phi dt
+\f{\rho^2 }{\Delta}dR^2 +\rho^2 d\theta^2+\left( R^2+b^2+\f{2R_gRb^2 \sin^2\theta}{\rho^2 }\right) \sin^2\theta d\phi^2
\\
\label{kerr}
\end{eqnarray}
where
\begin{equation}
\rho^2 =R^2+b^2 \cos^2\theta, \\  \,\,\,\,\Delta=R^2-2R_gR+b^2 .
\label{k2}
\end{equation}
Here $b$ is a parameter which is related to the angular momentum ($J$)
and mass ($M$) of the Kerr spacetime \cite{Kato_S_1990} as $b=J/(Mc)$. The dimensionless
Kerr parameter $a$ is defined as $a=b/R_g=J/(McR_g)$ \cite{Scheuer_P_A_G_1996}. The gravitational
radius $R_g$ is defined as $R_g=GM/c^2$ where $G$ is the Newton's constant and $c$ is the speed of light in vacuum.\\

In BH ($0 < a\leq 1$), the outer and inner horizons are located
at $r_{\pm}=R_g(1 \pm \sqrt{1-a^2})$ and boundaries of outer and inner ergoregions are
located at $r^e_{\pm}=R_g (1\pm \sqrt{1 -a^2 \cos^2\th})$. Therefore, $r_+ (\equiv R_{\rm h})$ and $r_-$
coincide with $r_+^e$ and $r_-^e$, respectively, at the pole. For $a > 1$, $r_{\pm}$
become imaginary, i.e., horizons do not exist. Thus, the ``singularity'' could be
observed from infinity, in principle, and is regarded as the Kerr NaS or Kerr superspinar.
\\

As the Kerr spacetime is stationary and axisymmetric, the angular momentum ($L_p$) and energy ($E_p$) per unit mass of a test particle orbiting in the Kerr spacetime should be conserved. For the prograde rotation, one can write \cite{Shapiro_et_al_1983}
\begin{eqnarray}
E_p=c^2\f{R^2-2R_g R+aR_g \sqrt{R R_g}}
{R\left(R^2-3R_g R+2aR_g\sqrt{R R_g}\right)^{1/2}}
\label{Ep}
\end{eqnarray}
and,
\begin{eqnarray}
L_p = c\f{\sqrt{R R_g} \left(R^2-2aR_g\sqrt{R R_g}+a^2 R_g^2 \right)}
{R \left(R^2-3R_g R+2aR_g\sqrt{R R_g}\right)^{1/2}}=\eta \sqrt{GMR}
\label{Lp}
\end{eqnarray}
where $\eta$ is a function of $a, R_g,$ and $R$,
\begin{equation}
\eta = \frac{1-2a\left(\frac{R_{g}}{R}\right)^\frac{3}{2}+a^2\left(\frac{R_{g}}{R}\right)^2}{\sqrt{1-3\left(\frac{R_{g}}{R}\right)+2a\left(\frac{R_{g}}{R}\right)^\frac{3}{2}}}.
\label{eta}
\end{equation}
At large distances ($R \gg R_{g}$) from a slowly spinning Kerr BH, $\eta \rightarrow 1$, closely approximating the Schwarzschild case \cite{Shapiro_et_al_1983}.

\begin{figure}[ht]
\centering
\begin{subfigure}[]{\includegraphics[scale=0.76]{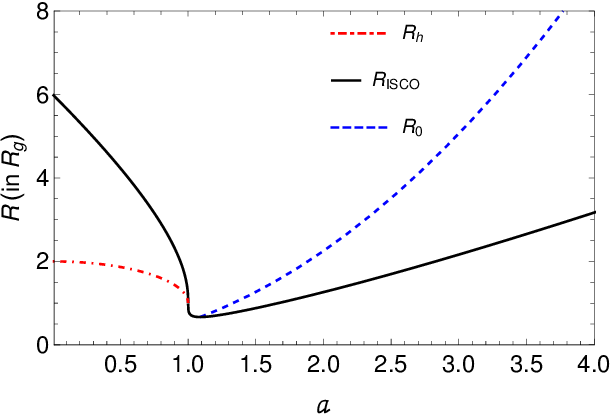}}
\end{subfigure}
% \hspace{1cm}
\begin{subfigure}[]{\includegraphics[scale=0.37]{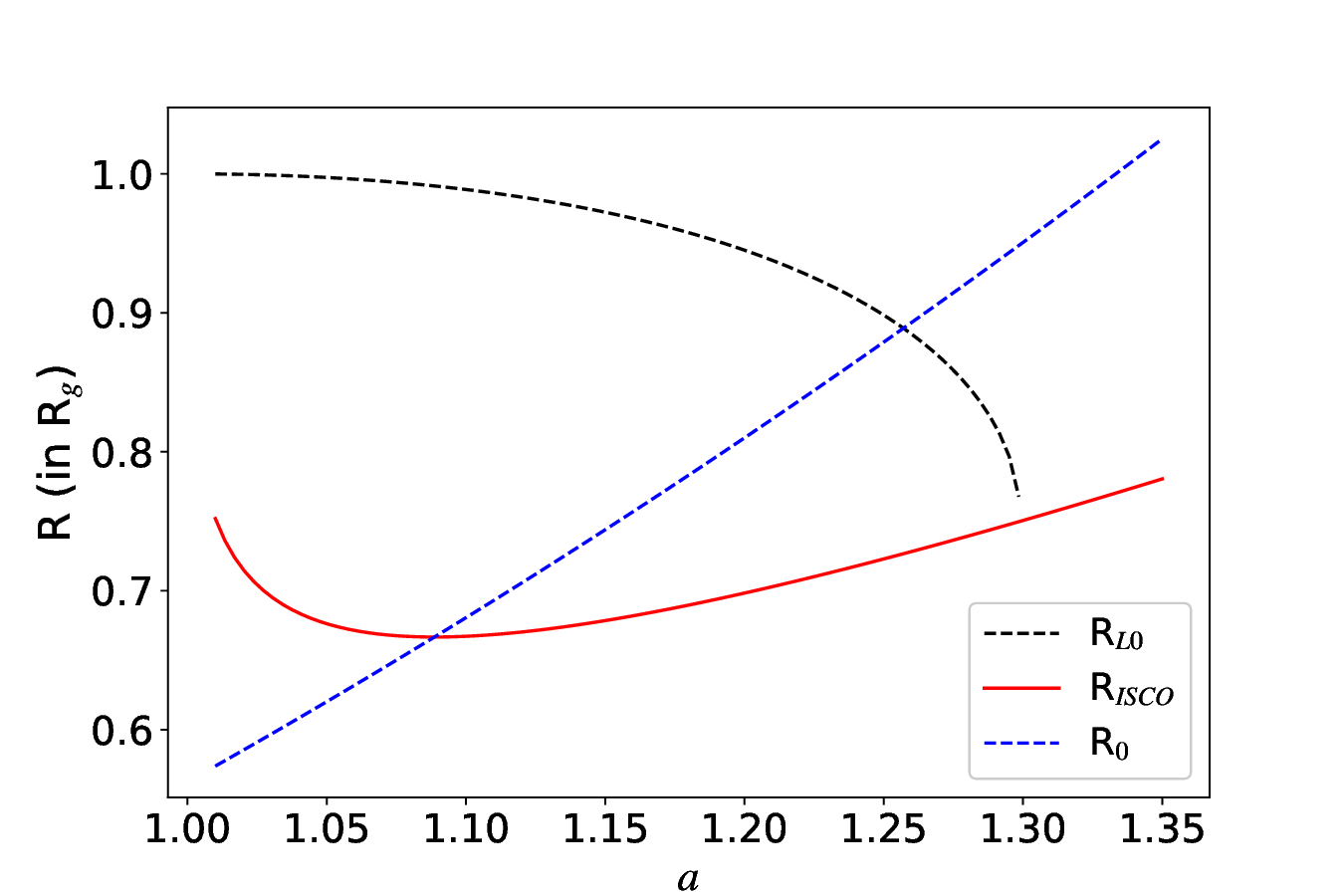}}
\end{subfigure}
\caption{\label{isco}Panel (a) corresponds to the three radial quantities (in unit of $R_g$) for prograde orbits, namely the ISCO radius ($R_{\rm ISCO}$), the horizon radius ($R_{\rm h}$), and the radius at which the LT precession frequency ($\O_{\rm LT}$) vanishes ($R_{0}$), plotted as the functions of Kerr parameter $a$ \cite{Chakraborty_C_2017_1}. Panel (b) shows the variation of $R_{\rm ISCO}$, $R_{\rm 0}$, and the radius ($R_{\rm L0}$) at which specific angular momentum $L$ of the test particle becomes zero, as a function of Kerr parameter $a$. For further discussion, see Sec. \ref{subsection_Kerr} and \ref{subsection_anomalous}. }
\label{varn_of_radii}
\end{figure}

For prograde rotation, the radius of the innermost stable circular orbit (ISCO) of the Kerr spacetime is defined as \cite{Uhikata_2021},
\begin{eqnarray}
\f{R_{\rm ISCO}}{R_g} = 3 + H(a) - \sqrt{6 + 2a^2 - \f{a^4 - 10a^2 + 9}{G(a)} - G(a) + \f{16a^2}{H(a)}},
\label{RISCO}
\end{eqnarray}
where $G(a) = (27 - 45a^2 + 17a^4 + a^6 + 8a^3 |a^2 - 1|)^{\f{1}{3}}$, and $H(a) = \sqrt{3 + a^2 + \f{a^4 - 10a^2 + 9}{G(a)} + G(a)}$. \\

Figure \ref{varn_of_radii} shows the variation of $R_{\rm ISCO}$ for different values of $a$. It is clear that $R_{\rm ISCO}$ has its lowest value at a particular value of $a$, i.e., $a = \sqrt{32/27}\approx 1.089$.
In a recent work~\cite{Mum24}, several interesting properties of Kerr NaSs were investigated. In particular, it was shown that \(R_{\rm ISCO} < R_g\) for \(1 < a < 5/3\). It was further demonstrated that every Kerr BH with a given ISCO radius admits a corresponding super-extremal Kerr NaS with the same ISCO location, characterized by a spin parameter in the range \(5/3 < a < 9\) and considering both prograde and retrograde rotations. Possible implications of such super-extremal configurations for accretion disk structure and emergent X-ray spectra were also discussed \cite{Mum24}.
A further detailed inspection shows that Eq.~(\ref{Lp}) becomes zero at a specific radius, denoted by \( R = R_{L0} \), for \( 1 < a < 1.3 \), which lies within the Kerr NaS regime. As illustrated in panel~(b) of Fig.~\ref{varn_of_radii}, \( R_{L0} \) is found to be greater than \( R_{\rm ISCO} \) in this range of \( a \). Consequently, stable circular orbits with negative angular momentum can exist within the interval \( R_{\rm ISCO} \leq R \leq R_{L0} \). This feature may be attributed to the manifestation of repulsive gravitational effects in the spacetime~\cite{Pugliese_2011}. It is further observed that the separation between \( R_{L0} \) and \( R_{\rm ISCO} \) gradually decreases as \( a \) increases from 1 to 1.3.

\subsection{Anomalous behavior of Lense-Thirring precession \label{subsection_anomalous}}
Among the three fundamental frequencies related to the prograde accretion disk,   the Keplerian frequency $\O$ and vertical epicyclic frequency $\O_{\th}$ are derived for the Kerr metric \cite{Okasaki_A_T_1987,Kato_S_1990} as
\begin{eqnarray}
\O &=& \f{\sqrt{GM}}{R^{\f{3}{2}}+a R_g^{\f{3}{2}}}
=\sqrt{\f{GM}{R^3}}.\f{1}{1+ a\left(\f{R_g}{R}\right)^{\f{3}{2}}}=\O_s.\c^{-1},
\label{Omega}
\\
\O_{\th} &=& \O \left(1-\f{4aR_g^{\f{3}{2}}}{R^{\f{3}{2}}}
+\f{3a^2 R_g^2}{R^2} \right)^{\f{1}{2}}
\label{fr3},
\end{eqnarray}
where
\begin{eqnarray}
\O_s\equiv \sqrt{\f{GM}{R^3}} \,\,\,\,\,\,\, {\rm and} \,\,\,\,\,\,\,\,
\c \equiv \c (R) = 1+ a\left(\f{R_g}{R}\right)^{\f{3}{2}}.
\end{eqnarray}
Note that the parameter $\c$ indirectly accounts for the spin ($a$) of the Kerr collapsed object in the later sections we will discuss. In contrast, $\c$ does not appear in the expressions for a slowly spinning Kerr BH, as $\c \to 1$ in that limit.
However, the orbital/nodal plane precession ($\O_{\rm nod}$ or $\Omega_{\rm LT}$) frequency which arises due to the LT effect \cite{Lense_1918}, is defined as $\O_{\rm LT}\equiv (\O -\O_{\rm \theta})$, i.e.,
\begin{eqnarray}
\O_{\rm LT}=\O_{\rm s}.\c^{-1}.\left[1-\left(1-\f{4aR_{g}^{\f{3}{2}}}{R^{\f{3}{2}}}
+\f{3a^2R_{g}^2}{R^2} \right)^{\f{1}{2}}\right].
\label{lt}
\end{eqnarray}
In the slow-spin approximation $(a \ll 1)$, Eq. (\ref{lt}) reduces to \cite{Banerjee_2019, Chakraborty_C_2017}
\begin{eqnarray}
\O_{\rm LT}|_{a \ll 1} \approx \f{2acR_g^2}{R^3}.
\label{ltsl}
\end{eqnarray}

An intriguing feature that emerges is that the LT precession frequency (Eq. \ref{lt}) vanishes at a radius $R=R_{0}$, where $R_0=9/16~a^2R_{g}=0.5625 ~a^2 R_{g}$. Furthermore, $\Omega_{\rm LT}$ becomes negative for $R < R_0$, indicating that accreting material begins to orbit in the retrograde direction in the region $R_{\rm ISCO} \leq R < R_0$ \cite{Chakraborty_C_2017_1}. Figure \ref{varn_of_radii} exhibits the variation of horizon radius $R_{\rm h}$, $R_{\rm ISCO}$, and $R_{\rm 0}$ as a function of $a$. It is clear that when $a>1.089$ (i.e., Kerr NaS), the $R_{\rm 0}$ values are greater than $R_{\rm ISCO}$, which means that there are orbits at which $\Omega_{\rm LT}$ is negative. Note that $R_0$ is feasible only for the Kerr NaS with $a>1.089$, whereas $R_0$ does not appear for a Kerr BH.

\section{Formalism \label{section_formalism}}
Let us consider a geometrically thin (i.e., $H/R\ll1$ where $H$ is the disk thickness and $R$ is the radial distance from the central object) accretion disk in the viscous regime (i.e., $\alpha > H/R$ where $\alpha$ is the Shakura-Sunyaev parameter) \cite{Shakura_1973}.
The material of the accretion disk is centrifugally supported primarily in the circular orbits with angular velocity $\O(R)$. We divide the disk into circular rings having a width $\delta R$. We define $\Sigma(R,t)$ as the surface density, and $V_{\rm R}(R,t)$ as the radial velocity on each annulus of the disk.  At each radius $R$, the disk has a unit tilt vector $\boldsymbol{l}(t,R)$. We assume the disk at each radius to be thin in the direction of $\boldsymbol{l}$. In addition, we consider that the tilt angle at each annulus is small such that the tilt vector can be approximated as $\boldsymbol{l} \simeq (l_x,l_y,1)$. The spin of the central collapsed object of mass $M$ and Kerr parameter $a$ is along the $z$-axis, and the accretion disk is tilted with respect to this. The mass conservation equation under this small tilt approximation is \cite{Papaloizou_1983, Pringle_1992} written as:
\begin{equation}
\frac{\partial \Sigma}{\partial t} = -\frac{1}{R}\frac{\partial}{\partial R}\left(R \Sigma V_{\rm R}\right).
\label{mass_conservation}
\end{equation}

The local angular momentum density of the disk around the Kerr spacetime can be written as $\S R^2 \eta \O_s \boldsymbol{l}$ \cite{Shapiro_et_al_1983}.
At large distances ($R \gg R_{g}$), or, for a slowly spinning Kerr BH, $\eta \rightarrow 1$, closely approximating the Schwarzschild case considered in \cite{Papaloizou_1983, Pringle_1992, Scheuer_P_A_G_1996}. Since our aim is to construct the tilted disk in the full Kerr spacetime, we accordingly retain the term $\eta$.
The angular momentum of accreting material contained in an annulus between $R+\f{1}{2}\delta R$ and $R-\f{1}{2}\delta R$ is $\delta \boldsymbol{l}= 2 \pi R. \delta R . \Sigma R^2 \eta \Omega_{\rm s} \boldsymbol{l}$.
Thus, from the conservation of angular momentum for an annulus of width $\delta R$, centered at radius $R$, one can write,

\begin{align}
\frac{\partial}{\partial t}(2\pi R^3 \Sigma \Omega_{\rm s} \eta. \delta R. \boldsymbol{l}) = \left(2\pi R^3. \Sigma \Omega_{\rm s} \eta. V_{\rm R} \boldsymbol{l}\right)|_{\rm R-\frac{1}{2}\delta R} - \left(2\pi R^3. \Sigma \Omega_{\rm s} \eta. V_{\rm R} \boldsymbol{l}\right)|_{\rm R+\frac{1}{2}\delta R}
+ \textbf{G} \left(R+\frac{1}{2}\delta R\right) - \textbf{G} \left( R-\frac{1}{2}\delta R\right).
 \label{angular_momentum_conservation}
\end{align}

Here, $\textbf{G}(R,t)$ is the viscous torque acting on one disk annulus due to the other. There are two kinds of viscous torques acting on the accretion disk: $\textbf{G}_1,$ which is the $(R, \phi)$ component of viscous torque, acting perpendicular to the disk,  and $\textbf{G}_2$ which is the $(R, z)$ component of torque, acting in the plane of the disk \cite{Papaloizou_1983, Nixon_2016}:
\begin{equation}
 {\bf G_1} = 2\pi R \nu_1 \Sigma R \Omega' R \boldsymbol{l},
 \label{G01}
\end{equation}
and
\begin{equation}
{\bf G_2} = \pi R \nu_2 \Sigma \Omega R^2 \frac{\partial \boldsymbol{l}}{\partial R}.
\label{G02}
\end{equation}

Here, prime ($'$) denotes the differentiation with respect to $R$. $\nu_1$ is the viscosity associated with azimuthal shear and $\nu_2$ is the viscosity associated with the vertical shear \cite{Papaloizou_1983}, the more details of which is given in Sec. \ref{subsection_viscous torque}. In this paper, we assume $\nu_1$ and $\nu_2$ are independent of $R$. The ratio between $\nu_{\rm 2}$ and $\nu_{\rm 1}$ is called viscous anisotropy. For small amplitude warp, it is related to $\alpha$ by \cite{Ogilvie_1999}
\begin{equation}
 \frac{\nu_{\rm 2}}{\nu_{\rm 1}}=\frac{1}{2\alpha^2}.\frac{4\left(1+7\alpha^2\right)}{4+\alpha^2}.
 \label{viscosityratio}
\end{equation}

Thus one can write angular momentum conservation equation by taking $\delta R \to 0$ in Eq. (\ref{angular_momentum_conservation}),

\begin{equation}
R\frac{\partial}{\partial t} \left(\Sigma R^2 \Omega_{\rm s} \eta \boldsymbol{l}\right) + \frac{\partial}{\partial R}\left(\Sigma \Omega_{\rm s} \eta V_{\rm R} R^3 \boldsymbol{l}\right) = \frac{\partial}{\partial R}\left(\nu_1 R^3 \Sigma \Omega' \boldsymbol{l}\right) + \frac{\partial}{\partial R}\left(\frac{1}{2}\nu_2 \Sigma R^3 \Omega \frac{\partial \boldsymbol{l}}{\partial R}\right).
\label{ang_mom_conservation_1}
\end{equation}

Using Eq. (\ref{mass_conservation}) and (\ref{ang_mom_conservation_1}), we derive the radial velocity ($V_{\rm R}$) of the disk material as,
\begin{equation}
V_R = \frac{\frac{\partial}{\partial R}\left(\nu_1 R^3 \Sigma \Omega'\right)}{\Sigma R \frac{\partial}{\partial R}\left(R^2 \Omega_{\rm s} \eta\right)}
\label{radial_velocity}
\end{equation}
for the small tilt approximation, i.e., we have dropped the term containing $|\frac{\partial \boldsymbol{l}}{\partial R}|^2$ following \cite{Papaloizou_1983}.

Let us define ${\bf L}= \Sigma R^2 \Omega_{\rm s} \eta \boldsymbol{l}$, which is the local angular momentum density in the disk \cite{Pringle_1992}. This means that the annulus between $R-\f{1}{2}\delta R$ and $R+\f{1}{2}\delta R$ has total mass $M=2\pi \Sigma R\delta R$ and total angular momentum $\delta {\bf L}=2\pi {\bf L}R\delta R$. Now, by using Eq. (\ref{radial_velocity}), we can write the equation governing ${\bf L}$ as
\begin{equation}
\frac{\partial \boldsymbol{L}}{\partial t} = \frac{1}{R} \frac{\partial}{\partial R} \left[-V_R R \boldsymbol{L} + \frac{\nu_1}{\eta}\left(\frac{R\Omega'}{\Omega_{\rm s}}\right)\boldsymbol{L} + \frac{1}{2}\nu_2 \frac{R \chi^{-1}}{\eta}|\boldsymbol{L}|\frac{\partial \boldsymbol{l}}{\partial R}\right].
\label{ang_mom_conservation_3}
\end{equation}

Equation (\ref{ang_mom_conservation_3}) describes the specific angular momentum subjected to internal torques only. In a steady state, to capture the general relativistic effect in Kerr spacetime, one needs to add the external torque due to the LT effect,

\begin{equation}
\frac{\partial \boldsymbol{L}}{\partial t} = \boldsymbol{\Omega}_{\rm LT} \times \boldsymbol{L}
\end{equation}
following \cite{Scheuer_P_A_G_1996, Banerjee_2019}.
Thus, the equation of specific angular momentum in a steady--state disk can be written as

\begin{equation}
 \frac{1}{R} \frac{\partial}{\partial R} \left[\frac{-\frac{\partial}{\partial R}\left(-\frac{3}{2} \nu_1 \chi^{-2} \frac{L}{\eta}\right)}{\Sigma \frac{\partial}{\partial R}\left(R^2 \Omega_{\rm s}\eta\right)} \boldsymbol{L} + \frac{\nu_1}{\eta}\left(\frac{R\Omega'}{\Omega_{\rm s}}\right)\boldsymbol{L} + \frac{1}{2}\nu_2 \frac{R \chi^{-1}}{\eta}L\frac{\partial \boldsymbol{l}}{\partial R}\right] + \Omega_{\rm LT} \times \boldsymbol{L} = 0,
 \label{ang_mom_conservation_4}
\end{equation}
where we use $L = |\boldsymbol{L}|$ and substitute the expression of $V_R$ from Eq. (\ref{radial_velocity}) in terms of $L$. Taking the scalar product of $\boldsymbol{l}$ with Eq. (\ref{ang_mom_conservation_4}), we obtain

\begin{equation}
 \frac{1}{R} \frac{\partial}{\partial R} \left[\frac{\frac{\partial}{\partial R}\left(\frac{3}{2} \nu_1 \chi^{-2} \frac{L}{\eta}\right)}{\Sigma \frac{\partial}{\partial R}\left(R^2 \Omega_{\rm s}\eta\right)} L - \frac{3}{2} \frac{\nu_1}{\eta} \chi^{-2} L\right] = 0.
 \label{ang_mom_conservation_5}
\end{equation}

Here, we ignore the term containing $|\frac{\partial \boldsymbol{l}}{\partial R}|^2$ due to small tilt approximation following \cite{Scheuer_P_A_G_1996}. Assuming $\nu_1$ is independent of $R$ \cite{Scheuer_P_A_G_1996, Chakraborty_C_2017, Banerjee_2019}, we obtain the general expression for $L$ by solving Eq. (\ref{ang_mom_conservation_5}),

\begin{equation}
L = \eta \chi^2 \left(C_2 \eta \sqrt{R} - 2 C_1\right),
\label{L}
\end{equation}
where $C_2$ and $C_1$ are integral constants. For $a \ll 1$, Eq. (\ref{L}) reduces to Eq. (4) of \cite{Scheuer_P_A_G_1996}.
In the case of a Keplerian thin disk around a Kerr collapsed object, we can write the specific angular momentum density as \cite{Shapiro_et_al_1983}
\begin{equation}
L = \Sigma \sqrt{GMR} \eta(a,R).
\label{L_shapiro}
\end{equation}

Using Eqs. (\ref{L_shapiro}) and (\ref{L}) we can write

\begin{equation}
 C_{\rm 2} = \eta^{-1} \left(\Sigma \sqrt{GM} \chi^{-2} + 2 C_{\rm 1} R^{-\frac{1}{2}}\right).
\label{C2}
\end{equation}

On applying the boundary condition $\Sigma \to \Sigma_{\rm \infty}$ as $R \to \infty$ \cite{Scheuer_P_A_G_1996} we derive $C_{\rm 2}$ as

\begin{equation}
 C_{\rm 2} = \sqrt{GM} \Sigma_\infty.
\label{C2_1}
\end{equation}

This expression is the same as that given in \cite{Banerjee_2019} [see Eq. (15)]. Using Eq. (\ref{C2_1}) and the boundary condition $\Sigma (R_{\rm in}) = \Sigma_{\rm in}$ with $\eta(R_{\rm in}) = \eta_{\rm in}$, we derive $C_{\rm 1}$ as

\begin{equation}
C_{\rm 1} = \frac{1}{2} \sqrt{GMR_{\rm in}} \left[\eta_{\rm in}\Sigma_\infty - \Sigma_{\rm in} \chi_{\rm in}^{-2}\right],
\label{C1}
\end{equation}

where

\begin{eqnarray}
\chi_{\rm in} = 1+ a\left(\f{R_g}{R_{\rm in}}\right)^{\f{3}{2}}.
\end{eqnarray}

The expression of $C_{\rm 1}$ is different from that of \cite{Banerjee_2019} (see Eq. (16) of \cite{Banerjee_2019}). Here $R_{\rm in}$ corresponds to the inner edge radius of the accretion disk, which is taken as identical to the ISCO radius (i.e., $R_{\rm in} \equiv R_{\rm ISCO})$ in this formulation. Here $C_{\rm 2}$ depends on $\Sigma_{\rm \infty}$, which provides the contribution from the outer part of the accretion disk. On the other hand, $C_{\rm 1}$ depends on both $\Sigma_{\rm in}$ and $\Sigma_{\rm \infty}$, which deliver the contribution of both inner and outer parts of the accretion disk. This indicates that $C_{\rm 1}$ is associated with the inner edge boundary condition, which was dropped in \cite{Scheuer_P_A_G_1996} in their calculation. Substituting $C_{\rm 1}$ and $C_{\rm 2}$ in Eq. (\ref{L}), we obtain

\begin{equation}
L = \sqrt{GMR} \eta \chi^2 \left[\eta \Sigma_\infty-\sqrt{\frac{R_{\rm in}}{R}}\left[\eta_{\rm in}\Sigma_\infty-\Sigma_{\rm in}\chi_{\rm in}^{-2}\right]\right].
\label{L1}
\end{equation}

This is the final expression of the angular momentum per unit area of the accretion disk,  which is not only a function of the mass of the Kerr spacetime but also a function of its spin. However, using Eqs. (\ref{L_shapiro}) and (\ref{L1}), we get the final expression of surface density as

\begin{equation}
\Sigma(R) = \chi^2 \left[\eta \Sigma_\infty - \sqrt{\frac{R_{\rm in}}{R}}\left(\eta_{\rm in} \Sigma_\infty - \Sigma_{\rm in} \chi_{\rm in}^{-2}\right)\right].
\label{sdf}
\end{equation}
One can check that Eq. (\ref{L1})
and Eq. (\ref{sdf}) reduce to Eq. (17) and Eq. (18) of \cite{Banerjee_2019}, respectively. for the slowly spinning Kerr BH, as $\chi$, $\chi_{\rm in}$, $\eta$, and $\eta_{\rm in}$ tends to $1$ for $a \ll 1$.

\section{Derivation of Tilt equation for full Kerr metric \label{Section_derivation}}

Substituting the expression of $L$ (see Eq. \ref{L}) in Eq. (\ref{ang_mom_conservation_4}), one obtains
\begin{eqnarray}
\frac{1}{R}\frac{\partial}{\partial R}\left[3 \nu_1 C_1 \boldsymbol{l}
+\frac{1}{2} \frac{\chi^{-1}}{\eta} \nu_2 R L \frac{\partial \boldsymbol{l}}{\partial R}\right]+\boldsymbol{\Omega}_{\rm LT}\times\boldsymbol{L}=0.
\label{tiltnd}
\end{eqnarray}
One can now decouple this equation for $x$ and $y$ components of the tilt vector (i.e., $l_x$ and $l_y$),
\begin{equation}
\frac{1}{R}\frac{\partial}{\partial R}\left[3\nu_{\rm 1} C_{\rm 1} l_x + \frac{1}{2} (\chi^{-1}\eta^{-1}) \nu_2 R L \frac{\partial l_x}{\partial R}\right] = \Omega_{\rm LT} L l_y,
\label{tilt_x}
\end{equation}
and
\begin{equation}
\frac{1}{R}\frac{\partial}{\partial R}\left[3\nu_{\rm 1} C_{\rm 1} l_y + \frac{1}{2} (\chi^{-1}\eta^{-1}) \nu_2 R L \frac{\partial l_y}{\partial R}\right] = -\Omega_{\rm LT} L l_x .
\label{tilt_y}
\end{equation}
In Cartesian coordinates, with the BH at the origin and the $z$-axis along $\boldsymbol{\Omega}_{\rm LT} = (0, 0, \Omega_{\rm LT})$ and taking the unit tilt vector as $\boldsymbol{l} = (l_x, l_y, 1)$, the cross product evaluates to $\boldsymbol{\Omega}_{\rm LT} \times \boldsymbol{L} = \left(-\Omega_{\rm LT} L\, l_y,\; \Omega_{\rm LT} L\, l_x,\; 0\right)$
\cite{Scheuer_P_A_G_1996, Banerjee_2019}. Equation (\ref{tiltnd}) [or Eq. (\ref{tilt_x}) and (\ref{tilt_y})] characterizes the warped accretion disk in Kerr spacetime. This equation encompasses the whole accretion disk, which is valid for all values of $a>0$. We solve this full warped disk equation [Eq. (\ref{tiltnd})] numerically with suitable boundary conditions to compute the tilt profile (change in tilt angle with radial distance) of the disk. Here $\beta = \sqrt{l_x^2+l_y^2}$, which is the tilt angle and $\gamma = \tan^{-1}(l_y/l_x)$ is the twist angle of the accretion disk \cite{Pringle_1992, Banerjee_2019}. For the simplicity in calculations, we derive the dimensionless version of $L$ by taking $R_{g}$ as the length scale (i.e., $R \to R/R_{g}$) and $C_{\rm 1}$ as the angular momentum scale,
\begin{eqnarray}
L \rightarrow L/C_1 = \eta \chi^2 \left(C \eta \sqrt{R} - 2 \right)
\label{LC1}
\end{eqnarray}
where

\begin{equation}
C = \frac{C_2}{C_1} = \frac{2 \chi_{\rm in}^2 z_{\rm in}}{\sqrt{R_{\rm in}}\left(\eta_{\rm in} z_{\rm in} \chi_{\rm in}^2 - 1\right)}
\label{C}
\end{equation}

with
\begin{eqnarray}
z_{\rm in} = \f{\S_{\infty}}{\S_{\rm in}}.
\end{eqnarray}

Thus, the dimensionless version of the final tilt equations for the full Kerr spacetime can be expressed as

\begin{eqnarray} \nonumber
R\frac{\partial^2 l_x}{\partial R^2} + \left[1 + \chi \eta \left\{\frac{n}{L} + R \left(\frac{1}{\chi \eta}\right)'\right\} + R\frac{L'}{L} \right] \frac{\partial l_x}{\partial R} &=& 2 \xi F l_y,
\\
R\frac{\partial^2 l_y}{\partial R^2} + \left[1 + \chi \eta \left\{\frac{n}{L} + R \left(\frac{1}{\chi \eta}\right)'\right\} + R\frac{L'}{L} \right] \frac{\partial l_y}{\partial R} &=& -2 \xi F l_x
\label{tilteq}
\end{eqnarray}
where, the prime denotes the differentiation with respect to $R$ and $\xi = c R_g / \nu_2$ \cite{Banerjee_2019}.  Note that although \(\xi\) is dimensionless, it depends on both \(M\) and \(\nu_2\). Consequently, the solution of the tilt equation (Eq.~\ref{tilteq}), and hence the resulting tilt profile, changes with variations in either parameter when the other is held fixed. Nevertheless, Eq.~(\ref{tilteq}) itself remains valid for arbitrary values of $M$, and the qualitative behavior of the tilt profiles remains broadly similar across different mass scales, provided the relevant physical parameters are scaled appropriately.

In Eq.~(\ref{tilteq}),
\begin{eqnarray}
F=\frac{\eta}{\sqrt{R}}\left[1-\left(1-\f{4a}{R^{\f{3}{2}}}
+\f{3a^2}{R^2} \right)^{\f{1}{2}}\right]
\end{eqnarray}
and $n=6 \nu_1/\nu_2$ which is inversely proportional to the viscous anisotropy $(\nu_2/\nu_1)$. One intriguing feature of the above expression is that the LHS of the above equations depends on the Kerr parameter $a$ (previously it was not \cite{Banerjee_2019}). Equation (\ref{tilteq}) reduces to Eq. (40) of \cite{Banerjee_2019} for the slow spin approximation.  \\

To solve the tilt equations, let us consider the boundary conditions \cite{Banerjee_2019}:
\begin{eqnarray}
l_x(R_{\rm in}) = \beta_i \cos(\gamma_i) \,\,\,\,\,\,\, {\rm ,} \,\,\,\,\,\,\,\,
l_y(R_{\rm in}) = \beta_i \sin(\gamma_i),
\label{BC1}
\end{eqnarray}
and
\begin{eqnarray}
l_x(R_{f}) = \beta_f \,\,\,\,\,\,\, {\rm ,} \,\,\,\,\,\,\,\,
l_y(R_{f}) = 0, \label{BC2}
\end{eqnarray}
where $\gamma_i$, $\beta_i$, $\beta_f$, and $R_{\rm f}$ are the twist angle at the inner boundary, the tilt angle at the inner edge and the outer edge of the disk, and outer edge radius of the disk,  respectively. Since the effect of LT precession is negligible at the outer edge of the disk,  we assume that the twist angle is zero there. Consequently $l_y$ is assumed to be zero at the outer edge of the accretion disk \cite{Banerjee_2019}. As mentioned in Sec. \ref{subsection_Kerr}, $R_{\rm in}$ is the ISCO radius of the Kerr spacetime.
We fix the values of \(\beta_f\) and \(\gamma_i\), and consider different representative values of \(\beta_i\).
Now the tilt equations (Eq. \ref{tilteq}) are subjected to the boundary conditions mentioned in Eqs. (\ref{BC1}) and (\ref{BC2}). We can plot the tilt profile of the accretion disk by solving this boundary value problem.

\subsection{Viscous torques acting on the accretion disk \label{subsection_viscous torque}}
The accretion disk that we consider is sufficiently viscous. As discussed in Sec. \ref{section_formalism}, there are two viscosity parameters associated with the disk: $\nu_1$ describing azimuthal shear and $\nu_2$ describing vertical shear. Accordingly, there are two viscous torques acting on the disk: $\textbf{G}_1$ acting perpendicular and $\textbf{G}_2$ acting in the plane of the disk.
By using $\Omega' = -\frac{3\chi^{-1}}{2}.\frac{\Omega}{R}$ and Eq. (\ref{L_shapiro}) we can rewrite Eq. (\ref{G01}) as \cite{Banerjee_2019}
\begin{equation}
\textbf{G}_1
= -3 \pi \nu_1 \chi^{-2} \eta^{-1} L(R) \boldsymbol{l}.
% ~=~-3\pi \nu_1 \left(C\eta \sqrt{R} - 2\right)\boldsymbol{l}.
\label{G1}
\end{equation}
Since both $L(R)$ and $\Sigma(R)$ are nonzero at the inner edge of the disk in our formulation [see Eqs.~(\ref{BC1}), (\ref{L1}), and (\ref{sdf}), and the discussion following Eq.~(\ref{C2_1})], it follows from the above expression that the $x$ and $y$ components of $\textbf{G}_1$ vanish at the inner boundary only when the tilt angle at the inner edge is zero, while the $z$-component of $\textbf{G}_1$ remains nonzero \cite{Banerjee_2019}. In this paper, we also consider the case in which the inner-edge tilt angle is nonzero. In such a scenario, the $x-$ and $y-$ components of $\textbf{G}_1$ likewise remain nonzero at the inner boundary \cite{Banerjee_2019}. \\

We next discuss the role of the inner edge tilt in determining the behavior of the torque $\textbf{G}_2$, which arises only in a warped disk.
Using Eq. (\ref{L_shapiro}) we can rewrite Eq. (\ref{G02}) as \cite{Banerjee_2019}
\begin{eqnarray}
\textbf{G}_{\rm 2} = \pi \nu_2 R L(R) \chi^{-1} \eta^{-1} \frac{\partial \boldsymbol{l}}{\partial R}.
% ~=~\pi \nu_2 R \chi \left(C \eta \sqrt{R}-2\right)\frac{\partial \boldsymbol{l}}{\partial R}.
\label{G2}
\end{eqnarray}

The $x$ and $y$ components of $\textbf{G}_2$ vanish at the inner edge only if the corresponding components of $\partial \boldsymbol{l}/\partial R$ vanish at the inner boundary. The radial derivative $\partial \boldsymbol{l}/\partial R$ is zero at $R_{\rm in}$ when the inner disk is aligned with the BH spin equatorial plane but remains nonzero for a misaligned inner disk, even in cases where the inner-edge tilt angle itself is zero.\\

We emphasize that the adopted inner boundary conditions in our calculations are quite general, and we do not impose any additional constraints at the inner edge--such as those considered in \cite{Pringle_1992, Lodato_G_2006, Lodato_2010}--to enforce a torque free inner boundary. The physical plausibility of a nonzero torque at the inner edge of the disk has been discussed extensively in the literature, including in \cite{Kulkarni_2011,Penna_2012,Zhu_2012,Mcclintock_2014, Agol_2000, Pot, Shafee_2008}.
In particular, accretion disks with nonzero central torque were investigated in~\cite{NixBa21}, where it was demonstrated that such torques can significantly modify the disk structure and energy dissipation profile. More recently,~\cite{MumBal23} analytically studied accretion flows across the ISCO and emphasized that finite stresses at the ISCO naturally lead to finite trans-ISCO velocity and nonvanishing thermodynamic quantities. The presence of finite stresses near the inner edge of the disk is also supported by GRMHD simulations of thin accretion flows discussed in~\cite{MumBal23}. Similar indications of nonzero stresses near the ISCO have additionally emerged from GRMHD simulations and thin-disk models incorporating finite inner-boundary stresses~\cite{Penna_2012}.
In addition, \cite{Banerjee_2019, Banerjee_2019_1} adopted inner disk boundary conditions that allow for a nonvanishing viscous torque at the inner edge in order to study the tilt profile and associated timescales of accretion disks around Kerr BHs within the slow-spin and weak-field approximations. More recently, \cite{Sen_2024} investigated the impact of a nonzero inner-edge torque on warped accretion disks in the Kerr-Taub-NUT spacetime.

At the same time, we also consider the alternative inner boundary condition \((\Sigma_{\rm in}\rightarrow 0)\) \cite{Drewes_2021, Lodato_2010}, corresponding to the standard zero-torque prescription (see \cite{NixBa21} for a discussion) commonly adopted in thin-disk accretion theory, where the plunging motion of matter inside the ISCO is assumed to strongly suppress the local viscous stresses~\cite{Shakura_1973,Novikov_1973}. Such a prescription has also been argued to improve consistency between one-dimensional warped-disk calculations and three-dimensional simulations in certain regimes~\cite{Drewes_2021, Lodato_2010}. Since both finite-stress and zero-torque inner-boundary conditions are physically motivated and remain actively discussed in the literature, we investigate both \(\Sigma_{\rm in} \neq 0\) and \(\Sigma_{\rm in} \rightarrow 0\) cases to examine the robustness of the resulting tilt profiles.

\section{Results and discussions \label{Section_results}}
In this section, we investigate the radial tilt structure of the accretion disk, namely the variation of the tilt angle with radial distance from the central collapsed object for different choices of the physical parameters. We further examine the radial extent of disk alignment and warping and how these properties depend on the relevant parameters. In addition, we analyze the behavior of the LT torque and its connection to the resulting tilt structure of the disk.

To study the tilt behavior, we adopt physically motivated parameter ranges appropriate for X-ray binaries. We consider a central collapsed object of mass \(10\,M_{\odot}\)~\cite{Fragos_2015}, while noting that the analysis remains applicable to other mass values as discussed below in Eq.~(\ref{tilteq}). The vertical shear viscosity parameter \(\nu_2\) is taken in the range \(10^{14}-10^{15}\,\mathrm{cm}^2\,\mathrm{s}^{-1}\)~\cite{Frank_2002}. The viscosity \(\nu_1\) is related to \(\nu_2\) through the viscosity parameter \(\alpha\) (or equivalently \(n\)). We adopt \(n\) in the range \(0.0012-0.94\), corresponding to \(\alpha \simeq 0.01-0.4\)~\cite{King_2007} (see Eq.~(\ref{viscosityratio})). The parameter \(z_{\rm in}\) is chosen within the range \(0.001 \leq z_{\rm in} \leq 0.75\), corresponding to \(\Sigma_{\rm in}=10^5\,\mathrm{g\,cm^{-2}}\) and \(\Sigma_{\infty}=(10^3-0.75\times10^5)\,\mathrm{g\,cm^{-2}}\)~\cite{Frank_2002,Shakura_1973}. We also consider the limiting case \(\Sigma_{\rm in}\rightarrow 0\)~\cite{Drewes_2021,Lodato_2010}, as discussed in Sec.~\ref{subsection_viscous torque}. The inner-edge twist angle \(\gamma_i\) and the outer-edge tilt angle \(\beta_f\) are fixed at \(5^\circ\) and \(10^\circ\), respectively, while the inner-edge tilt angle \(\beta_i\) is varied between \(0^\circ\) and \(10^\circ\) for illustration. Since we consider only prograde motion and do not impose any approximation on the spin parameter \(a\), its allowed range is \(0<a<\infty\). For the adopted parameter values, the total disk mass,$
M_{\rm disk}=2\pi\int_{R_{\rm in}}^{R_{\rm f}}R\,\Sigma(R)\,dR, $
is estimated to lie in the range \((10^{-8}-10^{-6})\,M_\odot\).

We perform a detailed analysis of the radial tilt profile, i.e., the variation of the tilt angle with radial distance \(R\), as a function of the parameters \(a\), \(\nu_2\), \(\beta_i\), \(M\), and \(n\). The tilt structure is governed primarily by the competition between the LT torque and the viscous torques. While the LT torque depends mainly on \(M\) and \(a\), the viscous torques are determined by \(\nu_2\) and \(n\).

\subsection{Tilt profiles for black holes} \label{subsection_tiltprofile_BH}

\begin{figure}[ht]
\centering
\subfigure[$\beta_i=0^\circ$]{
\includegraphics[scale=0.3]{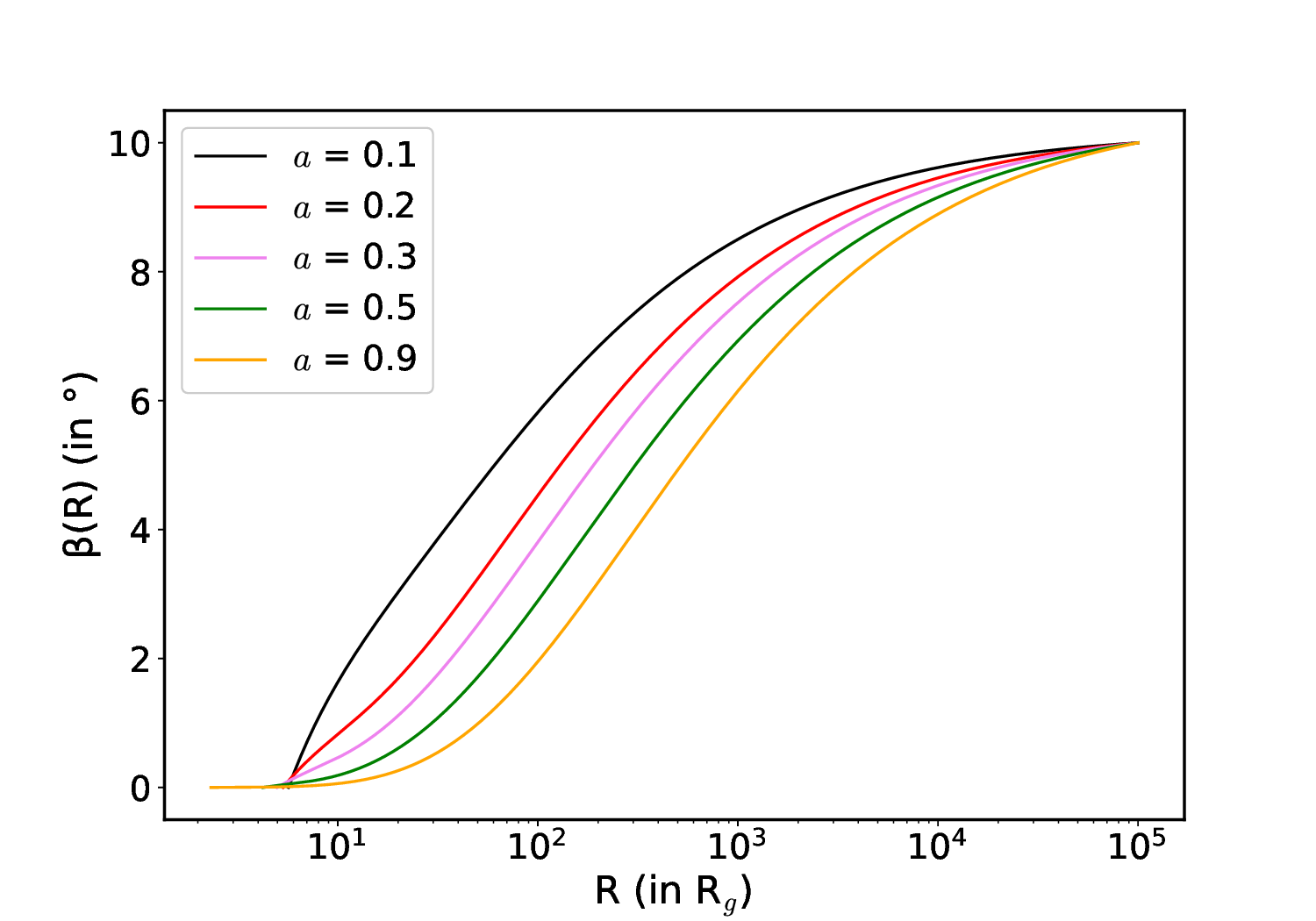}
\label{fig:sub-a}}
\hspace{1cm}
\subfigure[$\beta_i=5^\circ$]
{\includegraphics[scale=0.3]{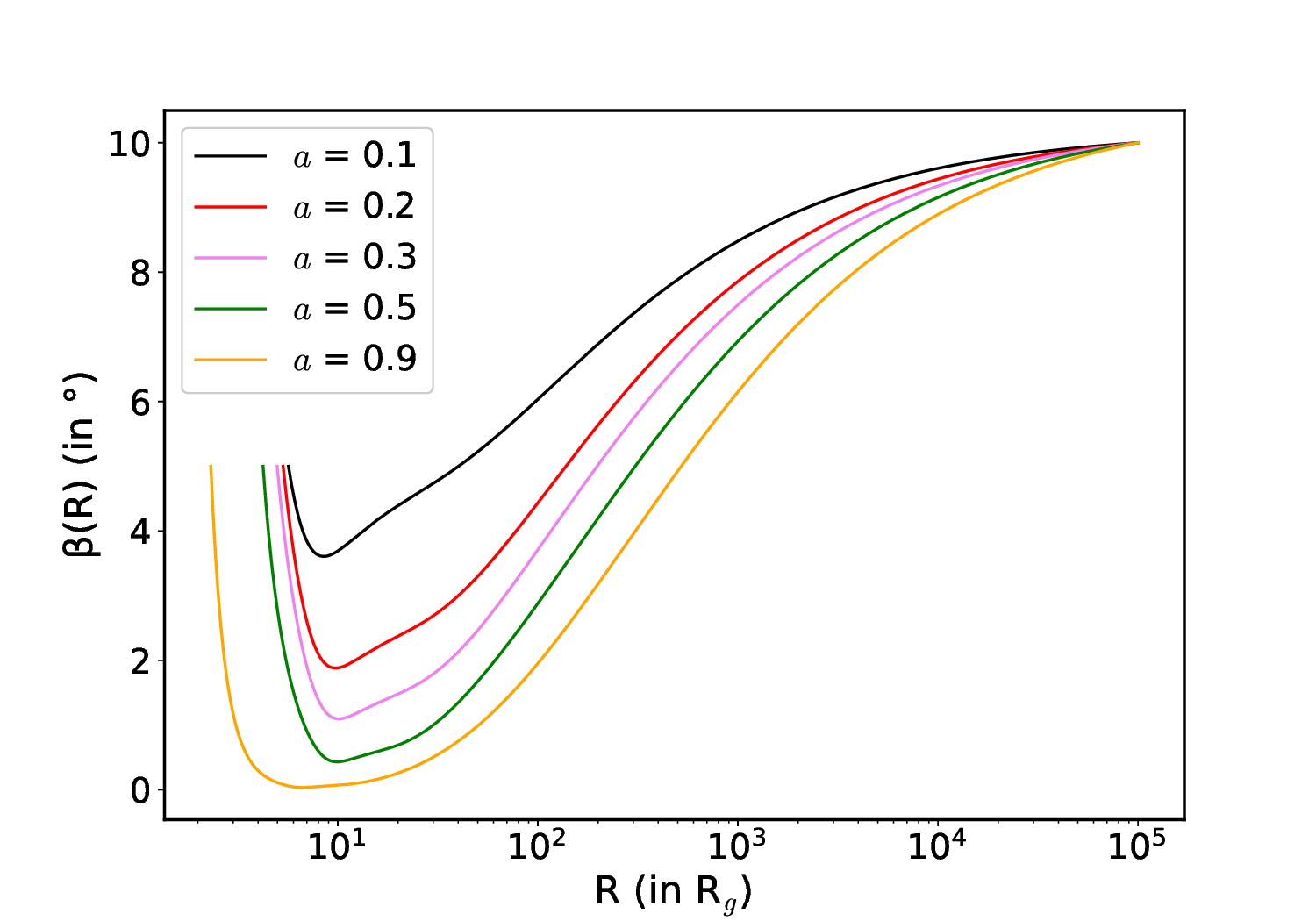}
\label{fig:sub-b}}
\caption{Radial tilt profiles of the accretion disk for different values of $a$ for a Kerr BH of mass $M=10 M_{\odot}$ with $\nu_2 = 10^{15}~\mathrm{cm}^2\mathrm{s}^{-1}$, $n = 0.25$, and $z_{\mathrm{in}} = 0.75$. The figure reveals that, under realistic parameter conditions, the inner portion of the disk can retain a substantial misalignment, implying that complete alignment is not ensured even in a steady-state configuration. See Sec. \ref{subsection_tiltprofile_BH} for details.}
\label{plot22}
\end{figure}

Figure \ref{plot22} presents the radial tilt profiles of an accretion disk around a Kerr BH for different values of the spin parameter $a$. The LT torque becomes stronger as the value of Kerr parameter $a$ increases, if we fix all the other parameters. Therefore, the expected tilt angle is smaller for higher values of $a$. The results are consistent with those reported in Fig. 7 of \cite{Banerjee_2019}. Panel (b) demonstrates that the degree of alignment increases with increasing spin, and the radial extent over which the disk remains aligned also broadens. Additionally, panel (a) indicates that even if the disk is assumed to be aligned at its inner edge (consistent with a torque-free boundary condition), it can become misaligned at slightly larger radii, depending sensitively on the value of the vertical shear viscosity $\nu_2$.
\begin{figure}[ht]
\centering
\begin{subfigure}[$a=0.3$]{\includegraphics[scale=0.3]{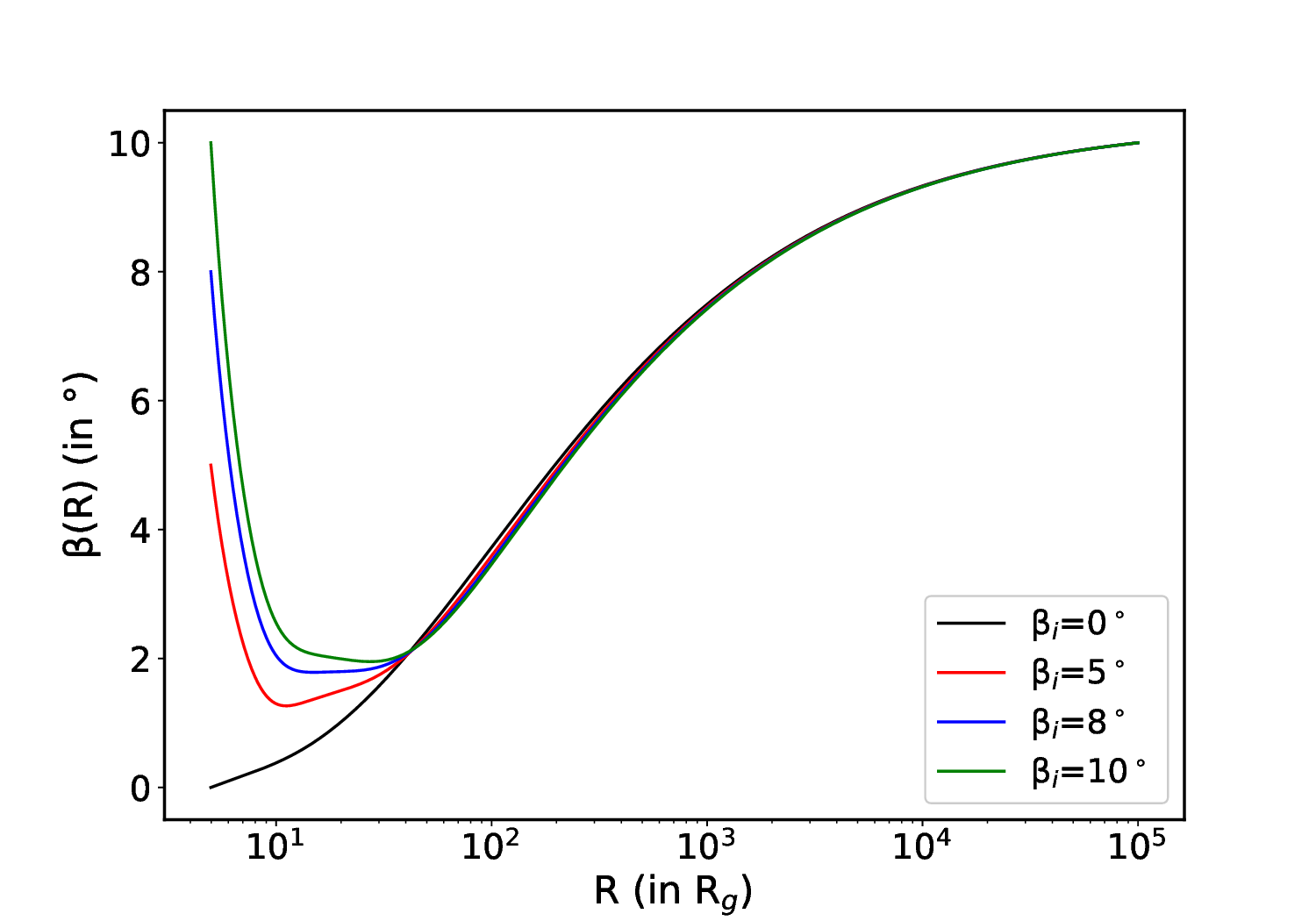}
\label{fig:sub-a}}
\end{subfigure}
\hspace{1cm}
\begin{subfigure}[$a=0.9$]{\includegraphics[scale=0.3]{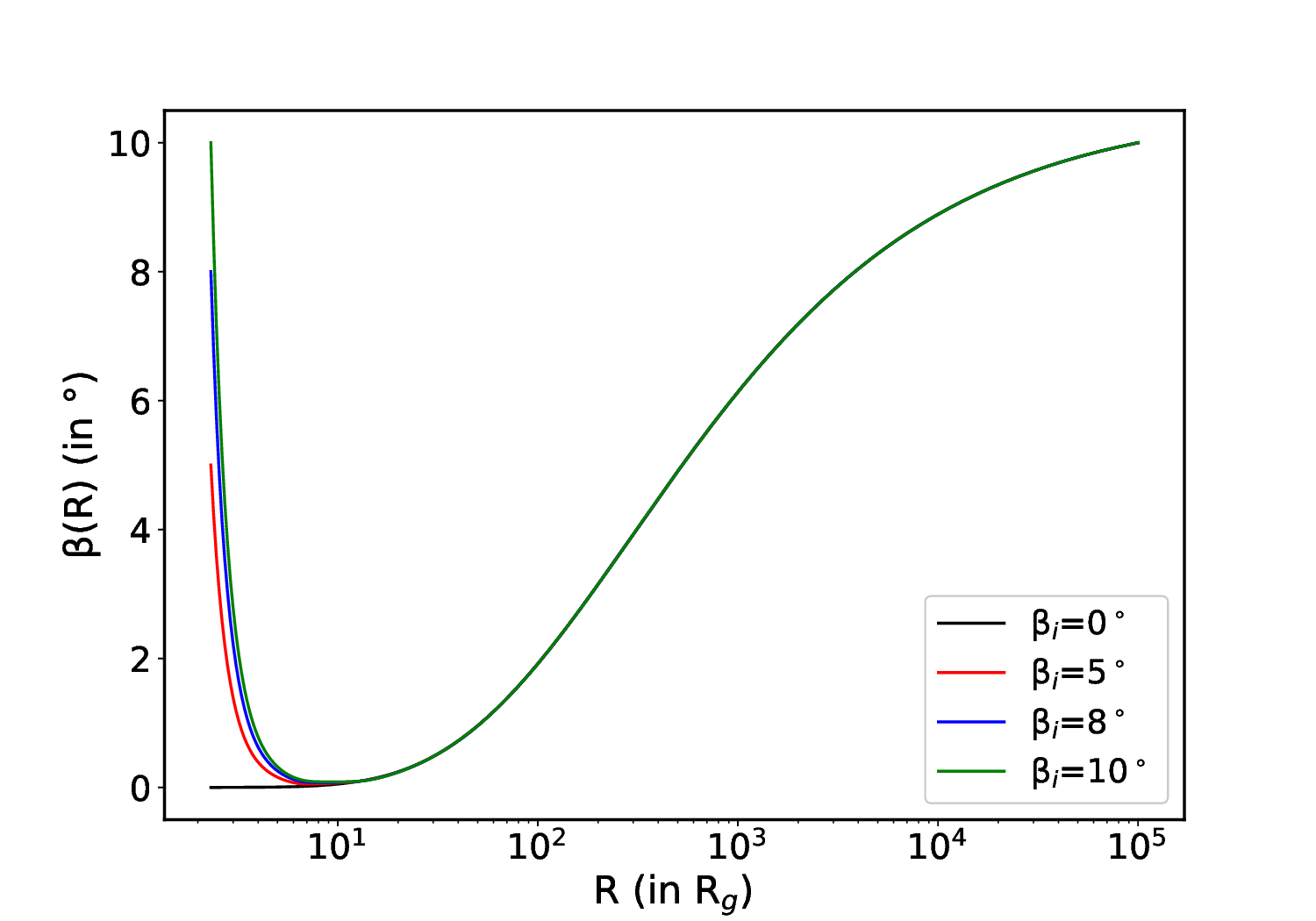}
\label{fig:sub-b}}
\end{subfigure}
\caption{Radial tilt profiles of the accretion disk are shown for different values of $\beta_i$ around a Kerr BH of $M = 10M_{\odot}$ with $\nu_2 = 10^{15}{~\mathrm{cm}^2\mathrm{s}^{-1}}$, $n = 0.25$, and $z_{\rm in} = 0.75$. As expected, the inner disk tilt increases with increasing $\beta_i$ and decreasing $a$. See Sec. \ref{subsection_tiltprofile_BH} for details.}
\label{plot3}
\end{figure}
Figure \ref{plot3} illustrates how the radial tilt profile behaves for different values of the inner tilt angle $\beta_i$ around a Kerr BH. Since the strength of the LT torque depends on the misalignment between the angular momentum of the disk and the BH, we expect an increase in LT torque for higher values of $\beta_i$ \cite{Banerjee_2019}. Therefore, the tilt profile becomes steep as $\beta_i$ increases. Since viscous torque dominates far away from the BH, the tilt profile is expected to be the same for all values of $\beta_i$ at large values of $R$ (i.e., $R\to R_{\rm f}$). From Fig. \ref{plot3}, it is clear that the influence of the LT torque near the inner edge of the disk becomes more pronounced with increasing $\beta_i$, highlighting the interplay between LT torque and viscous torques, as expected. Panel (a), which corresponds to the case of $a=0.3$, shows results consistent with Fig. 9 of \cite{Banerjee_2019}. In panel (b), representing $a=0.9$, the stronger LT effect due to the higher spin leads to a clear alignment of the disk in the inner regions.

\begin{figure}[ht]
\centering
\subfigure[$a=0.2$]{\includegraphics[scale=0.3]{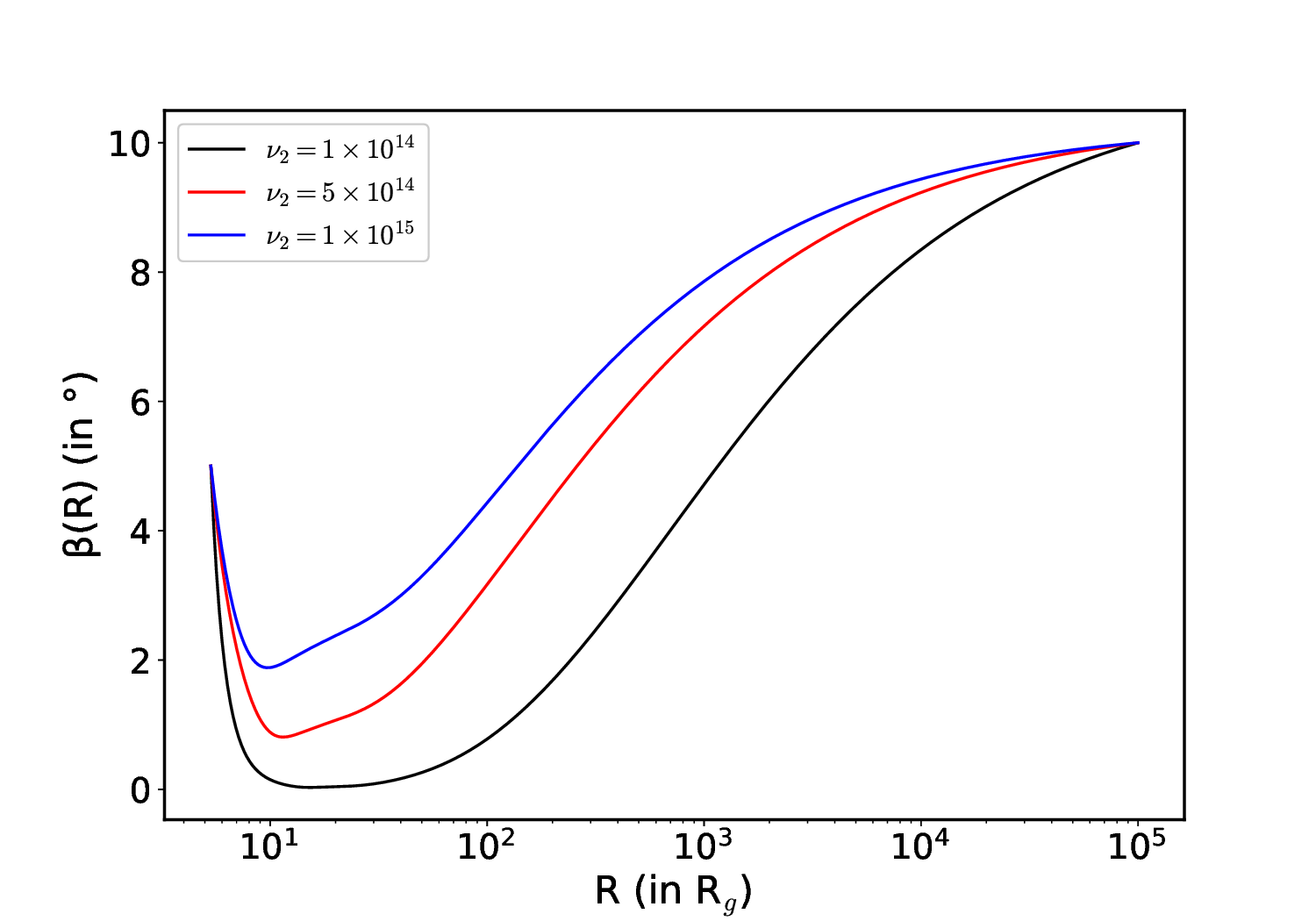}
\label{fig:sub-a}}
\hspace{1cm}
\subfigure[$a=0.9$]{\includegraphics[scale=0.3]{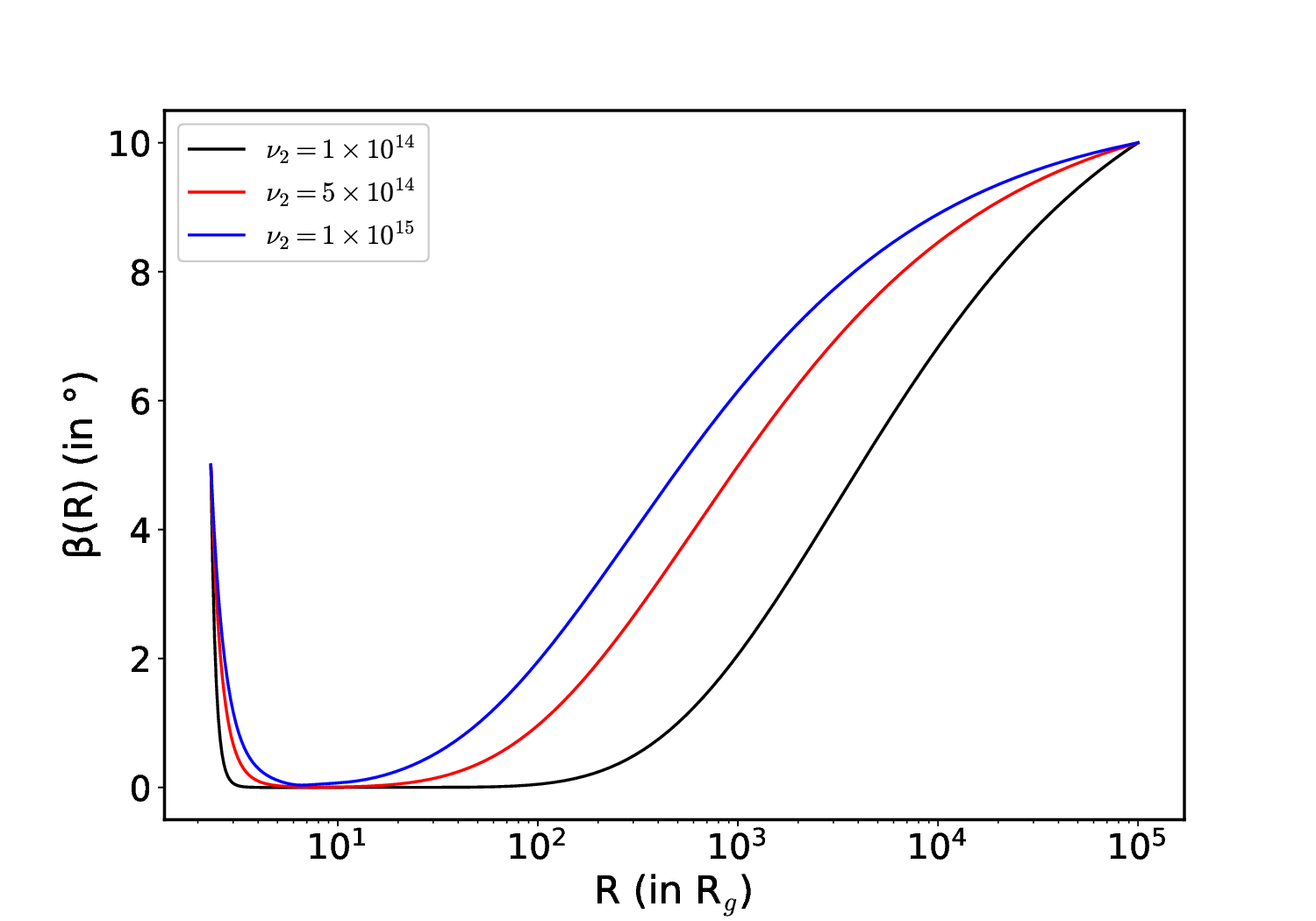}
\label{fig:sub-b}}
\caption{Radial tilt profiles are shown for different values of $\nu_2$ (in ${\rm cm^2\,s^{-1}}$) around a Kerr BH of $M = 10M_{\odot}$ with $n = 0.25$, $\beta_i = 5^\circ$, and $z_{\rm in} = 0.75$. The figure indicates that the inner disk exhibits a greater tilt with increasing $\nu_2$ and decreasing $a$. See Sec.~\ref{subsection_tiltprofile_BH} for details.}
\label{plot24}
\end{figure}

\begin{figure}[ht]
\centering
\subfigure[$\beta_i=0^\circ$]{\includegraphics[scale=0.3]{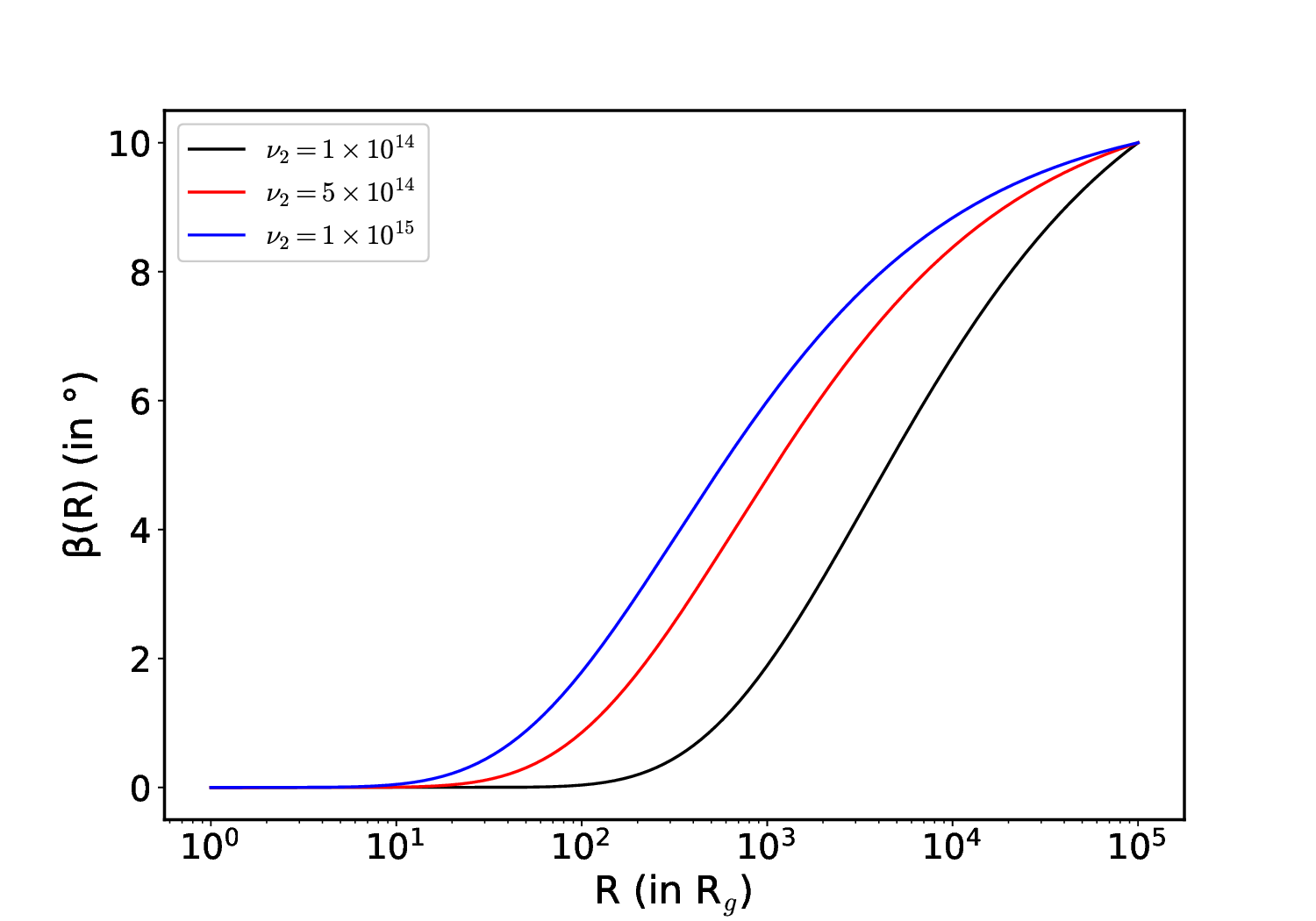}
\label{fig:sub-a}}
\hspace{1cm}
\subfigure[$\beta_i=5^\circ$]{\includegraphics[scale=0.3]{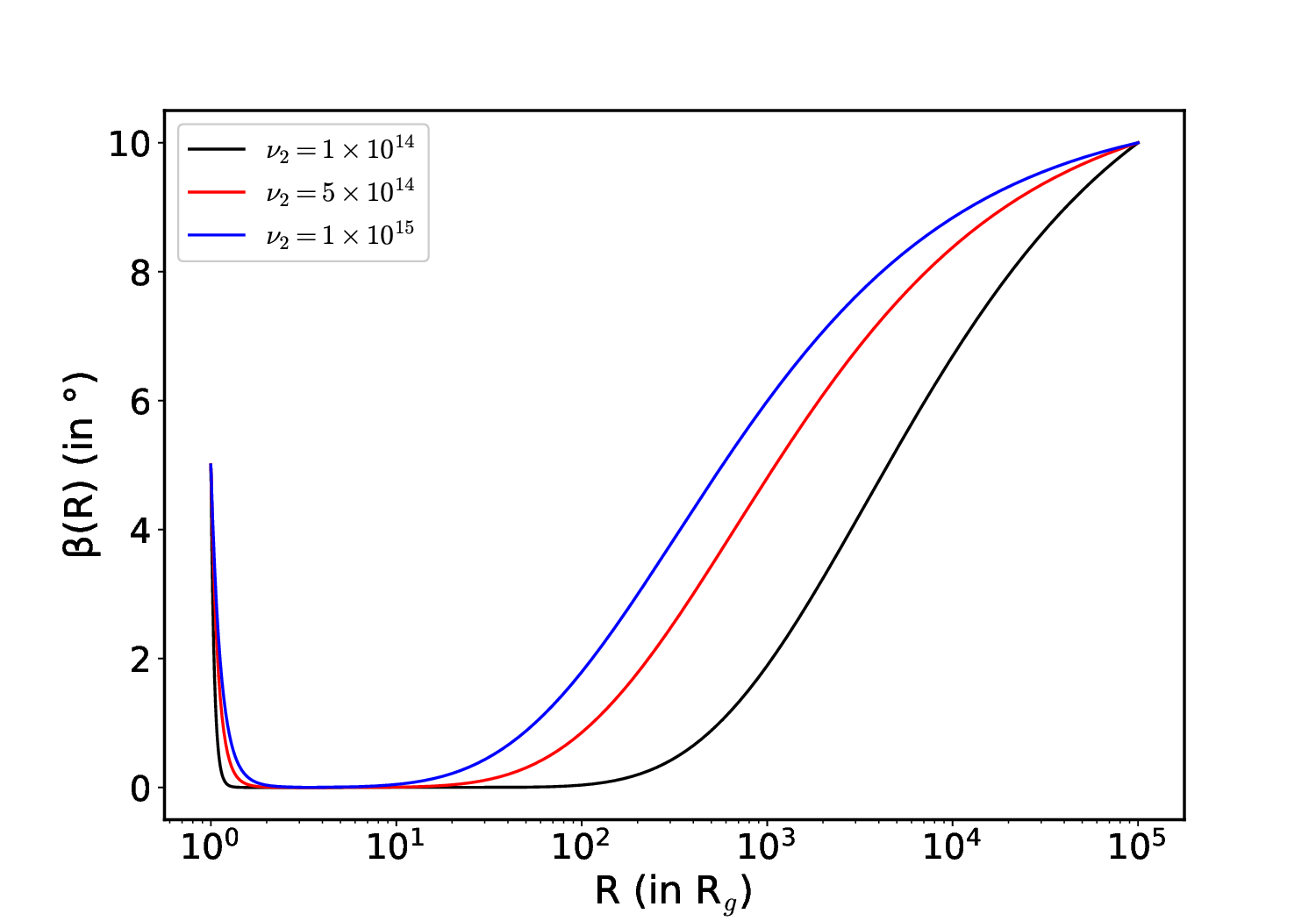}
\label{fig:sub-b}}
\caption{Similar to Fig. \ref{plot24} but for $a=1$, i.e., extremal Kerr BH.}
\label{plot6}
\end{figure}
Figure~\ref{plot24} illustrates the dependence of the radial tilt profile on the vertical shear viscosity parameter $\nu_2$. Panel (a), corresponding to the case $a=0.2$, shows qualitative agreement with Fig. 8 of \cite{Banerjee_2019}, while panel (b) represents the higher spin case of $a=0.9$. In both cases, a reduction in $\nu_2$ enhances the tendency of the disk to align, indicating that weaker vertical viscous stresses allow the LT torque to dominate more effectively and drive alignment in the inner regions.
Figure \ref{plot6} shows the variation of tilt profile for different values of $\nu_{2}$ for $\beta_i=0^\circ$ and $\beta_i=5^\circ$, respectively, for an extremal Kerr BH. The radial extent of alignment is highest for $a=1$ in the case of a Kerr BH. Still, the shape of the tilt profile has a similar shape as that given in Fig. \ref{plot24}.

\subsection{{Tilt profiles for naked singularities and their distinguishing features from black holes}
\label{subsection_tiltprofile_NaS}}

\begin{figure}[ht]
\centering
\subfigure[$\beta_i=0^\circ$]{\includegraphics[scale=0.3]{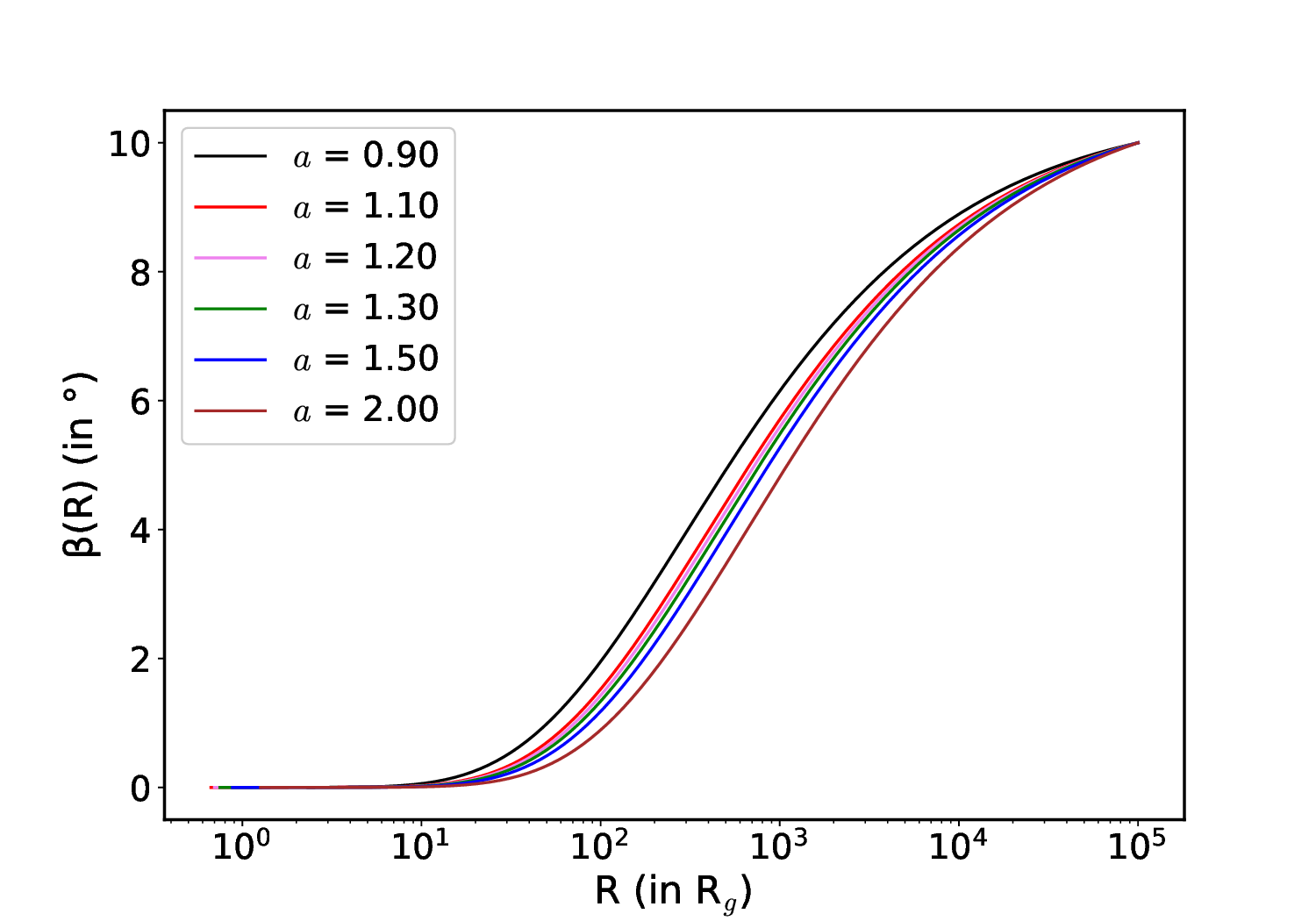}}
\hspace{1cm}
\subfigure[$\beta_i=5^\circ$]{\includegraphics[scale=0.3]{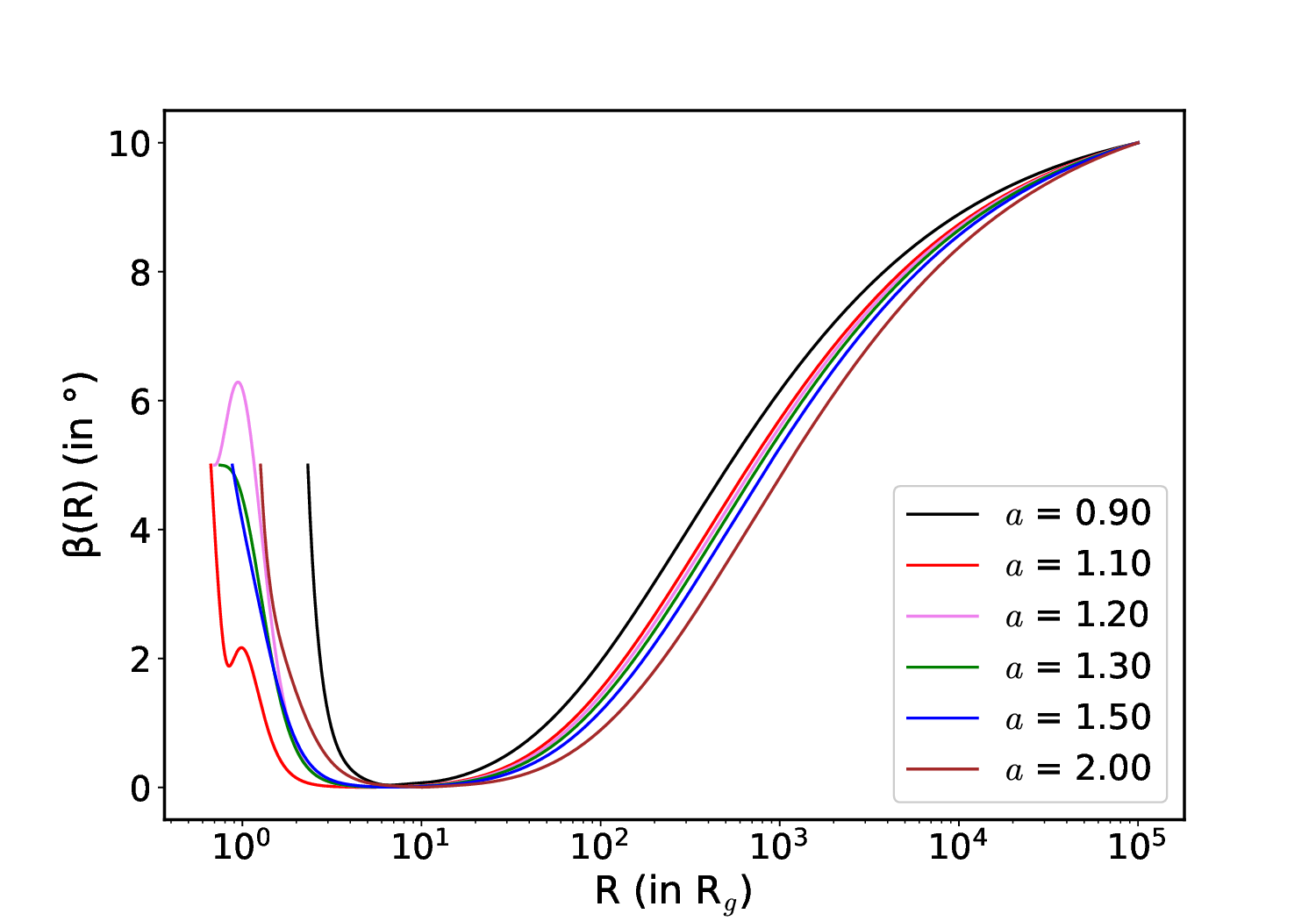}}
\caption{Radial profiles of the disk tilt angle for a Kerr NaS of $M=10M_{\odot}$ are shown for different values of the spin parameter $a$, with the $a=0.9$ case (Kerr BH) included for comparison. The other parameters used are $n=0.25$, $\nu_2 = 10^{15}~\mathrm{cm}^2\mathrm{s}^{-1}$, and $z_{\mathrm{in}}=0.75$. The black and red curves correspond to the Kerr BH ($a=0.9$, i.e., $a=1-0.1$) and the Kerr NaS ($a=1.1$, i.e., $a=1+0.1$), respectively. For Kerr NaSs with $1<a<1.3$, distinct hump-like features appear in the inner region of the disk within the diffusive regime considered here. See Sec.~\ref{subsection_tiltprofile_NaS} for further details.}
\label{plot11}
\end{figure}

Figure \ref{plot11} presents the radial tilt profiles for the Kerr NaS cases, with the $a=0.9$ Kerr BH included for comparison. A clear distinction between the BH and NaS profiles is visible in panel (b). The NaS case shows a noticeably steeper tilt gradient in the inner disk region compared to the BH. Moreover, the radial extent over which the disk achieves alignment is larger for the NaS, consistent with theoretical expectations.

Most importantly, for a nonzero inner edge tilt angle, the tilt profile of the NaS with $1 < a < 1.3$ exhibits a distinct hump (and a corresponding dip) at a radius $R > R_{\rm in}$. This feature appears only in the innermost region of the accretion disk and is characteristic of the NaS case within the diffusive regime. Notably, when $\beta_i=0^\circ$ [panel (a)], the hump is absent. For Kerr NaSs with $1 < a < 1.3$, such humps may provide a potentially useful indicator of Kerr NaSs within the diffusive regime considered here. For instance, the black and red curves correspond to the Kerr BH ($a=0.9$, i.e., $a=1-0.1$) and the Kerr NaS ($a=1.1$, i.e., $a=1+0.1$), respectively. Although both cases deviate by the same amount from the extremal value ($a=1$), the NaS curve exhibits a distinct hump, whereas the BH curve does not.

\begin{figure}[ht]
\centering
\begin{subfigure}[]{\includegraphics[scale=0.3]{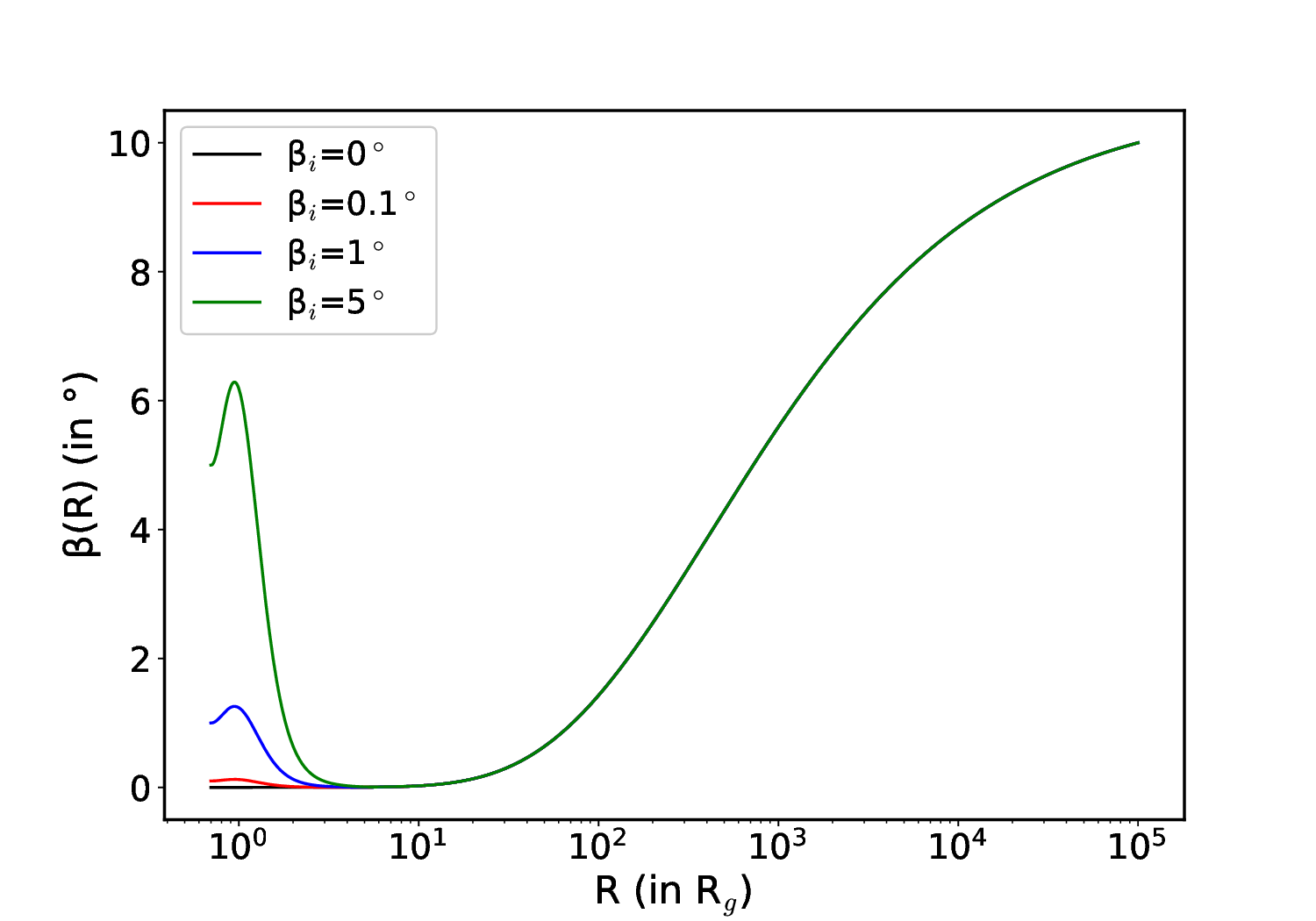}
\label{fig:sub-a}}
\end{subfigure}
\hspace{1cm}
\begin{subfigure}[]{\includegraphics[scale=0.3]{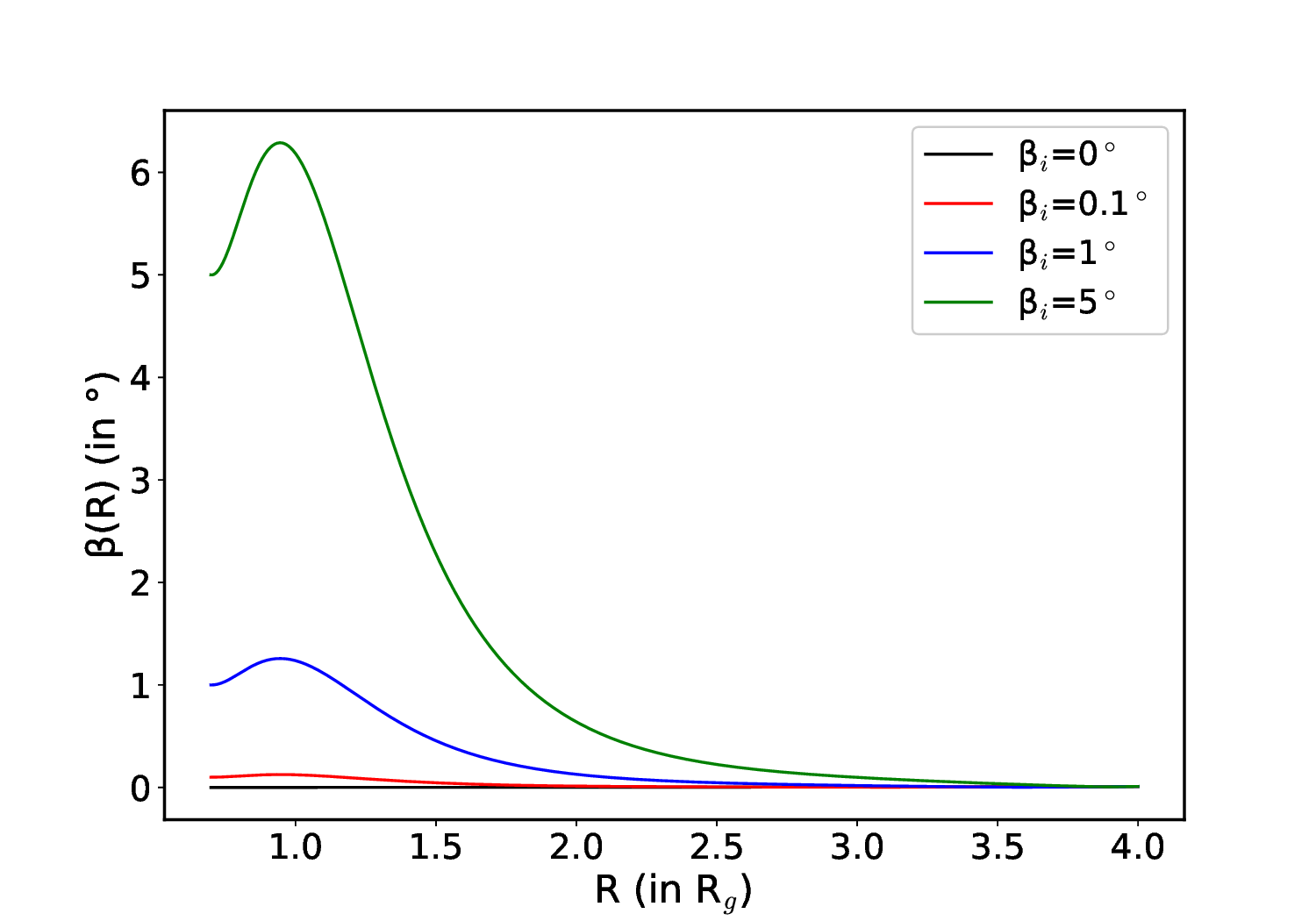}
\label{fig:sub-b}}
\end{subfigure}
\caption{Radial tilt profile of the accretion disk for different values of $\beta_i$, for a Kerr NaS of $M=10 M_{\rm \odot}$ and $a=1.2$. The other parameters are $\nu_2 = 10^{15} \rm cm^2 s^{-1}$, $n = 0.25,$ and $z_{\rm in} = 0.75$. Panel (b) shows a magnified view of the inner part of the disk shown in panel (a). This figure shows that the height of the inner hump increases with increasing the value of $\beta_i$. For further details, see Sec. \ref{subsection_tiltprofile_NaS}.}
\label{plot21}
\end{figure}

As mentioned above, the hump in the tilt profile arises only for Kerr NaSs with $1<a<1.3$ with a nonzero $\beta_i$. It occurs at a particular radius where the specific angular momentum of the accretion disk, $L$, vanishes [see Eqs. (\ref{Lp}) and (\ref{L})]. The prominence of the hump and dip features increases with higher values of $\beta_i$, as evident from Fig. \ref{plot21}, which shows the radial tilt profile of a Kerr NaS of $a=1.2$ for different values of $\beta_i$. Notably, the hump persists even for a very small inner tilt angle such as $\beta_i = 0.1^\circ$. However, we denote this specific radius as $R_{L0}$ [see the discussion following Eq. (\ref{RISCO})] where $L$ vanishes. Since the LT torque $\tau_{\rm LT}$ depends on $L$, it vanishes at $R_{L0}$. However, the viscous torques acting on the disk ($\textbf{G}_1$ and $\textbf{G}_2$) remain finite and significant at this radius. In the absence of LT torque, these strong viscous torques can induce the observed hump through vertical and radial perturbations of the disk.

The location $R_{L0}$ depends solely on the Kerr parameter $a$, which is an intrinsic property of the spacetime. In an earlier work, \cite{Pugliese_2011} discussed the existence of orbits with $L=0$ around Kerr NaSs with $1<a\leq1.3$  and showed that such orbits arise due to repulsive gravitational effects in the region near the singularity. Interestingly, in this range of $a$, we find that the specific angular momentum becomes negative in the region $R_{\rm in}<R<R_{L0}$, implying counter-rotation of the disk matter, while for $R>R_{L0}$, the disk is corotating. Thus, the accretion disk can be divided into two regions--an inner counter-rotating zone and an outer corotating zone--separated by $R_{L0}$, where the hump in the tilt profile appears.

\begin{figure}[ht]
\centering
\subfigure[]{\includegraphics[scale=0.3]{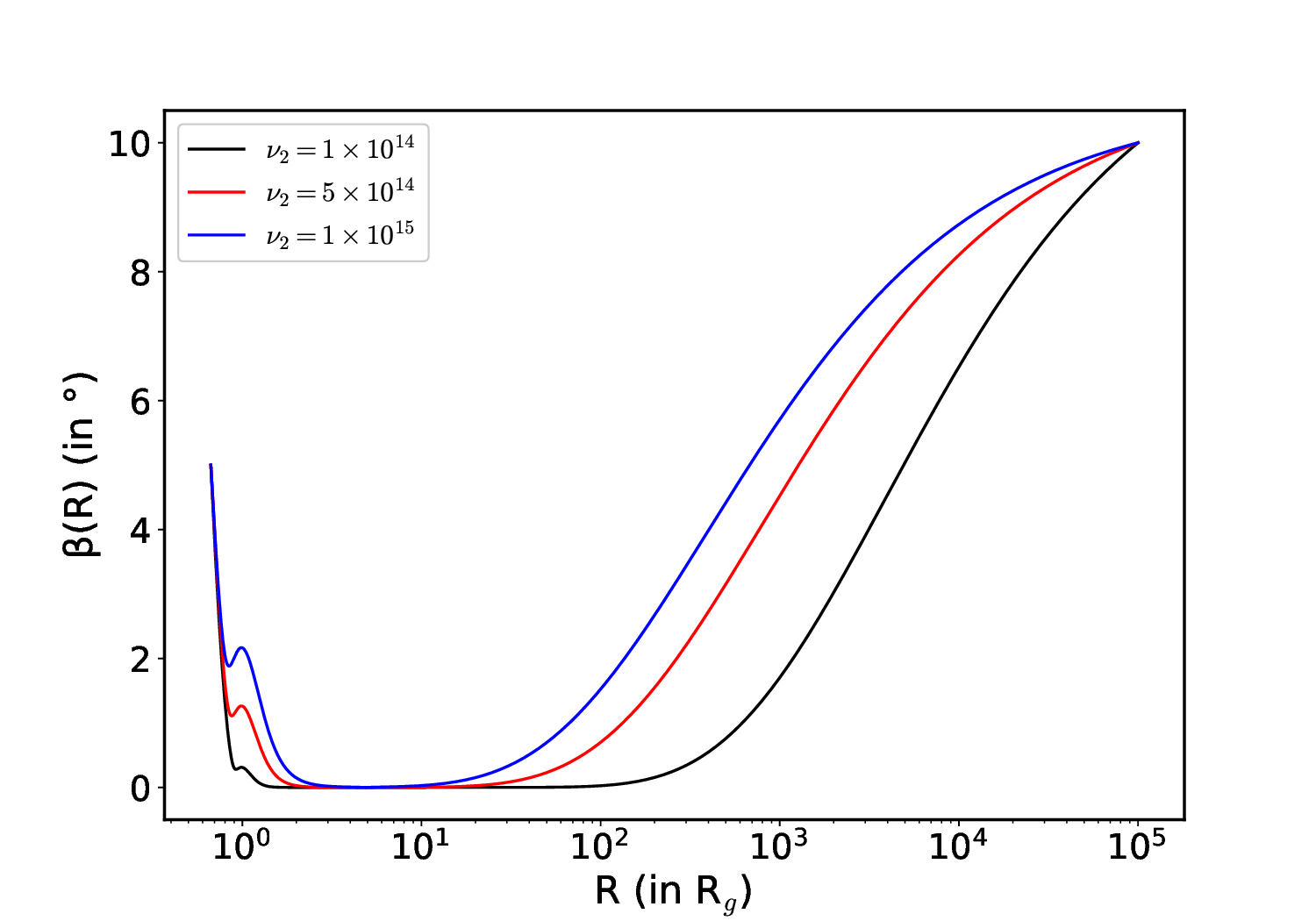}
\label{fig:sub-a}}
\hspace{1cm}
\subfigure[]{\includegraphics[scale=0.3]{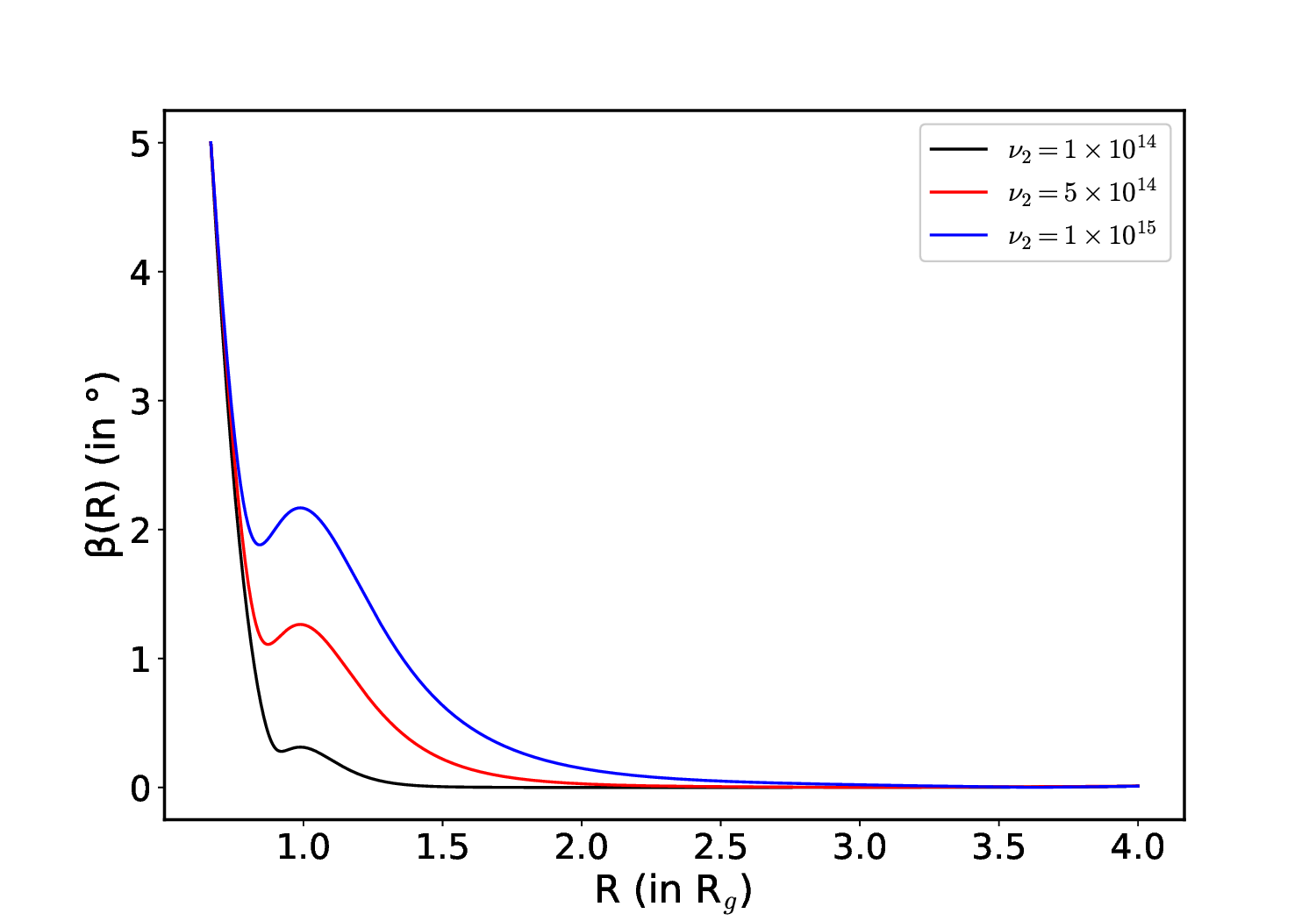}
\label{fig:sub-b}}
\caption{Similar to Fig. \ref{plot21} but for $a=1.1$ and different values of $\nu_2$. }
\label{plot7}
\end{figure}

Panel (a) of Fig.~\ref{plot7} illustrates the variation of the tilt profile with different values of the viscosity parameter $\nu_2$ for a spin parameter $a=1.1$. The results indicate that the degree of alignment increases as $\nu_2$ decreases, implying more efficient communication of the warp in the outer region for lower viscous diffusion. The distinct hump is observed in all cases (except for $\beta_i = 0,$ as discussed before), whose radial location remains nearly unchanged with respect to both $\nu_2$ and $\beta_i$. The zoomed view in panel (b) highlights this feature more clearly, showing that although the amplitude of the hump varies slightly with $\nu_2$, its position is largely insensitive to changes in the viscosity parameter.

\begin{figure}[ht]
\centering
\subfigure[$\beta_i=0^\circ$]{\includegraphics[scale=0.3]{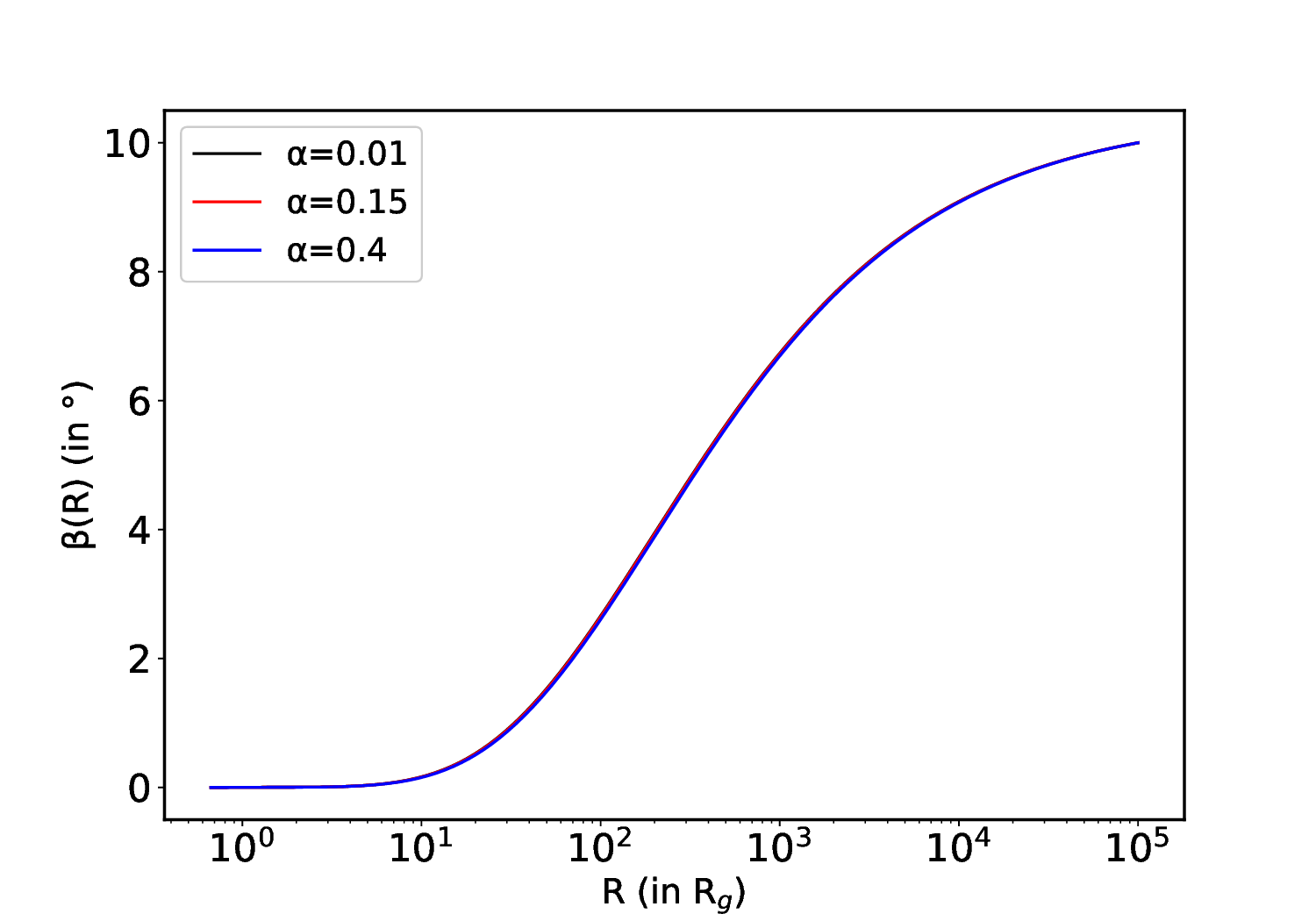}
\label{fig:sub-a}}
\hspace{1cm}
\subfigure[$\beta_i=5^\circ$]{\includegraphics[scale=0.3]{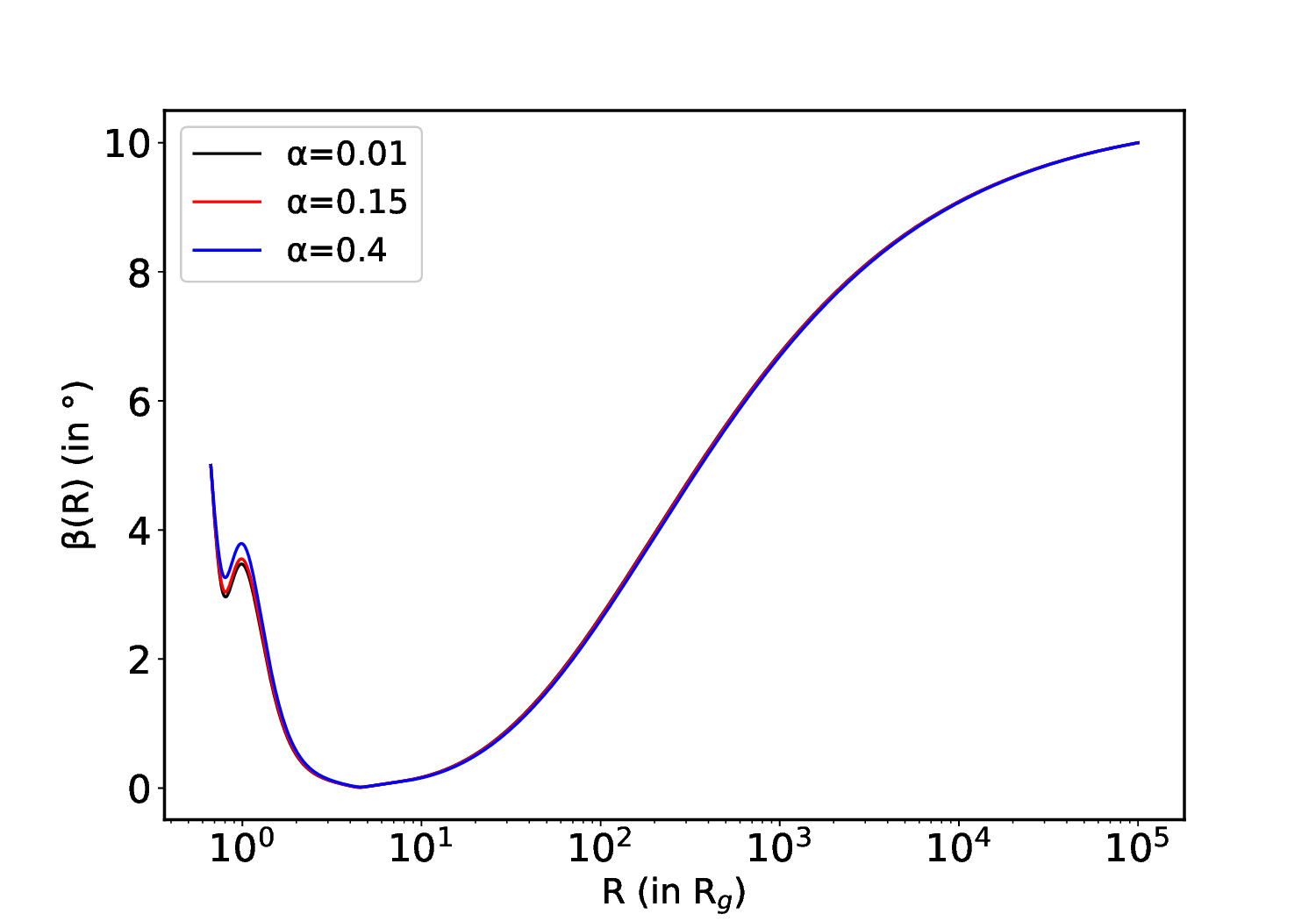}
\label{fig:sub-b}}
\caption{Radial tilt profile of the accretion disk around a Kerr NaS of $M = 10\,M_\odot$ and $a = 1.1$, shown for different values of $\alpha$, with
$\nu_2 = 10^{15}\mathrm{cm^2\,s^{-1}}$ and $z_{\rm in} = 0.75$.
For details, see Sec.~\ref{subsection_tiltprofile_NaS}.}
\label{plot25}
\end{figure}

Figure \ref{plot25} illustrates the variation of the tilt profile with different values of the viscosity parameter $\alpha$ (or equivalently $n$) for a Kerr NaS with spin $a = 1.1$. It is evident that when the initial tilt $\beta_i \neq 0^\circ$, the magnitude of the hump developed in the inner portion of the disk increases with $\alpha$. The effect is most prominent for $\alpha = 0.4$, indicating that stronger viscous coupling enhances the local misalignment between the disk plane and the BH spin axis. For lower values of $\alpha$, the profile remains comparatively smoother. Here we fix the vertical shear viscosity $\nu_2$, which controls warp diffusion and vary $\alpha$ only through its effect on the azimuthal viscosity $\nu_1$. For fixed $\nu_2$, the torque associated with warp diffusion ($\mathbf{G}_2$) remains unchanged. The increasing $\nu_1$ enhances the azimuthal shear torque ($\mathbf{G}_1$) which acts perpendicular to the disk plane. Thus at $R_{L0}$ where LT torque is zero and $\mathbf{G}_2$ is fixed, this increase in $\mathbf{G}_1$ can enhance the magnitude of the hump.

\begin{figure}[h!]
\centering
\begin{subfigure}[$\nu_{\rm 2} = 5\times10^{14}$ cm$^2$ s$^{-1}$]{\includegraphics[scale=0.3]{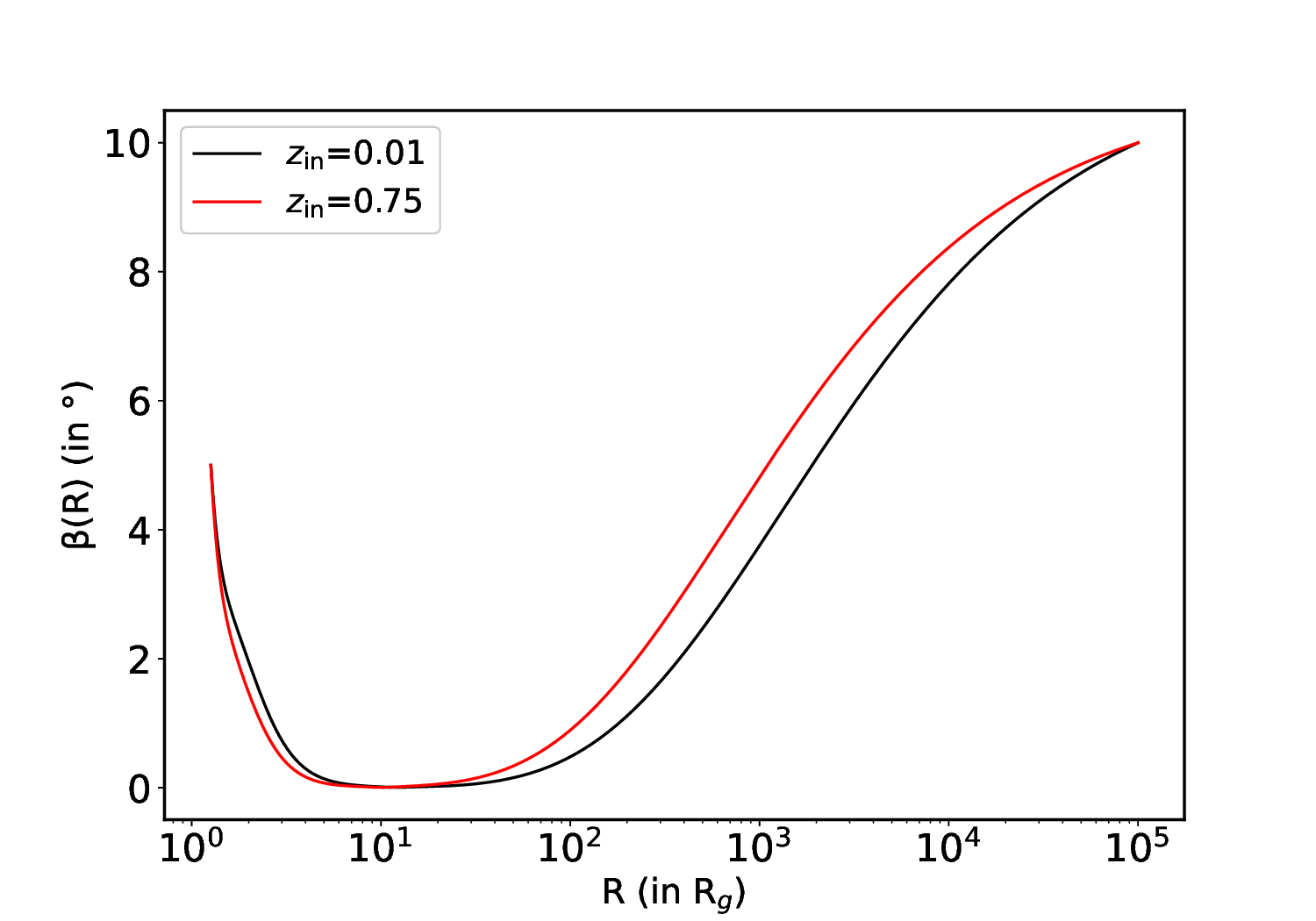}
\label{fig:sub-a}}
\end{subfigure}
\hspace{1cm}
\begin{subfigure}[$\nu_2 = 10^{15}$ cm$^2$ s$^{-1}$]{\includegraphics[scale=0.3]{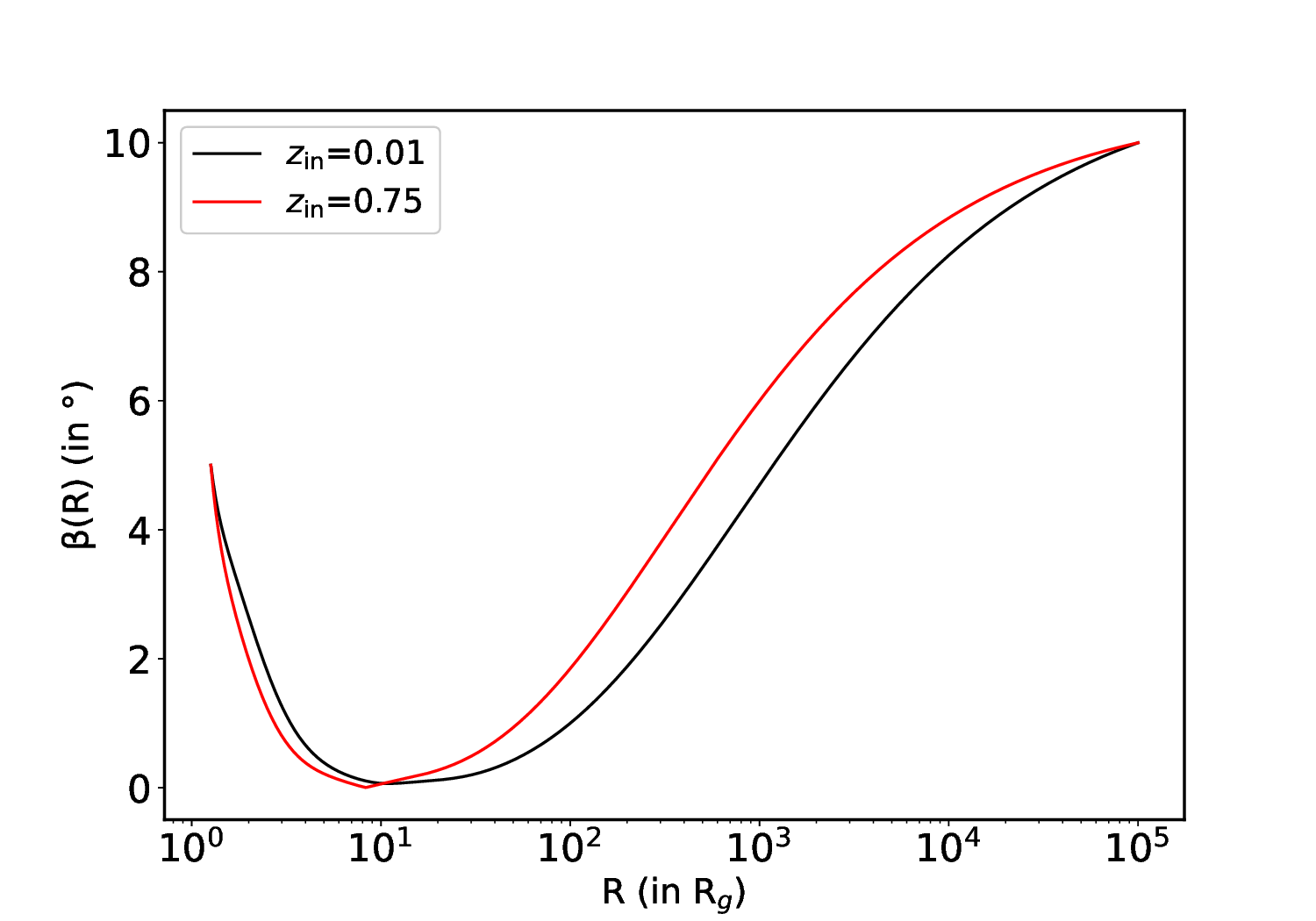}
\label{fig:sub-b}}
\end{subfigure}
\caption{Radial tilt profile of the accretion disk around a Kerr NaS of $M = 10\,M_{\odot}$ and $a = 2$ with $\beta_i=5^\circ$ and $n = 0.25$ shown for different values of $z_{\rm in}$ and $\nu_2$. See Sec.~\ref{subsection_tiltprofile_NaS} for further details.}
\label{plot18}
\end{figure}

Figure~\ref{plot18} shows the dependence of the radial tilt profile on the surface density parameter $(z_{\rm in})$ and viscosity for a Kerr NaS with spin parameter $a = 2$. Panels~(a) and~(b) correspond to $\nu_2 = 5\times10^{14}\,{\rm cm^2\,s^{-1}}$ and $\nu_2 = 10^{15}\,{\rm cm^2\,s^{-1}}$, respectively, with each panel displaying the profiles for $z_{\rm in}=0.75$ and $z_{\rm in}=0.01$. In both cases, the alignment extends over a larger radial range for the disk with a lower surface density parameter ($z_{\rm in}=0.01$). Since the viscous torque is proportional to the surface density [see Eqs. (\ref{sdf}),(\ref{G1}) and (\ref{G2})], a smaller $z_{\rm in}$ results in reduced $\Sigma$, thereby weakening the viscous coupling and allowing the LT torque to dominate over a broader region. Thus, the surface density profile critically influences the competition between viscous and LT torques and, consequently, the overall shape of the tilt distribution.

\begin{figure}[ht]
\centering
\subfigure[$R_{\rm in}=1.26 R_g$]{
\includegraphics[scale=0.3]{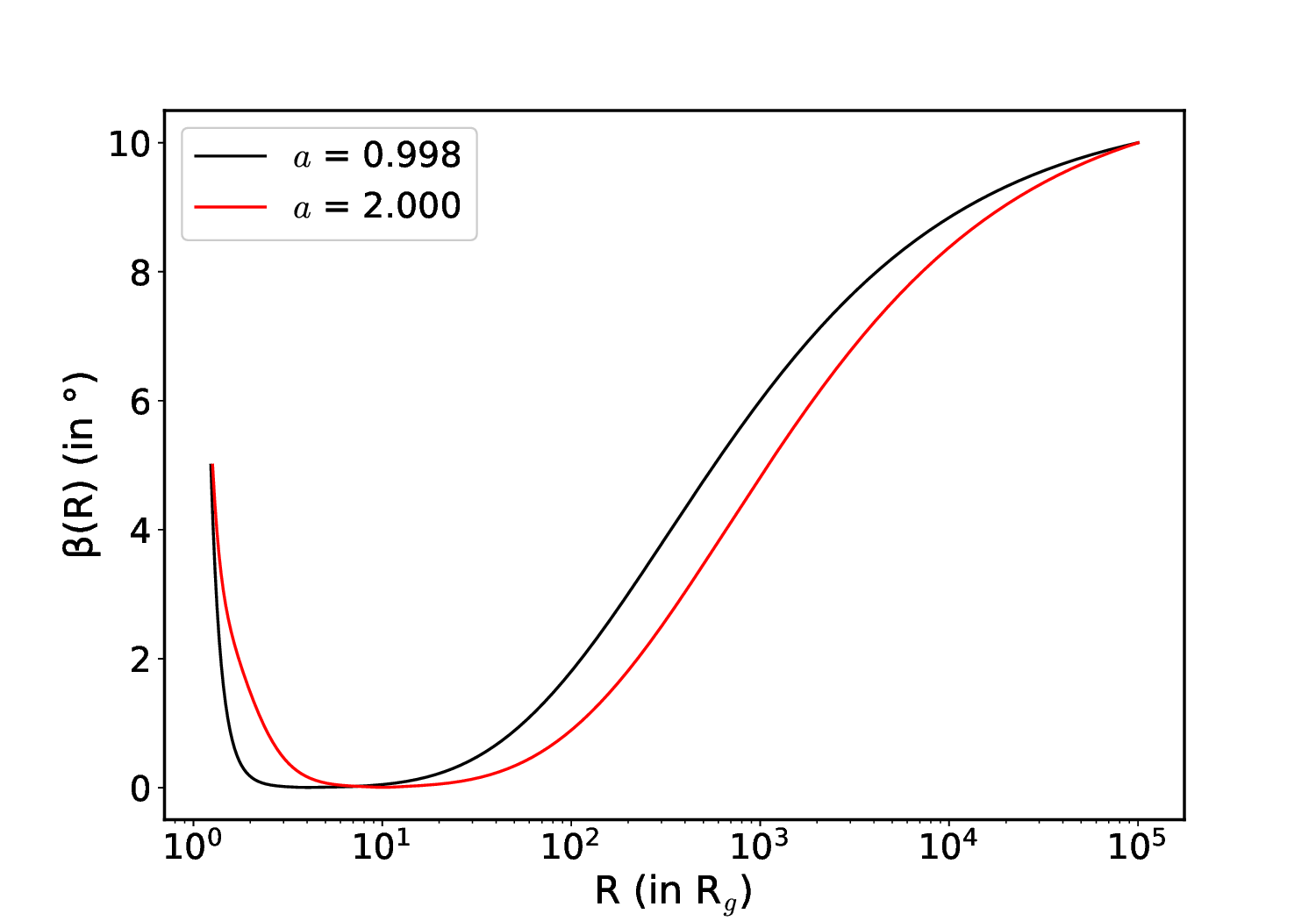}
\label{fig:sub-a}}
\hspace{1cm}
\subfigure[$R_{\rm in}=1.93 R_g$]
{\includegraphics[scale=0.3]{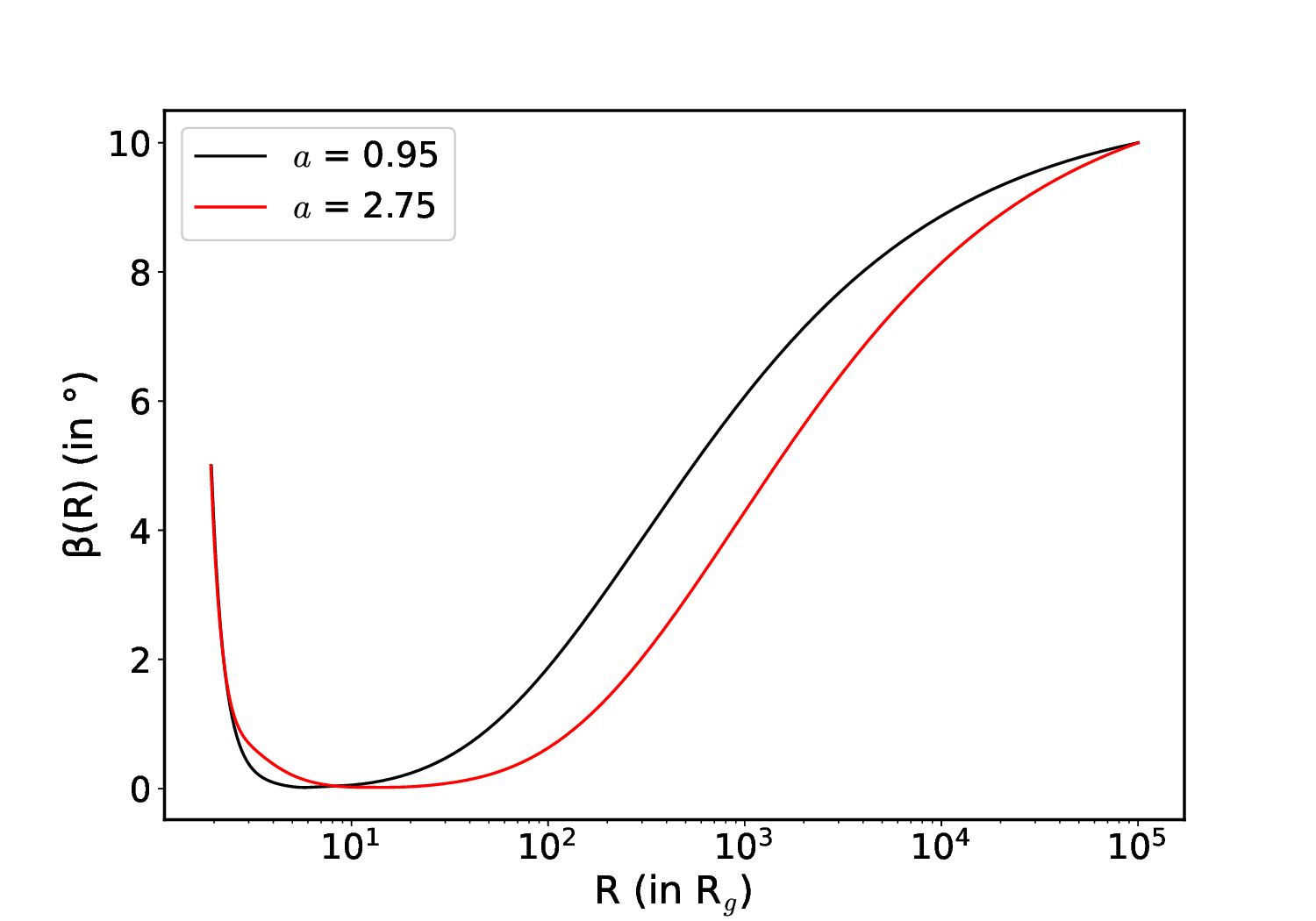}
\label{fig:sub-b}}
\caption{Comparison of the radial tilt profiles of accretion disks around a
Kerr BH and a Kerr NaS of
$M = 10\,M_{\odot}$, $\nu_2 = 10^{15}~\mathrm{cm}^2\,\mathrm{s}^{-1}$,
$\beta_i = 5^{\circ}$, $n = 0.25$, and $z_{\mathrm{in}} = 0.75$ with identical ISCO radii ($R_{\rm in}$) but different $a$.
The figure demonstrates that the tilt profiles remain
clearly distinct, revealing an observable difference between the Kerr BH and the Kerr NaS in the diffusive regime, even if the ISCO radii are the same. For more details, see Sec.~\ref{subsection_tiltprofile_NaS}.}
\label{pn}
\end{figure}

For spin parameters \( a > 1.3 \), circular orbits with \( L \leq 0 \) no longer exist; consequently, the characteristic hump (and accompanying dip) in the tilt profile also disappears. Nevertheless, the tilt profile itself could potentially serve as a diagnostic to distinguish between a Kerr BH and a Kerr NaS in the diffusive regime, as discussed below.

For instance, Fig.~\ref{pn} presents a comparison of the radial tilt profiles for accretion disks around a Kerr BH and a Kerr NaS that share the same ISCO radius. This setup enables a direct evaluation of how differences in the underlying spacetime geometry affect the disk’s tilt structure, independent of the location of the inner edge. Despite having identical \( R_{\rm in} \), the two disks display distinctly different tilt behaviors.

For the Kerr BH ($0 < a < 1$), the tilt angle $\beta(R)$ first decreases smoothly and monotonically with increasing radius. The disk aligns progressively toward the equatorial plane of the BH, and the alignment occurs over a relatively narrow radial extent. The behavior is consistent with the dominance of the LT precession in the inner region, counteracted by viscous diffusion at larger radii.

In contrast, for the Kerr NaS ($a > 1$), the tilt profile follows a distinctly different trend even though the disk begins at the same ISCO radius. The overall alignment occurs over a larger radial extent, and the inner region exhibits a steeper decline of the tilt angle. This arises from the stronger LT effects and modified precession dynamics in the Kerr NaS, which enhance the coupling between the LT and viscous torques. Consequently, the disk morphology is more sensitive to the spacetime spin structure in the NaS regime.

The key inference from Fig.~\ref{pn} is that equating the ISCO radius does not yield identical tilt behavior for a Kerr BH and a Kerr NaS, even in the outer regions of the disk. Despite having the same inner-edge boundary conditions, exact expression of the LT precession frequency and the specific angular momentum distribution maintain a clear distinction between the two cases across both the inner and outer disk regions. Consequently, the tilt profile retains sensitivity to the underlying spacetime geometry and may potentially help distinguish Kerr BHs from Kerr NaSs when combined with additional observational constraints.

\begin{figure}[ht]
\centering
\subfigure[$\beta_i=0^\circ$]{
\includegraphics[scale=0.3]{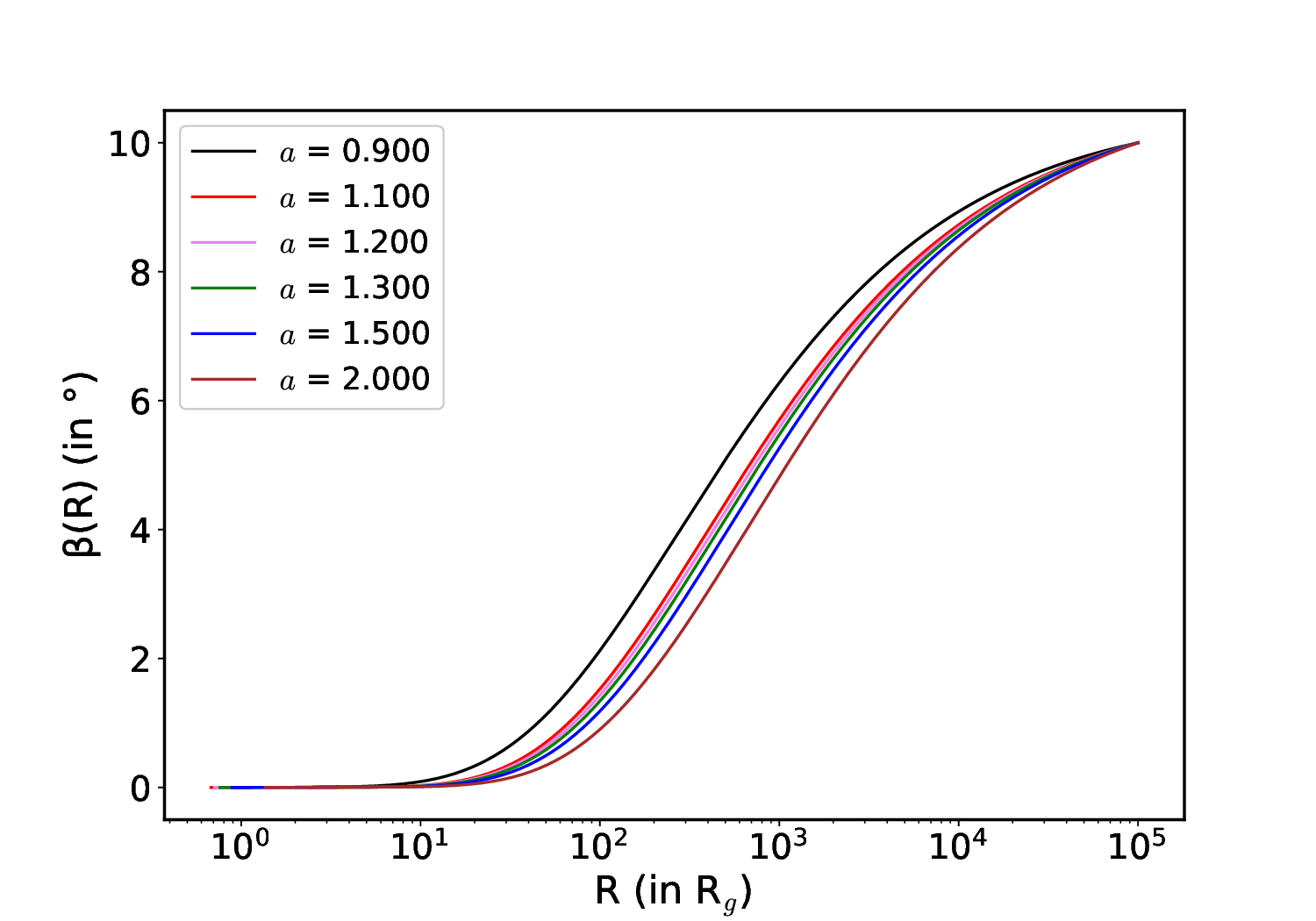}
\label{fig:sub-a}}
\hspace{1cm}
\subfigure[$\beta_i=5^\circ$]
{\includegraphics[scale=0.3]{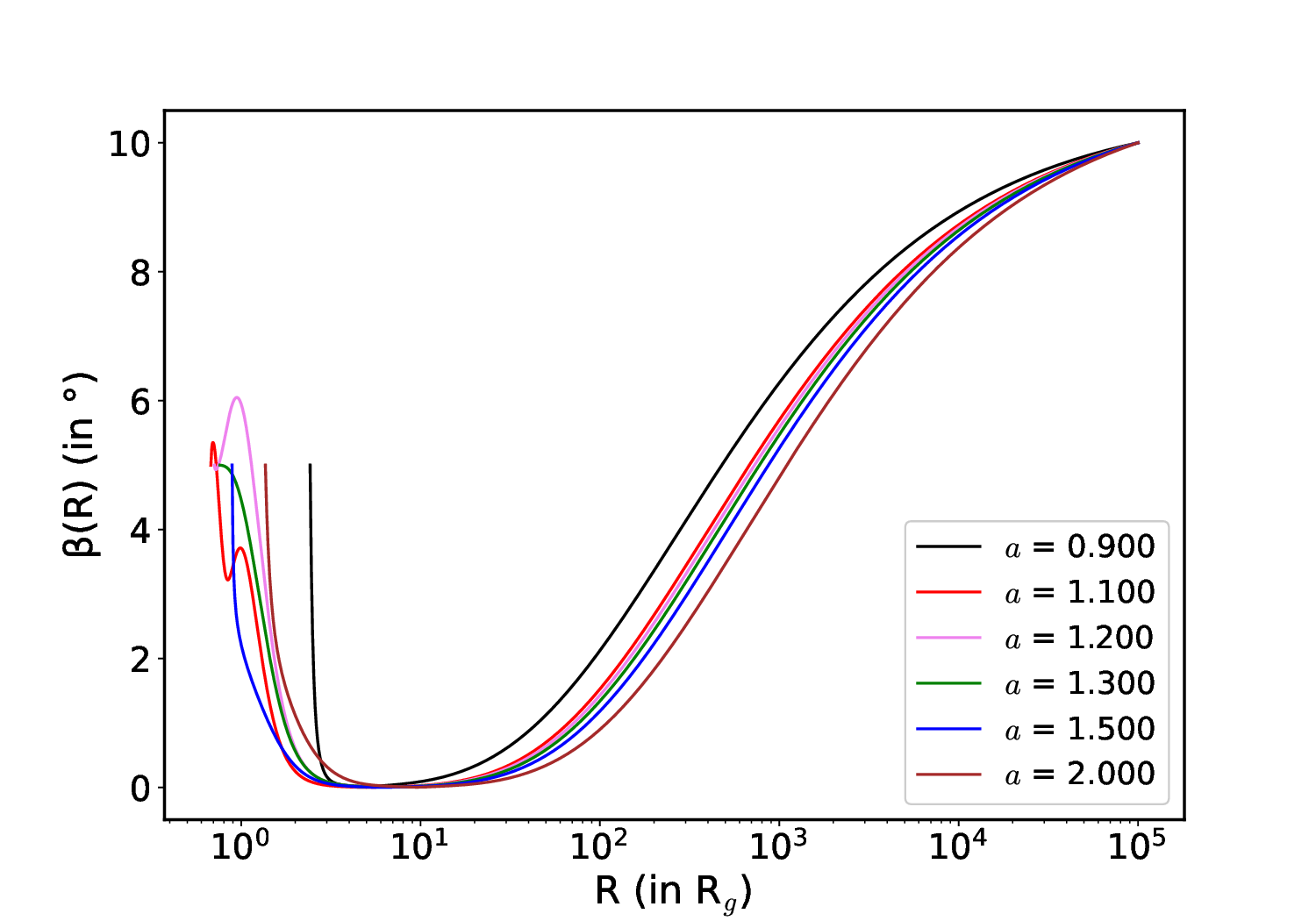}
\label{fig:sub-b}}
\caption{Radial profiles of the disk tilt angle for a Kerr NaS with \(M=10M_{\odot}\) are shown for different values of \(a\), with the \(a=0.9\) Kerr BH case included for comparison for \(\Sigma_{\rm in}\rightarrow 0\). The other parameters are \(n = 0.25\) and \(\nu_2 = 10^{15}~\mathrm{cm}^2\,\mathrm{s}^{-1}\). See Sec.~\ref{subsection_tiltprofile_NaS} for further details.}
\label{plot26}
\end{figure}

The alternative inner boundary condition \(\Sigma_{\rm in}\to0\), discussed in Sec.~\ref{subsection_viscous torque}, corresponds to the limit \(C \rightarrow 2/(\eta_{\rm in}\sqrt{R_{\rm in}})\) in Eq.~(\ref{C}), yielding \(L|_{R=R_{\rm in}} \rightarrow 0\) (see Eq.~\ref{LC1}). 
The resulting tilt profiles are shown in Fig.~\ref{plot26}. Panel~(a) corresponds to \(\beta_i=0^\circ\), where the overall tilt profile remains qualitatively similar to the \(\Sigma_{\rm in}\neq0\) case [see panel~(a) of Fig.~\ref{plot11}]. Panel~(b) shows the \(\beta_i=5^\circ\) case, demonstrating that the hump--dip structure persists even in the \(\Sigma_{\rm in}\to0\) limit, although the detailed behavior close to the inner edge differs from the nonzero \(\Sigma_{\rm in}\) case [see panel~(b) of Fig.~\ref{plot11}]. These results indicate that the choice of the inner boundary condition mainly affects the immediate vicinity of the inner edge, while the global tilt structure and the main qualitative features of the solutions remain largely preserved.
\\

 A possible concern is that the hump-like structure disappears in the limiting case $\beta_i=0$, which may arise if the inner disk undergoes strong BP alignment. GRMHD simulations by \cite{Liska2019, Liska2021} indeed demonstrated substantial inner-disk alignment for very thin accretion disks ($H/R\sim 0.015-0.05$) around a rapidly spinning BH $(a=0.9375)$ evolving in the diffusive regime. Therefore, in some accretion states, the inner tilt may become sufficiently small (or even zero) such that the hump-like feature discussed here could be strongly suppressed or absent.
At the same time, the degree of inner alignment is expected to depend sensitively on the disk thickness, viscosity, magnetic stresses, cooling prescription, and BH spin. In particular, \cite{Zhuravlev2014} showed that warped disks may exhibit incomplete inner alignment even in the diffusive regime (see Appendix B of \cite{Zhuravlev2014}) for slow-spin BHs, with the tilt relaxing toward the equatorial plane without vanishing completely. Moreover, the GRMHD simulations discussed above were performed for Kerr BHs, whereas the hump-like structures identified in the present work arise primarily in NaSs, where the inner spacetime geometry and accretion-flow dynamics may differ significantly from the standard BH case. Therefore, although the special case $\beta_i=0$ would eliminate the hump feature, finite residual inner-disk misalignment may still remain physically plausible in such systems. Our solutions should thus be interpreted as relevant to accretion flows in which a nonzero inner tilt survives in the innermost region. We also note that the hump-like feature persists even for relatively small nonzero values of $\beta_i$ (see Fig.~\ref{plot21}).

It is important to note that hump-like features, often referred to as ``tilt oscillations'', have also been reported for Kerr BHs in the bending-wave regime ($\alpha < H/R$) \cite{Lubow_2002, Nealon_2015, Drewes_2021, Ivanov_1997}, where they arise due to wave-mediated warp propagation and depend sensitively on the properties of the accretion disk. In contrast, in the diffusive regime considered in this work, the hump feature appears to be closely connected to the spacetime properties and is associated with the vanishing of the angular momentum. The distinction between the wave-like and diffusive regimes is set by the relation between the viscosity parameter $\alpha$ and the disk aspect ratio $H/R$. It is suggested that $\alpha$ is measurable through modeling of outburst light curves and viscous timescales within the disk instability framework \cite{King_2007, lasota_2001, schreiber_2003, hameury_1998, lasota_2008}. The disk aspect ratio $H/R$ can be calculated via eclipse mapping \cite{wood_1989, baptista_2000}, or from modeling of X-ray transient outbursts \cite{dubus_2001}.
In this context, if a given system is observationally constrained to lie in the diffusive regime, the presence of hump-like features in the tilt profile may provide supportive evidence for Kerr NaS configurations. However, hump-like features alone cannot serve as unique identifiers. Moreover, observational constraints on \(\alpha\) and \(H/R\) may not always be sufficiently conclusive to unambiguously distinguish between the two regimes. Nevertheless, the behavior of the tilt profile in the diffusive regime considered here may still offer useful insight into the underlying spacetime geometry when interpreted together with additional observational information.

\subsection{Comparison between exact and approximate formulations
\label{subsection_comparison}}

\begin{figure}[h!]
\centering
\begin{subfigure}[]{\includegraphics[scale=0.3]{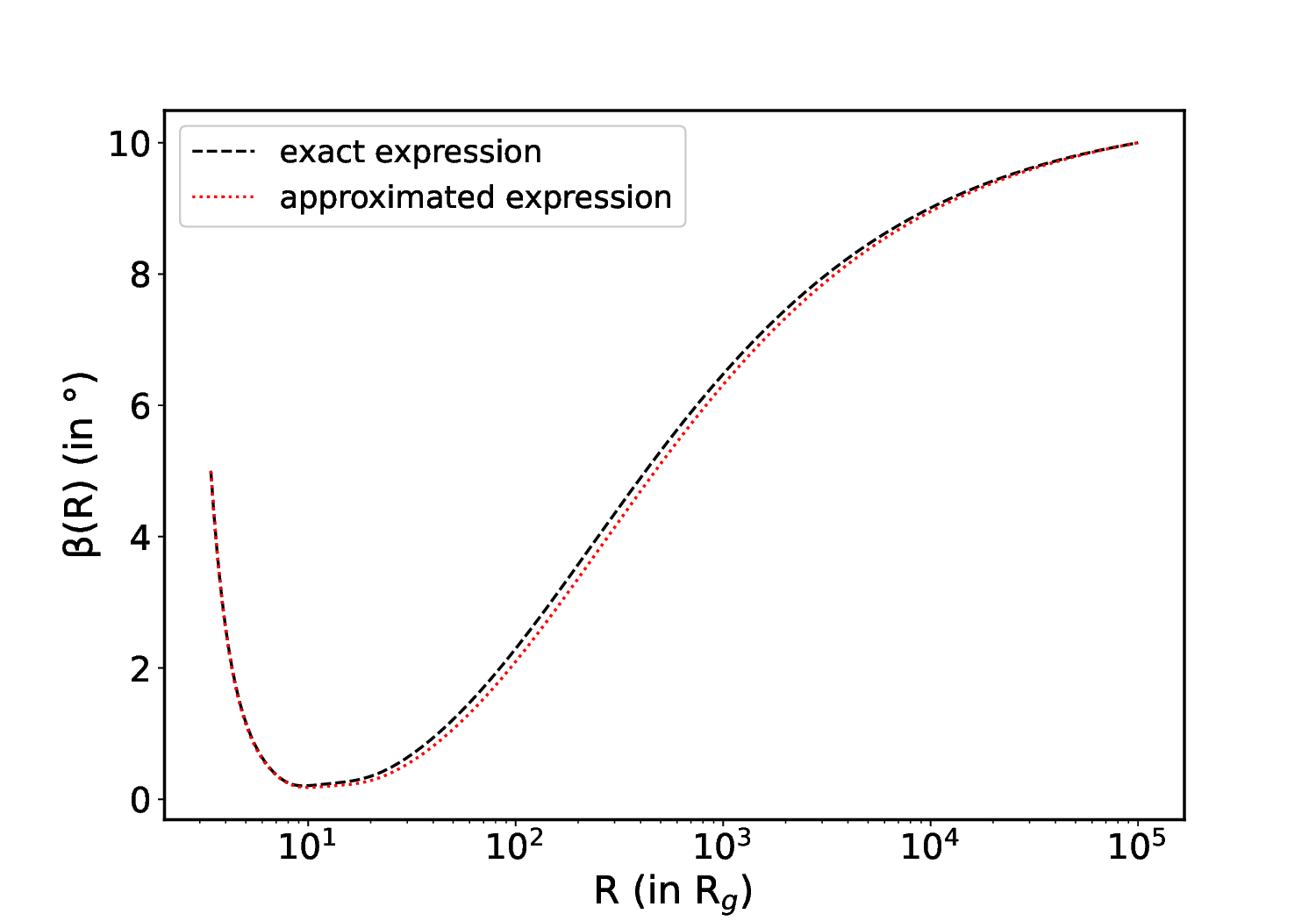}
\label{fig:sub-a}}
\end{subfigure}
\hspace{1cm}
\begin{subfigure}[]{\includegraphics[scale=0.3]{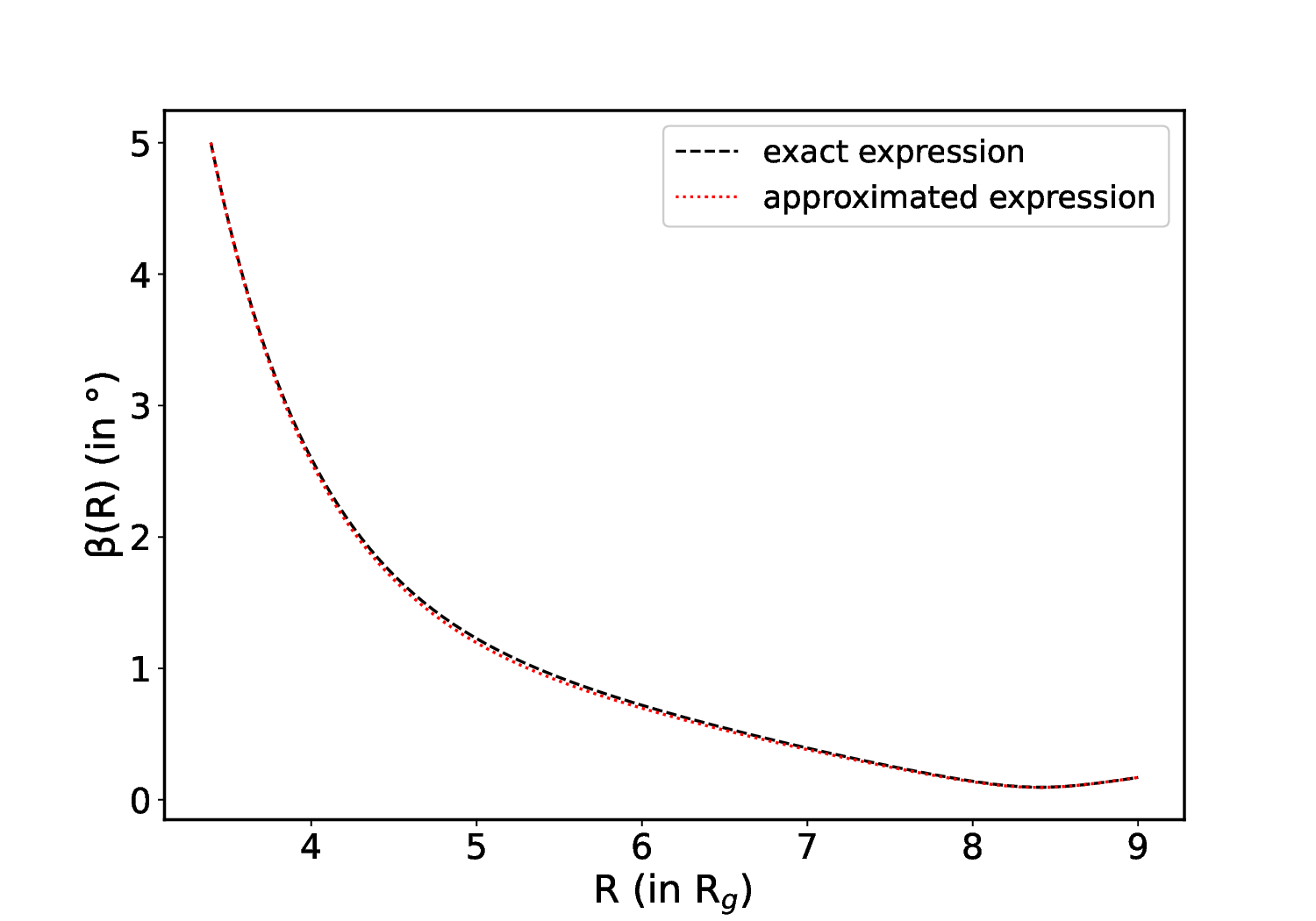}
\label{fig:sub-b}}
\end{subfigure}
\caption{Radial profiles of the accretion disk tilt angle are shown for a Kerr BH of $M=10 M_{\rm \odot}$ and $a=0.7$. Panel (b) provides a magnified view of the inner disk shown in panel (a). Other parameters are $\nu_2 = 10^{15}$ cm$^2$ s$^{-1}$, $n = 0.25$, and $z_{\text{in}} = 0.75$. This figure shows that the maximum deviation between the two expressions/approaches is $\sim 2.92\%$ (negligibly small) at $R \sim 4.8R_{g}$ for $a=0.7$. See Sec. \ref{subsection_comparison} for further details.}
\label{plot1}
\end{figure}

\begin{figure}[h!]
\centering
\begin{subfigure}[]{\includegraphics[scale=0.3]{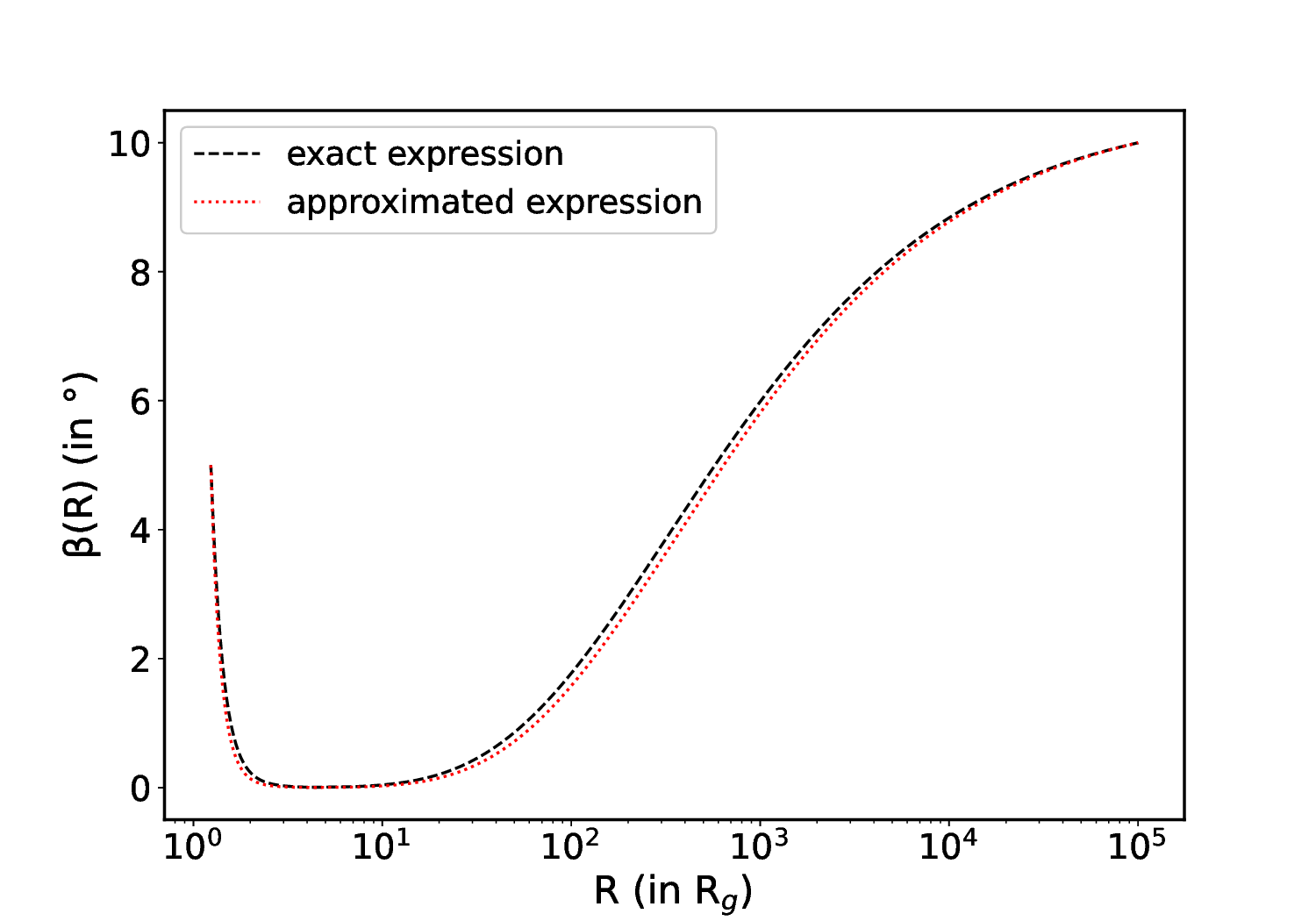}
\label{fig:sub-a}}
\end{subfigure}
\hspace{1cm}
\begin{subfigure}[]{\includegraphics[scale=0.3]{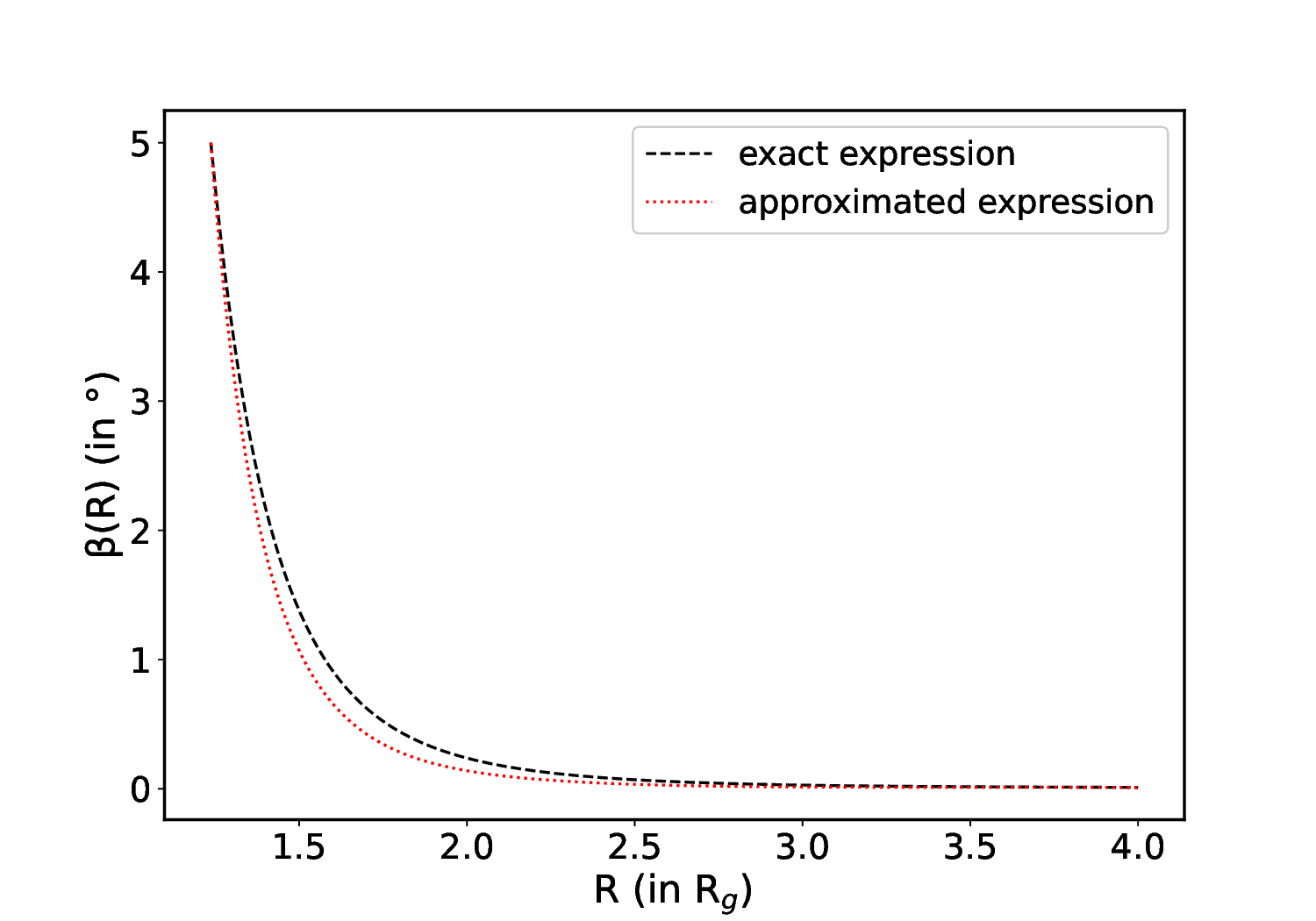}
\label{fig:sub-b}}
\end{subfigure}
\caption{Similar to Fig. \ref{plot1} but for $a=0.998$. The figure shows that the deviation between the two expressions/approaches grows markedly in the inner disk region with increasing spin. In this case, the maximum deviation is $\sim 62.5\%$ at $R \sim 1.7R_{g}$.}
\label{plot2}
\end{figure}

We compare the radial tilt profiles obtained using the approximate formulation of \cite{Banerjee_2019} with the exact expression derived in this work. Figures~\ref{plot1} and \ref{plot2} illustrate the comparison for $a=0.7$ and $a=0.998$, respectively. The deviation between the two approaches grows markedly in the inner disk region with increasing spin. For instance, the maximum deviation is $\sim 2.92\%$  at $R \sim 4.8R_{g}$ for $a=0.7$, while it increases to $62.5\%$ at $R \sim 1.7R_{g}$ for $a=0.998$. This demonstrates that the exact formulation is crucial for accurately describing the tilt profile in the strong-gravity regime and for rapidly spinning objects. As $a$ increases, the LT effect becomes stronger, extending the radial region of alignment.
In the outer disk, where the gravitational field is weak, both formulations converge, as the slow-spin approximation [Eq. (\ref{ltsl})] of LT frequency remains valid even for high-spin cases at large radii (see Sec. 4.4 of \cite{Sen_2024} for a detailed study). However, for the inner disk---particularly for Kerr NaSs ($a>1$), where $R_{\rm in}$ can fall below $R_{g}$---the slow-spin approximation fails, making the exact treatment indispensable.

Note that the inner boundary condition is intentionally held fixed in Figs.~\ref{plot1} and \ref{plot2}, as our aim is not to model the self-consistent alignment of the innermost disk with increasing spin but rather to isolate the effect of the Kerr spacetime on the global tilt profile for a given misaligned inner flow. Consequently, the tilt offset at the ISCO is not expected to vary with spin. Although the inner tilt is fixed by construction, the underlying LT torque increases with spin, leading to a stronger global warp response of the disk at higher spin.

Earlier works \cite{Banerjee_2019, Caproni_2006, Fragile_P_C_2007, Fragile_2005, Martin_2008_1, P_Natarajan_1999, Banerjee_2019_1, Nelson_2000, Li_2015, Li_2013} often applied the approximate LT frequency even for near-extremal Kerr BHs ($a \to 1$), although it is strictly valid for $a/R \ll 1$. Our analysis shows that for small $a$, both approaches yield nearly identical tilt profiles, while deviations become significant for $a \gtrsim 0.8$ and diminish at large $R$ due to the $1/R^3$ dependence of the LT effect. Therefore, the exact formalism is essential for modeling the inner accretion disk and for exploring systems with arbitrary spin values.

\subsection{Alignment radius and warp radius \label{subsection_alignment_and_warp}}

To further characterize the structure of tilted accretion disks, we examine two key spatial scales: the alignment radius and the warp radius. These serve as important diagnostics of the relative strength and spatial influence of the LT and viscous torque effects within the disk. From Secs. \ref{subsection_tiltprofile_BH} and \ref{subsection_tiltprofile_NaS}, it is clear that, even if there is a nonzero $\beta_i$, the accretion disk can gradually align as $R$ increases, depending on the values of different parameters.

The alignment radius, $R_{\rm A}$, is generally defined as the radial distance from the central collapsed object up to which the disk’s tilt angle drops below $0.01^\circ$ \cite{Banerjee_2019}. In addition, the warp radius $R_{\rm W}$ was introduced as the radial location interior to which the LT torque dominates \cite{Banerjee_2019}. According to \cite{Scheuer_P_A_G_1996, Lodato_G_2006, Sen_2024}, the warp radius is formally defined as the radius at which the LT precession timescale $(1/\Omega_p)$ becomes equal to the timescale for warp diffusion, $(R^2/\nu_2)$.
In many earlier studies, the alignment radius $R_{\rm A}$ is often approximated as being equal to the warp radius $R_{\rm W}$ \cite{P_Natarajan_1998, Martin_R_G_2008}. However, it was shown that in practice, the disk can remain significantly tilted even at $R_{\rm W}$ \cite{Banerjee_2019}. The relation between $R_{\rm A}$ and $R_{\rm W}$ was estimated as $R_{\rm A} = 0.165 R_{\rm W}$ in \cite{P_Natarajan_1999} and later refined in \cite{Banerjee_2019} to $R_{\rm A}=0.094 R_{\rm W}$.
\\

As shown in Fig.~\ref{plot3}, the radial position where the inner disk aligns with the equatorial plane is found to be almost insensitive to the initial tilt angle $\beta_i$, particularly in the case of the Kerr NaS (see Fig.~\ref{plot21}). Moreover, for appropriate parameter choices, the LT torque can dominate over the viscous torque within a certain radial interval (see Sec.~\ref{subsection_LTtorque} for details). Based on these findings, we propose a refined definition of the alignment radius: instead of a single characteristic value, we define an alignment region, bounded by two radii, $R_{\rm A1}$ and $R_{\rm A2}$, within which the tilt angle satisfies $\beta \leq 0.01^\circ$. The radial extent of this alignment region can thus be expressed as $R_{\rm A21} \equiv R_{\rm A2} - R_{\rm A1}$.

Now, according to \cite{Scheuer_P_A_G_1996}, the warp radius ($R_{\rm W}$) can be obtained by solving the following equation:
\begin{equation}
\frac{\c(R) R^{\frac{3}{2}}}{\sqrt{GM}\left[1-\sqrt{1-4a\left(\frac{R_g}{R}\right)^{\frac{3}{2}}+3a^2\left(\frac{R_g}{R}\right)^2}\right]} = \frac{R^2}{\nu_2}
\label{eqn_r_warp}
\end{equation}
for $R = R_{\rm W}$. The warp radius corresponding to different values of the Kerr parameter $a$ can be obtained by numerically solving the above equation. This procedure yields two distinct nonzero solutions, denoted as $R_{\rm W1}$ and $R_{\rm W2}$, with $R_{\rm W1} < R_{\rm W2}$. The computed values of these radii for various $a$ are listed in Tables~\ref{table1} and~\ref{table2}, assuming a fixed set of parameters.

For a Kerr BH, the smaller solution $R_{\rm W1}$ always lies within the innermost stable circular orbit ($R_{\rm in}$) and is therefore physically irrelevant. However, in the case of a Kerr NaS, there exists a threshold spin value, $a = 1.089$ \cite{Chakraborty_C_2017_1}, beyond which $R_{\rm W1}$ becomes physically meaningful, i.e., $R_{\rm W1} > R_{\rm in}$. Notably, this threshold is independent of the viscosity parameter $\nu_2$. Interestingly, $a = 1.089$ corresponds to the minimum possible ISCO radius for prograde accretion (see Fig.~\ref{varn_of_radii}). Hence, for $a > 1.089$, both inner and outer warp radii are physically relevant.

The computed warp radii are consistent with the radial tilt profiles obtained from our analysis. For example, Fig.~\ref{plot11} illustrates the tilt profiles of a Kerr NaS for various spin values, where the locations of disk alignment and warping agree well with the corresponding values in Table~\ref{table1}. From Tables~\ref{table1} and \ref{table2}, it is evident that both $R_{\rm W1}$ and $R_{\rm W2}$ increase monotonically with the spin parameter $a$.

It is important to note that the characteristic radii $R_{\rm W1}$ and $R_{\rm W2}$, derived from Eq.~(\ref{eqn_r_warp}), are mathematically independent of the boundary conditions used in the tilt equations. However, the actual manifestation of an inner warp in the tilt profile depends sensitively on the choice of boundary conditions. In particular, a nonzero torque at the inner edge ($\beta_i \neq 0^\circ$) permits the formation of an observable inner warp for both Kerr BHs and Kerr NaSs, while a zero-torque condition ($\beta_i = 0^\circ$) suppresses its appearance.

\begin{table}[h!]
\centering
\begin{tabular}{l c c c c c c c}
\hline
$a$ & $R_{\rm in}(R_{g})$ & $R_0(R_{g})$ & $R_{\rm W1}(R_{g})$ & $R_{\rm W2}(R_{g})$ & $R_{\rm A1}(R_{g})$ & $R_{\rm A2}(R_{g})$ & $R_{\rm A21}(R_{g})$ \\
\hline

0.99 & 1.45 & -- & -- & 80.8 & 3.5 & 9.0 & 5.5 \\

0.999 & 1.18 & -- & -- & 81.5 & 2.5 & 9.0 & 6.5 \\

0.9999 & 1.07 & -- & -- & 81.5 & 2.3 & 8.5 & 5.7 \\

0.99999 & 1.03 & -- & -- & 81.5 & 2.2 & 8.0 & 5.8 \\

1.0 & 1.00 & -- & -- & 81.5 & 2.0 & 9.0 & 7.0 \\

1.000001 & 0.98 & 0.56 & -- & 81.5 & 2.3 & 9.5 & 7.2 \\

1.00001 & 0.96 & 0.56 & -- & 81.5 & 2.3 & 9.5 & 7.2 \\

1.0001 & 0.93 & 0.56 & -- & 81.5 & 2.3 & 9.7 & 7.4 \\

1.001 & 0.86 & 0.56 & -- & 81.6 & 2.1 & 11.0 & 8.9 \\

1.01 & 0.75 & 0.57 & -- & 82.3 & 2.1 & 10.0 & 7.9 \\

1.02 & 0.71 & 0.59 & -- & 83.1 & 2.1 & 10.0 & 7.9 \\

1.03 & 0.70 & 0.60 & -- & 83.9 & 2.1 & 10.0 & 7.9 \\

1.04 & 0.68 & 0.61 & -- & 84.6 & 2.2 & 10.0 & 7.8 \\

1.06 & 0.67 & 0.63 & -- & 86.2 & 2.4 & 10.0 & 7.6 \\

1.08 & 0.667 & 0.66 & 0.68 & 87.7 & 2.4 & 10.0 & 7.6 \\

$\sqrt{32/27} \approx 1.089$ & 2/3 & 0.67 & 0.69 & 88.4 & 2.4 & 12.0 & 9.6 \\

1.1 & 0.67 & 0.68 & 0.71 & 89.3 & 2.4 & 12.0 & 9.6 \\

1.2 & 0.70 & 0.81 & 0.84 & 97.0 & 3.5 & 12.0 & 8.5 \\

1.4 & 0.81 & 1.10 & 1.15 & 112.1 & 3.7 & 12.0 & 8.3 \\

1.6 & 0.95 & 1.44 & 1.50 & 127.1 & 4.5 & 14.5 & 10.0 \\

1.8 & 1.10 & 1.82 & 1.90 & 141.9 & 4.8 & 17.0 & 17.0 \\

2.0 & 1.26 & 2.25 & 2.34 & 156.5 & 6.0 & 17.1 & 11.1 \\

2.5 & 1.70 & 3.51 & 3.68 & 192.2 & 7.0 & 23.0 & 16.0 \\
\hline
\end{tabular}

\caption{The ISCO radius $R_{\rm in}$, vanishing LT radius $R_0$, warp radii $R_{\rm W1}$ and $R_{\rm W2}$, alignment radii $R_{\rm A1}$ and $R_{\rm A2}$, and the radial extent of alignment $R_{\rm A21}$ are computed for a Kerr collapsed object of mass $M = 10M_{\odot}$, considering $\nu_{2} = 1 \times 10^{15}$ cm$^2$ s$^{-1}$, $\beta_i = 5^\circ$, and $z_{\rm in} = 0.75$, for various values of the Kerr parameter $a$. See Sec.~\ref{subsection_alignment_and_warp} for details.}
\label{table1}
\end{table}

\begin{table}[h!]
\centering
\begin{tabular}{l c c c c c c c}
\hline
$a$ & $R_{\rm in}(R_{g})$ & $R_0(R_{g})$ & $R_{\rm W1}(R_{g})$ & $R_{\rm W2}(R_{g})$ & $R_{\rm A1}(R_{g})$ & $R_{\rm A2}(R_{g})$ & $R_{\rm A21}(R_{g})$ \\
\hline

0.99 & 1.45 & -- & -- & 165.9 & 2.4 & 21.0 & 18.6 \\

0.999 & 1.18 & -- & -- & 167.4 & 1.9 & 21.0 & 19.1 \\

0.9999 & 1.07 & -- & -- & 167.5 & 1.8 & 21.0 & 19.2 \\

0.99999 & 1.03 & -- & -- & 167.6 & 1.7 & 21.0 & 19.3 \\

1.0 & 1.00 & -- & -- & 167.6 & 1.6 & 21.0 & 19.4 \\

1.000001 & 0.98 & 0.56 & -- & 167.6 & 1.6 & 23.0 & 21.4 \\

1.00001 & 0.96 & 0.56 & -- & 167.6 & 1.6 & 23.0 & 21.4 \\

1.0001 & 0.93 & 0.56 & -- & 167.6 & 1.6 & 23.0 & 21.4 \\

1.001 & 0.86 & 0.56 & -- & 167.7 & 1.5 & 23.5 & 22.0 \\

1.01 & 0.75 & 0.57 & -- & 169.2 & 1.5 & 23.0 & 21.5 \\

1.02 & 0.71 & 0.59 & -- & 170.8 & 1.5 & 23.0 & 21.5 \\

1.03 & 0.70 & 0.60 & -- & 172.4 & 1.5 & 23.0 & 21.5 \\

1.04 & 0.68 & 0.61 & -- & 174.0 & 1.6 & 23.5 & 21.9 \\

1.06 & 0.67 & 0.63 & -- & 177.3 & 1.7 & 23.5 & 21.8 \\

1.08 & 0.667 & 0.66 & -- & 180.5 & 1.7 & 23.5 & 21.8 \\

$\sqrt{32/27} \approx 1.089$ & 0.667 & 0.67 & 0.68 & 182.0 & 1.8 & 23.5 & 21.7 \\

1.1 & 0.67 & 0.68 & 0.69 & 183.7 & 1.9 & 23.5 & 21.6 \\

1.2 & 0.70 & 0.81 & 0.82 & 199.8 & 2.5 & 26.0 & 23.5 \\

1.4 & 0.81 & 1.10 & 1.12 & 231.8 & 2.7 & 28.0 & 25.3 \\

1.6 & 0.95 & 1.44 & 1.47 & 263.5 & 3.2 & 32.0 & 28.8 \\

1.8 & 1.10 & 1.82 & 1.86 & 295.0 & 3.7 & 35.0 & 31.3 \\

2.0 & 1.26 & 2.25 & 2.30 & 326.2 & 4.2 & 35.0 & 30.8 \\

2.5 & 1.70 & 3.51 & 3.59 & 403.1 & 5.6 & 40.0 & 34.4 \\
\hline
\end{tabular}
\caption{Similar to Table~\ref{table1} but for $\nu_{2}=5\times10^{14}$~cm$^2$~s$^{-1}$. See Sec.~\ref{subsection_alignment_and_warp} for details.}
\label{table2}
\end{table}

Tables \ref{table1} and \ref{table2} present the calculated values of the alignment radii ($R_{\rm A1}, R_{\rm A2}$) and the warp radii ($R_{\rm W1}, R_{\rm W2}$) for different spin parameters $a$, corresponding to $\nu_{2}=1 \times 10^{15}$ cm$^2$ s$^{-1}$ and $5 \times 10^{14}$ cm$^2$ s$^{-1}$, respectively. It is evident that both the radial extent of disk alignment and the region over which the LT torque dominates increase with increasing $a$. However, no clear linear relation between the alignment and warp radii, as suggested in \cite{Banerjee_2019}, can be established in the present analysis. Note that the values of $R_{\rm A21}$ in Table~\ref{table2} are larger than the corresponding values in Table~\ref{table1} for a given $a$, which is consistent with expectations since a lower value of $\nu_{2}$ leads to a greater alignment radius.

\subsection{Behavior of LT torque and viscous torques \label{subsection_LTtorque}}

\begin{figure}[ht]
\centering
\begin{subfigure}[$a=0.998$ with $R_{\rm in}=1.24R_g$]{\includegraphics[scale=0.3]{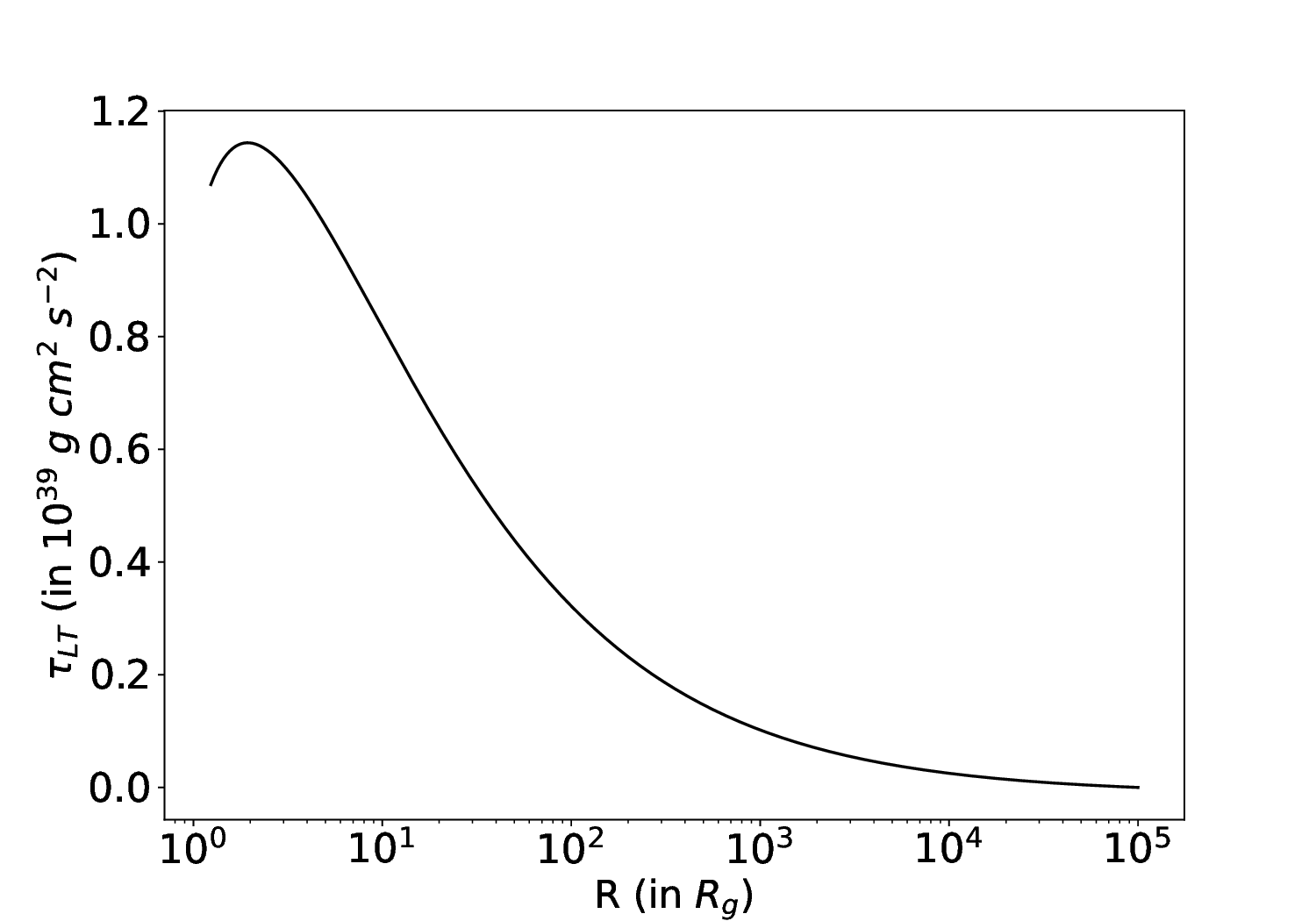}
\label{fig:sub-a}}
\end{subfigure}
\hspace{1cm}
\begin{subfigure}[$a=1.2$ with $R_{\rm in}=0.69R_g$]{\includegraphics[scale=0.3]{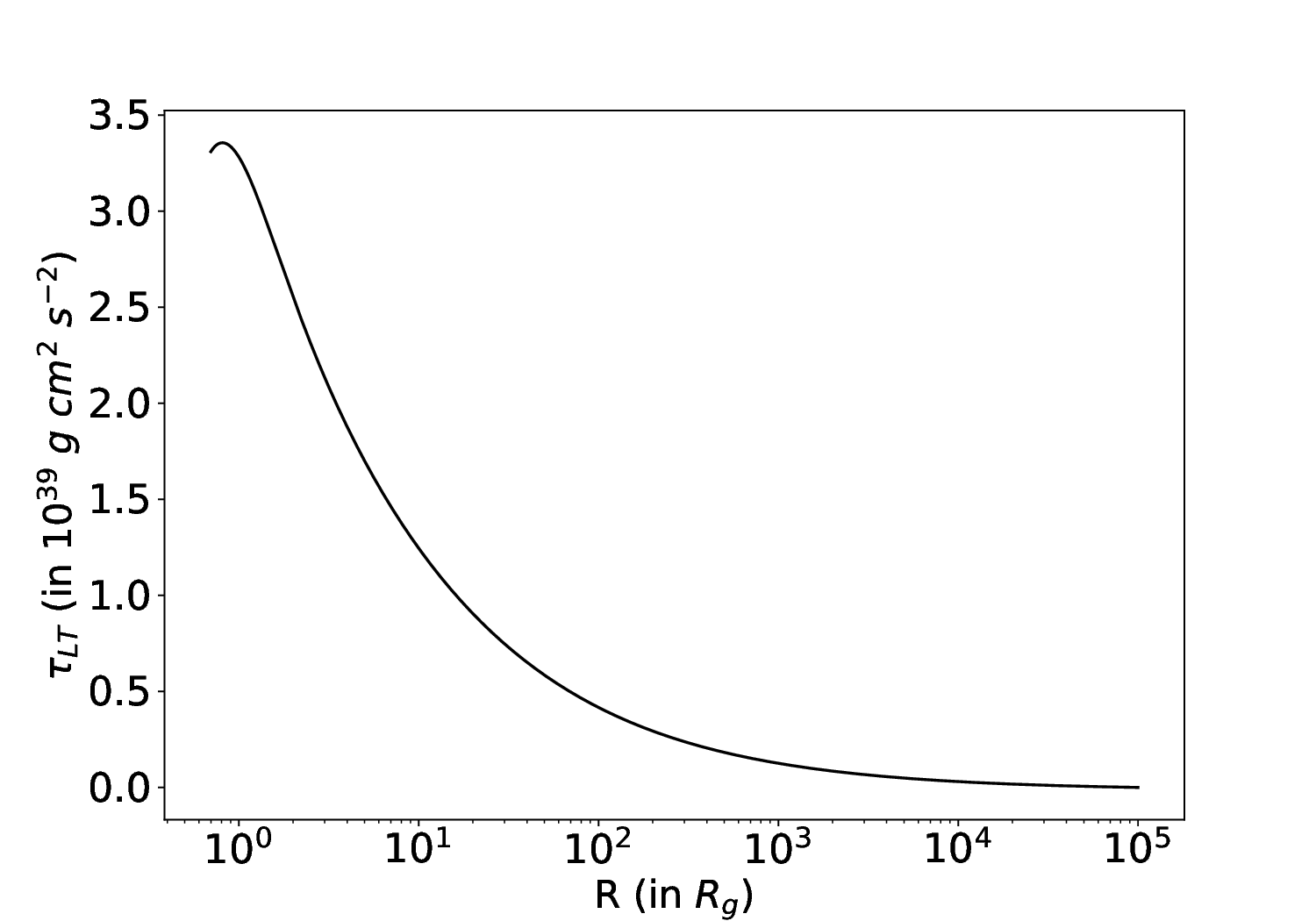}
\label{fig:sub-b}}
\end{subfigure}
\caption{Radial variation of $\tau_{\rm LT}$ obtained from the exact formalism for $M=10M_{\odot}$. The adopted parameters are: $\nu_2 = 10^{15}\,{\rm cm^2\,s^{-1}}$, $\Sigma_{\infty} = 0.75\times10^{5}\,{\rm g\,cm^{-2}}$, $\Sigma_{\rm in} = 10^{5}\,{\rm g\,cm^{-2}}$, $\beta_{\rm in} = 5^{\circ}$, and $z_{\rm in} = 0.75$. See Sec. \ref{subsection_LTtorque} for details.}
\label{plot8}
\end{figure}

Studying the radial dependence of the LT torque is essential for understanding the tilt profile obtained from the exact formulation.
We consider the accretion disk to be composed of a large number of infinitesimally thin annuli. The modulus of total LT torque ($ \tau_{\rm LT}$) acting on the entire disk can then be obtained \cite{Scheuer_P_A_G_1996} by integrating the torque contribution from each annulus,

\begin{eqnarray}
\tau_{\rm LT} &=& \bigg|\frac{d\mathbf{J}}{dt} \bigg| = \bigg| \int (\boldsymbol{\Omega}_{\rm LT} \times \mathbf{L})\, 2\pi R\, dR \bigg| \\
&=& \int \Omega_{\rm LT}(R)\, L(R)\, 2\pi R\, (-l_y,\, l_x,\, 0)\, dR \\
&=& 2\pi \int \Omega_{\rm LT}(R)\, L(R)\, R\, l \, dR
\label{ltt}
\end{eqnarray}
where $\mathbf{J} = (J_x,\, J_y,\, J_z)$ is the total angular momentum of the central collapsed object, and
$l = l_x + i l_y$ represents the complex tilt variable describing the inclination of the disk \cite{Scheuer_P_A_G_1996, Banerjee_2019_1}.

For a slowly spinning BH, an analytical expression for the LT torque can be obtained, as discussed in \cite{Scheuer_P_A_G_1996}.
However, for the general case, the evaluation of the LT torque requires numerical integration, which must be carried out for appropriately chosen parameter values.
By performing this integration from the outer to the inner radius ($R_{\rm f}$ to $R_{\rm in}$), we obtain the radial profile of the LT torque acting on the accretion disk.
Figure~\ref{plot8} shows the radial variation of $\tau_{\rm LT}$ for both a Kerr BH and a Kerr NaS.
As seen in the figure, the torque
$\tau_{LT}$ starts from a relatively low value at the inner radius $R_{\rm in}$, increases to a maximum at a larger radius, and then gradually decreases outward. The nonmonotonic behavior of the tilt profile arises as a response to the radial variation of the LT torque, together with the competing viscous torques within the disk. Note that the peak value of $\tau_{LT}$ increases with increasing spin parameter $a$.

Let us now examine the interplay between the LT torque and the viscous torques in the accretion disk. As discussed in Sec.~\ref{subsection_viscous torque}, two viscous torques act on a tilted disk \cite{Papaloizou_1983, Pringle_1992}: $\mathbf{G}_1$, associated with radial angular momentum transport, and $\mathbf{G}_2$, associated with vertical shear. In the diffusive regime, $\mathbf{G}_2$ typically provides the dominant contribution to the internal torque associated with warp evolution. The LT torque tends to align the disk with the equatorial plane, while $\mathbf{G}_2$ acts to redistribute and smooth the warp. Since $\mathbf{G}_2$ depends on $\partial \boldsymbol{l}/\partial R$, it can be evaluated by numerically solving the tilt equations (see Eq.~\ref{tilteq}). Both $\tau_{LT}$ and $\mathbf{G}_2$ are therefore sensitive to the adopted boundary conditions (see Eqs.~\ref{BC1} and \ref{BC2}).

\begin{figure}[ht]
\centering
\begin{subfigure}[]{\includegraphics[scale=0.3]{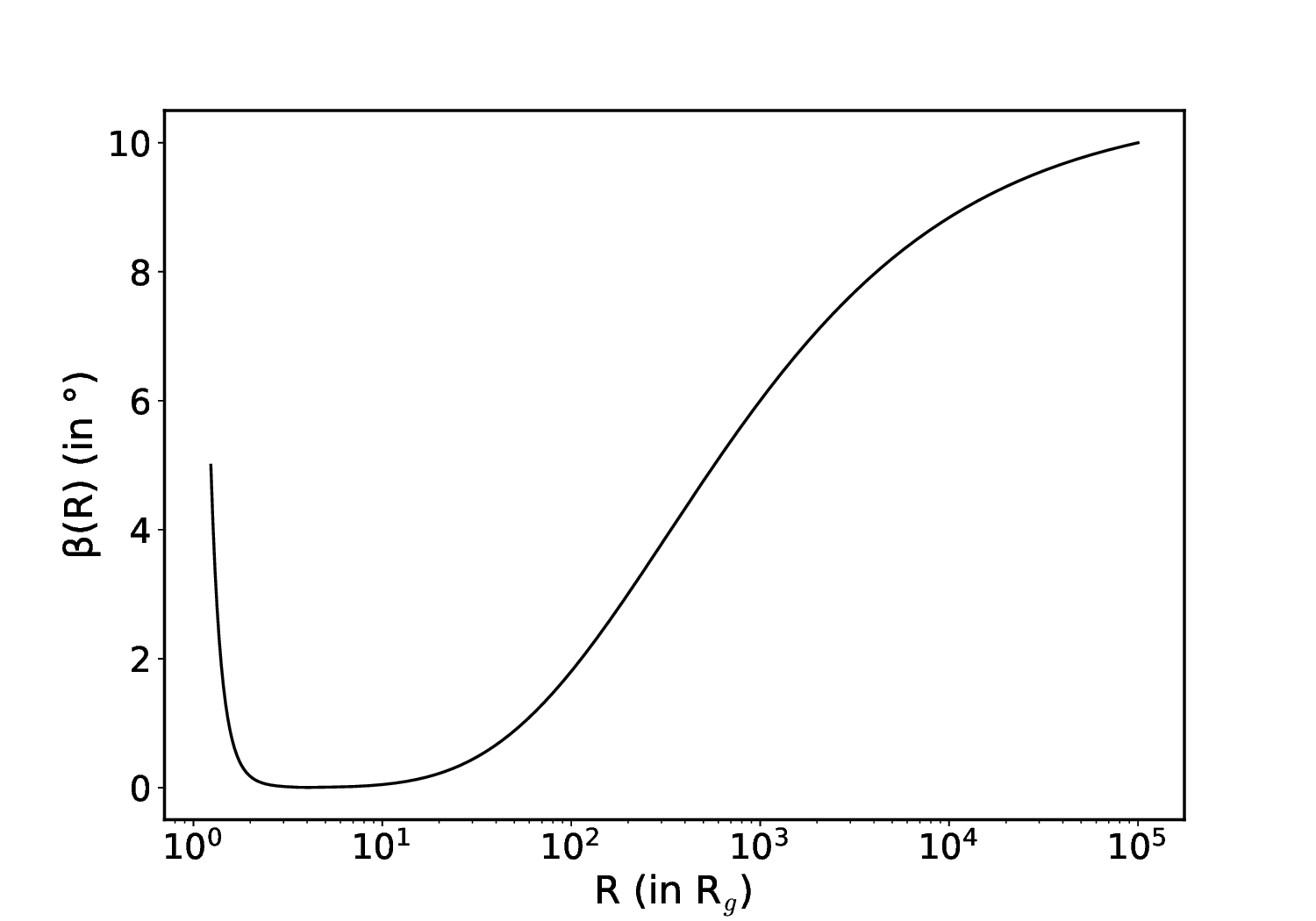}
\label{fig:sub-a}}
\end{subfigure}
\hspace{1cm}
\begin{subfigure}[]{\includegraphics[scale=0.3]{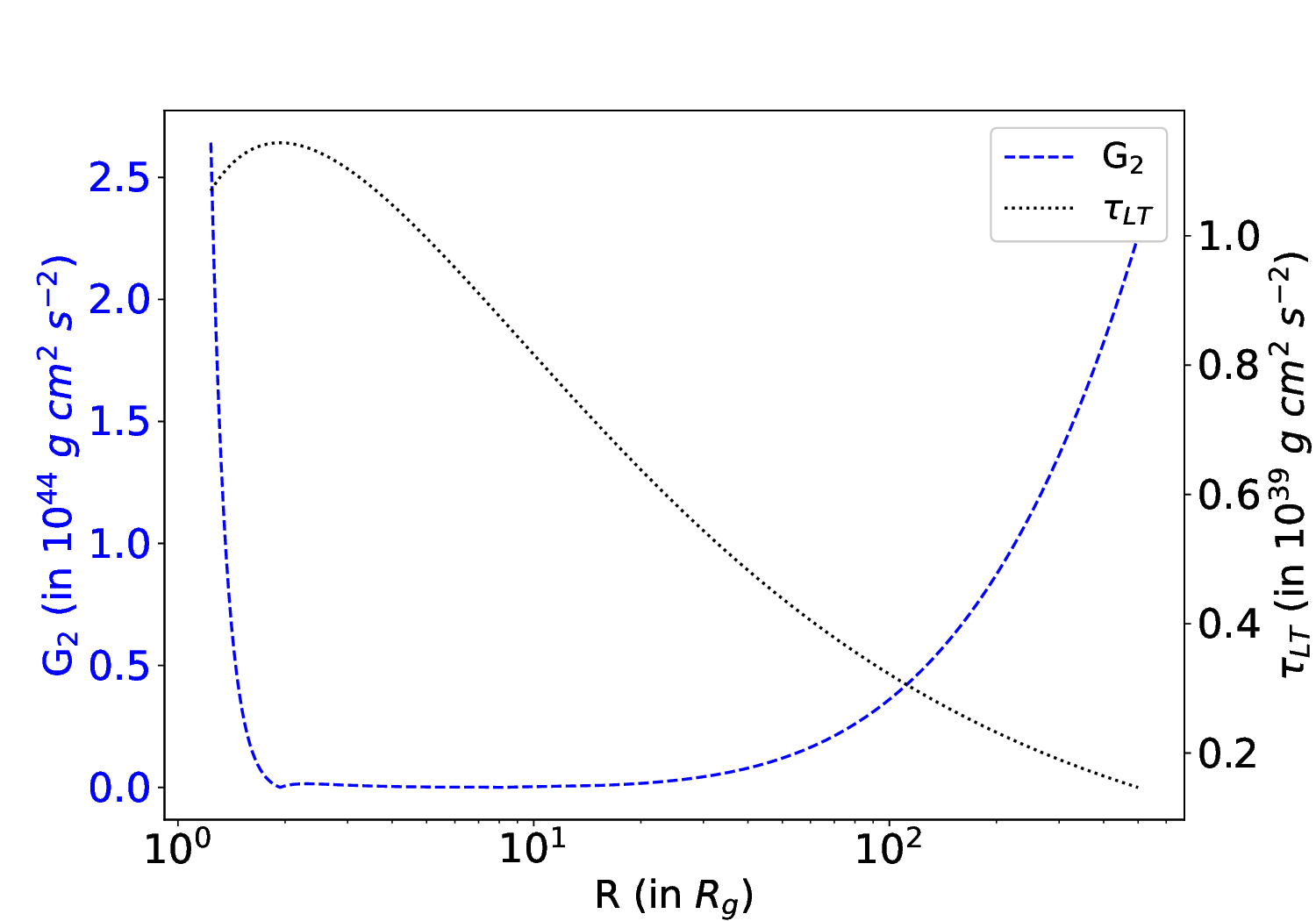}
\label{fig:sub-b}}
\end{subfigure}
\caption{Panel (a): Radial tilt profile of the accretion disk for a Kerr BH with $M = 10 M_{\odot}$ and $a = 0.998$, having $R_{\rm in} = 1.24 R_{g}$.
Panel (b): Enlarged view showing the variation of $\tau_{\rm LT}$ and $\mathbf{G}_2$ with radius $R$ for the same BH.
Adopted parameters: $\nu_2 = 10^{15}\,{\rm cm^2\,s^{-1}}$, $n = 0.25$, $z_{\rm in} = 0.75$, $\Sigma_{\rm in} = 10^5\,{\rm g\,cm^{-2}}$, and $\Sigma_{\infty} = 0.75\times10^{5}\,{\rm g\,cm^{-2}}$.
See Sec.~\ref{subsection_LTtorque} for details.}
\label{plot15}
\end{figure}

For illustration, Fig.~\ref{plot15} presents the results for a rapidly spinning Kerr BH with spin parameter $a = 0.998$. Panel (a) displays the radial profile of the disk tilt, showing that the inner part of the accretion disk progressively aligns with the BH’s equatorial plane between $R_{\rm A1} \sim 3 R_{g}$ and $R_{\rm A2} \sim 9 R_{g}$. Panel (b) depicts the corresponding radial variation of the LT torque ($\tau_{\rm LT}$) and the internal viscous torque ($\mathbf{G}_2$). Near the ISCO, the magnitude of $\mathbf{G}_2$ ($\sim 10^{44}\ {\rm g\,cm^2\,s^{-1}}$) is significantly larger (due to nonzero $\beta_i$) than that of $\tau_{\rm LT}$ ($\sim 10^{39}\ {\rm g\,cm^2\,s^{-1}}$). However, as the radius increases, $\mathbf{G}_2$ decreases rapidly, allowing the LT torque to dominate over a substantial radial range. From the plot, it is evident that $\tau_{\rm LT}$ exceeds $\mathbf{G}_2$ in the region $3 R_{g} \lesssim R \lesssim 9 R_{g}$. Within this range, the LT torque drives the disk toward alignment, whereas at larger radii, $\mathbf{G}_2$ regains dominance and the disk maintains a residual tilt. These results are consistent with the alignment radii listed in Table~\ref{table1}, confirming that a continuous interplay---or ``tug-of-war''---between the LT and viscous torques governs the gradual reduction of the disk tilt angle with increasing distance from the BH.

\begin{figure}[ht]
\centering
\begin{subfigure}[]{\includegraphics[scale=0.3]{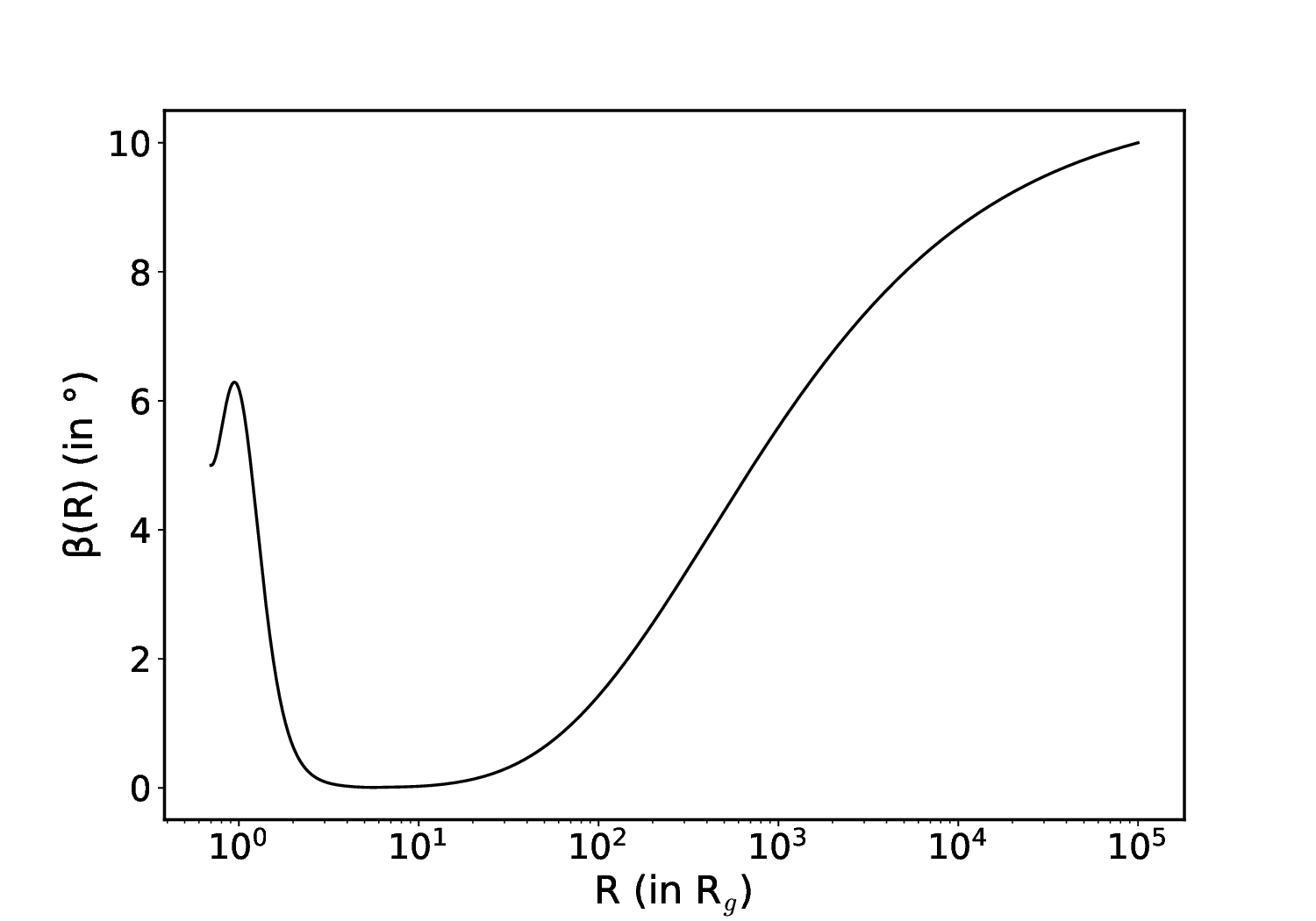}
\label{fig:sub-a}}
\end{subfigure}
\hspace{1cm}
\begin{subfigure}[]{\includegraphics[scale=0.3]{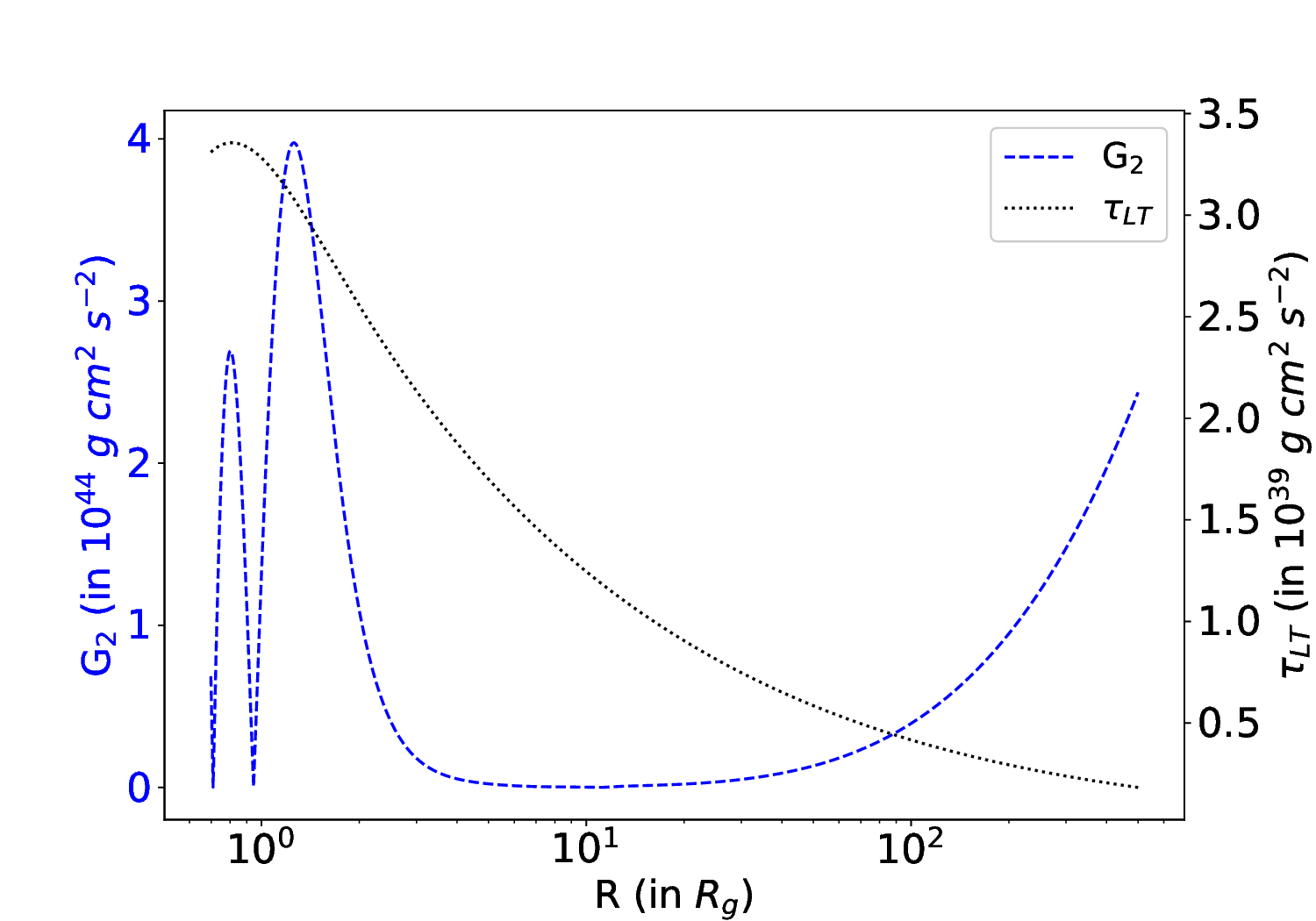}
\label{fig:sub-b}}
\end{subfigure}
\caption{Similar to Fig. \ref{plot15}, but for a Kerr NaS of $a=1.2$ with $R_{\rm in} = 0.69 R_{g}$.}
\label{plot16}
\end{figure}

Let us now focus on the case of a Kerr NaS characterized by a spin parameter $a = 1.2$, as illustrated in Fig.~\ref{plot16}. Panel (a) displays the radial variation of the disk tilt. The inner region of the disk exhibits a significant alignment with the equatorial plane, while beyond $R \sim 12 R_{g}$ the disk starts to misalign. The overall alignment region extends over a larger radial range than in the Kerr BH case, suggesting that the stronger frame-dragging in the NaS enhances the inner disk’s tendency toward coplanarity.  The minor hump (and dip) observed in the inner tilt profile is corroborated by corresponding variations in the torque profiles of $\mathbf{G}_2$ and $\tau_{\rm LT}$, as shown in panel (b).
In fact, panel (b) presents the radial profiles of $\tau_{\rm LT}$ and $\mathbf{G}_2$ for the same configuration. The dominance of the LT torque over the viscous torque term is clearly visible in the range $R \approx 4R_{g} - 12R_{g}$, facilitating efficient alignment within this zone. Beyond $R \sim 12R_{g}$, the viscous torque begins to dominate, leading to the saturation of the tilt angle. Consequently, the disk achieves alignment between $R_{\rm A1} \sim 4R_{g}$ and $R_{\rm A2} \sim 12R_{g}$, consistent with the alignment radii reported in Table~\ref{table1}. These results highlight that the enhanced spin of the Kerr NaS leads to a wider and more effective alignment zone compared to the corresponding Kerr BH configuration.

\section{Summary and CONCLUSION \label{Section_conclusion}}
This paper has presented several important and insightful
results, which we summarize as follows:

\begin{itemize}
\item Some of the qualitative features discussed here are consistent with our earlier work \cite{Banerjee_2019}, where the analysis was performed under slow-spin and weak-field approximations. In contrast, the present study is formulated entirely in the full Kerr spacetime using exact expressions for the angular momentum and the relevant dynamical frequencies (see Sec. \ref{section_Kerr}), without invoking any approximations. This enables us to derive tilt equations (Eq. \ref{tilteq}) that are valid for arbitrary spin and applicable to both Kerr BHs and Kerr NaSs within a unified framework.
The analytical expressions obtained in Sec. \ref{section_formalism} and Sec. \ref{Section_derivation} represent a general extension of our previous results, reducing to those of \cite{Banerjee_2019} in the appropriate limits. This formulation allows us to explore regimes, including high-spin Kerr BHs and $a>1$ (Kerr NaSs), which were not accessible in earlier studies, and to identify new features arising from the exact formulation.

\item The results confirm that viscous and LT torques act in concert to shape the global disk morphology, with the alignment region extending outward as the vertical viscosity decreases or the spin increases. Importantly, within realistic parameter regimes, the inner portion of the disk can remain appreciably misaligned, implying that complete alignment is not guaranteed even in steady state in the diffusive regime. We note that a misaligned inner accretion disk, manifested as hump-like features (tilt oscillations), has also been reported for Kerr BHs in the bending-wave (wave-like) regime \cite{Lubow_2002, Nealon_2015, Demianski_1997, Ivanov_1997, Drewes_2021}.

\item A particularly striking outcome arises in the Kerr NaS regime with $1<a<1.3$, where the tilt profile develops distinct hump--dip features near the radius $R_{L0}$ at which the specific angular momentum of the disk vanishes. This structure, absent in the Kerr BH case within the present framework, originates from the interplay between strong viscous torques and the vanishing of the LT torque, and provides a characteristic signature of the underlying spacetime.
The persistence of these features across a wide range of parameters suggests that the tilt distribution may offer a potential observational probe of the central object in the diffusive regime, particularly through future high-resolution X-ray and radio interferometric observations of warped inner disks. We note, however, that similar hump-like features (tilt oscillations) have also been reported for Kerr BHs in the bending-wave (wave-like) regime \cite{Lubow_2002, Nealon_2015, Demianski_1997, Ivanov_1997, Drewes_2021}. Therefore, such features alone do not provide a unique discriminator. As discussed in the last paragraph of Sec.~\ref{subsection_tiltprofile_NaS}, if independent observational constraints on the disk parameters ($\alpha$ and $H/R$) indicate that the system lies in the diffusive regime, the presence of these features may still serve as a useful diagnostic of the underlying spacetime.

\item Comparison with approximate formulations reveals that deviations from the exact treatment can exceed $60\%$ in the inner part of the disk for near-extremal spins, underscoring the necessity to develop the tilted disk formalism for the full Kerr metric. The present framework hence bridges the gap between slow-spin approximations and realistic high-spin systems, providing a rigorous foundation for studying warped and tilted accretion disks around compact objects of arbitrary spin.

\end{itemize}

The tilt of the inner accretion disk around a collapsed object can have a profound impact on both the spectral and timing properties of the emitted X-rays through LT precession, providing a powerful diagnostic of general relativistic effects in the strong-gravity regime. Observations of the Galactic BH source $\rm GRS~1915+105$ show that the $\rm Fe\,K\alpha$ line flux varies coherently with the phase of low-frequency quasiperiodic oscillations (QPOs) \cite{Miller_2005}, suggesting a direct coupling between the timing (QPO) and spectral ($\rm Fe\,K\alpha$) features---both likely driven by LT precession of a tilted inner disk. To explain this connection, \cite{Schnittman_2006} proposed a model involving an inclined, precessing ring of hot gas around a Kerr BH, which successfully reproduces observed correlations and constrains BH parameters. Subsequently,~\cite{Ingram_A_2016} reported that the centroid energy of the relativistic \(\rm Fe\,K\alpha\) line from H1743--322 exhibits systematic modulation with the QPO phase, which was interpreted in terms of LT precession of a tilted inner accretion flow within a truncated disk geometry~\cite{Ingram_A_2016_1}.

This study highlights the crucial role of exact relativistic effects in shaping the tilt morphology of accretion disks and clarifies the dynamical distinctions between Kerr BHs and NaSs in the diffusive regime. It opens a promising avenue for testing the cosmic censorship conjecture through precision measurements of disk warping and alignment in the strong-gravity regime. Moreover, as the results show that the inner disk can remain tilted over a wide and astrophysically realistic range of parameters, the derived radial tilt profiles provide a robust theoretical framework for interpreting observational signatures and probing the inner-disk geometry and spin structure of compact objects through high-resolution X-ray spectroscopy and timing. In particular, in the diffusive regime, the predicted misalignment and the characteristic hump--dip feature in the Kerr NaS regime could manifest as observable modulations in Fe K$\alpha$ line profiles, quasiperiodic oscillations, and X-ray polarization, offering a potential means to distinguish Kerr BHs from NaSs using \textit{IXPE} observations.
\\

\textbf{Acknowledgements:} KSS acknowledges the support of the DST-INSPIRE Fellowship , Govt. of India (Reference No: DST/INSPIRE/03/2023/003051, INSPIRE code: IF230417). This work is partially supported by the Dr. T. M. A. Pai Ph.D. Scholarship Program of the Manipal Academy of Higher Education (MAHE). CC acknowledges the support of MAHE. SB acknowledges financial support by the Fulbright-Nehru Academic \& Professional Excellence Award (Research), sponsored by the U.S. Department of State and the United States-India Educational Foundation (grant number: 3062/F-N APE/2024; program number: G-1-00005).  We sincerely thank the referees for the constructive comments, which have significantly improved the quality of the paper.

 \bibliography{ref}
 \bibliographystyle{apsrev4-2}

\end{document}